\RequirePackage{ifpdf,xcolor,tikz}

\documentclass[hyper,letterpaper]{JHEP3}

\pdfoutput=1
\graphicspath{{figures/}}

\usepackage{amsmath,amssymb,amsfonts,bm,empheq}
\usepackage{cite}
\usepackage{appendix}
\usepackage{amsmath}
\usepackage{youngtab}
\usepackage{ytableau}
\usepackage{graphicx}
\usepackage{mathtools}
\usepackage{float}
\usepackage{IEEEtrantools}
\usepackage[export]{adjustbox}
\usepackage{csquotes}
\usepackage{cancel}
\usepackage{tikz-cd} 
\usepackage{tkz-graph}
\usepackage{lscape}
\usepackage{comment}

\GraphInit[vstyle = Shade]
\tikzset{
  LabelStyle/.style = { rectangle, rounded corners, draw,
                        minimum width = 1em, fill = yellow!50,
                        text = red, font = \bfseries },
  VertexStyle/.append style = { inner sep=1pt,
                                font = \small},
  EdgeStyle/.append style = {->, bend left} }

\newcommand{\Mod}[1]{\ (\mathrm{mod}\ #1)}

\newtheorem{theorem}{Theorem}[section]

\newtheorem{corollary}[theorem]{Corollary}
\newtheorem{proposition}{Proposition}[section]
\newtheorem{definition}{Definition}[section]
\newtheorem{conjecture}{Conjecture}[section]

\title{Combinatorics of Nahm sums, quiver resultants and the K-theoretic condition}

\author{Dmitry Noshchenko$^{1}$
\\
$^1$ Faculty of Physics, University of Warsaw, ul. Pasteura 5, 02-093 Warsaw, Poland
\email{dsnoshchenko@gmail.com}}

\abstract{  
Algebraic Nahm equations, considered in the paper, are polynomial equations, governing the
$q\rightarrow 1$ limit of the $q$-hypergeometric Nahm sums.
They make an appearance in various fields: hyperbolic geometry, knot theory, quiver representation theory, topological strings
and conformal field theory. In this paper we focus primarily on Nahm sums and Nahm equations that arise in relation with symmetric quivers.
For a large class of them, we prove that quiver A-polynomials -- specialized resultants of the Nahm equations, are tempered (the so-called K-theoretic condition). This implies that they are quantizable. Moreover, we find that their face polynomials obey a remarkable combinatorial pattern. We use the machinery of initial forms and
mixed polyhedral decompositions to investigate the edges of the Newton polytope.
 We show that this condition holds for the diagonal quivers with adjacency matrix $C = \mathrm{diag}(\alpha,\alpha,\dots,\alpha),\ \alpha\geq 2$, and provide several checks for non-diagonal quivers. Our conjecture is that the K-theoretic condition holds for all symmetric quivers.
}

\begin{document}

\tikzset{every picture/.style={line width=0.75pt}} 

\section{Introduction}

Algebraic Nahm equations\footnote{Not to be confused with Nahm equations in gauge theory and differential geometry.} govern the $q\rightarrow 1$ limit of the $q$-hypergeometric Nahm sums, which arise in various fields: conformal field theory \cite{NRT}, quiver representation theory \cite{KS2008,KS2011,R}, hyperbolic geometry and ideal triangulations of 3-manifolds
\cite{DG,GZ}, knots-quivers correspondence \cite{KRSS1,KRSS2,EKL,EKL2, PSS} and topological strings \cite{PS, EKL,EKL2}. 
In the realm of quivers, the Nahm sums incarnate as the motivic Donaldson-Thomas (DT) generating series \cite{KS,KS2008,KS2011,R,E}:
\begin{equation}\label{quiver_series1}
    P_C(x_1,\dots,x_m) = \sum_{(d_1, \dots, d_m)\geq 0}\frac{(-q^{1/2})^{\sum_{i,j=1}^m C_{ij} d_i d_j}}{(q; q)_{d_1}\cdots (q;q)_{d_m}} x_1^{d_1} \cdots x_m^{d_m},
\end{equation}
where $C$ is symmetric matrix with integer entries, $q\in \mathbb{C}$ and $x_i$ are formal variables which commute with each other, and $(a;q)_n := \prod_{k=0}^{n-1}(1-aq^k)$ is the $q$-Pochhammer symbol.
If $C_{i,j}$ are non-negative,
the matrix $C$ is the adjacency matrix for some symmetric quiver. Otherwise we can apply the framing transformation $C \mapsto C + [f]$, where $[f]$ is a matrix with all values equal to $f\in \mathbb{Z}$, in order to get rid of the negative entries. It transforms the quiver series (\ref{quiver_series1}) in a simple way \cite{LNPS}.
For a curious reader, we sketch the derivation of (\ref{quiver_series1}) from the quiver representation theory in Section \ref{section_quiver_series}.
The crucial property of (\ref{quiver_series1}) is the following factorization\footnote{This formula is the cornerstone in \cite{KS2008}, which led to the mathematical theory of BPS invariants in 3d $\mathcal{N}=2$ theories, using quivers and their representations.} \cite{KS2008,R,E,PS}:
\begin{equation}
    P_C(x_1,\dots,x_m) = \prod_{(d_1,\dots,d_m) \neq 0}\prod_{j\in \mathbb{Z}}\prod_{k\geq 0} (1-q^{k+(j-1)/2}x_1^{d_1}\dots x_m^{d_m})^{\Omega_{d_1,\dots,d_m;j}}
\end{equation}
The exponents $\Omega_{d_1,\dots,d_m;j}$ are called the motivic DT invariants (or refined BPS invariants in physics), and were shown to be integers in \cite{E}. Consider the Laurent expansion at $q\rightarrow 1$ of the saddle point approximation to the logarithm of (\ref{quiver_series1}):
\begin{equation}\label{quiver_series_hbar}
    \left.\log P_c(x_1,\dots,x_m)\right|_{q=\mathrm{e}^{\hbar}\rightarrow 1} = \frac{1}{\hbar}S_0 + S_1 + \hbar S_2 + O(\hbar^2),
\end{equation}
where $S_i = S_i(x_1,\dots,x_m,z_1\dots,z_m)$ and $z_i := q^{d_i},\ i=1\dots m$. The (algebraic) Nahm equations arise from the critical points of the leading term (superpotential) in (\ref{quiver_series_hbar}): $\frac{\partial S_0}{\partial z_i} = 0$ implies
\begin{equation}\label{clNahm_eqs}
    F_i := z_i -1 + (-1)^{C_{i,i}}x_i \prod_{j=1}^m z_j^{C_{i,j}} = 0,\quad i=1\dots m
\end{equation}
 (see \cite{PS,LNPS} for the details). We add one extra equation:
\begin{equation}
F_0 := y -z_1 \dots z_m = 0,
\end{equation}
in order to introduce the quiver resultant $A(x_1,\dots,x_m,y) := \mathrm{res}_{z_1,\dots,z_m}(F_0,F_1,\dots,F_m)$. It is a unique (up to a sign) irreducible polynomial in the coefficients of (\ref{clNahm_eqs}), which vanishes whenever $F_0,F_1,\dots,F_m$ have a common root with respect to $z_i,\ i=1\dots m$. We will utilize a slightly refined version of the quiver resultant starting from Section \ref{section_the_main_conjecture}, which is useful for our combinatorial study.

Recall that the quiver A-polynomial is a two-variable specialization of the quiver resultant:
\begin{equation}
A(x,y) = A(\lambda_1 x,\dots,\lambda_m x,y),\quad \lambda_i \in \mathbb{C}\setminus \{0\}
\end{equation}
It has been introduced in \cite{PSS} and further studied in \cite{KRSS2} and \cite{LNPS}. Ultimately, it is a polynomial invariant of symmetric quivers.
Under a suitable choice of the quiver matrix $C$ and parameters $\lambda_i$, it can be related to augmentation variety or geometric A-polynomial for a knot \cite{KRSS2}. Also, from the mirror symmetry perspective, quiver A-polynomials may serve as the mirror curves (B-model) for some Calabi-Yau 3-folds (A-model). The case of strip geometries was studied in \cite{PS}, whereas the relation to Ooguri-Vafa large $N$ duality in \cite{EKL} and \cite{EKL2}.

Our object of interest is the Newton polygon $N(A)$, that is, the convex hull of all monomials of $A(x,y)$. 
%
We conjecture that $A(x,y)$ is tempered, i.e. all its face polynomials have roots only on the unit circle, for any symmetric quiver.
By a face polynomial we simply mean the sum all monomials in $A(x,y)$, which lie on a particular face of $N(A)$.
This is called the \emph{K-theoretic condition}, because of an elegant interpretation in terms of the group $K_2$ for a compact Riemann surface (in our case it is given by $A(x,y) = 0$) \cite{BRVD,RV,GS}. It turns out that this condition relates to quantization, modularity and integrality properties for $A(x,y)$. It is confirmed true for all knot A-polynomials \cite{CCGLS}, but, to our knowledge, has not been studied for quivers so far. In particular, it predicts the existence of a $q$-difference operator (``quantum curve'', or non-commutative A-polynomial \cite{GS}) which annihilates the associated partition function, and its $q\rightarrow 1$ limit gives back $A(x,y)$.

It is important to say a few words why we expect the conjecture to be true for all symmetric quivers. From \cite{EKL,EKL2,LNPS} we know that the quiver series (\ref{quiver_series1}) for any symmetric matrix $C$ with integer entries are annihilated by the quantized version of Nahm equations:
\begin{equation}\label{qNahm}
    \left(1-\hat{z}_i\right)P_C = \left((-1)^{C_{i,i}}x_i\prod_{j = 1}^m \hat{z}_j^{C_{i,j}}\right)P_C
\end{equation}
where $\hat{z}_i$ acts as follows: $\hat{z}_ix_j = q^{\delta_{i,j}}x_j$ ($\delta_{i,j}$ is the Kronecker delta). This can be re-written in operator form:
\begin{equation}\label{qNahm_operators}
    \hat{A}_i(x_1,\dots,x_m,\hat{z}_1,\dots,\hat{z}_m) P_C = 0,\ i=1\dots m
\end{equation}
It is therefore suggested that if we perform non-commutative elimination for the system (\ref{qNahm_operators}) with respect to $\hat{z}_1,\dots,\hat{z}_m$, we get a single $q$-difference operator $\hat{A}$,  which is a non-commutative polynomial in $x_1\dots,x_m,\widehat{y}$:
\begin{equation}\label{quantum_quiver_poly}
    \hat{A}(x_1,\dots,x_m,\hat{y}) P_C = 0,\quad \hat{y}:= \hat{z}_1 \dots \hat{z}_m
\end{equation}
Thus, the existence of a ``quantum hypersurface'' is expected
for any symmetric quiver\footnote{Note that the explicit calculations are hard to perform -- it has been achieved in fact only for a few families of quivers in \cite{LNPS}.}. The latter gives the quantum curve by setting $x_i = \lambda_ix$. However, it is not obvious at all if the formally constructed quantum curve from \cite{GS} would agree with the eliminant $\hat{A}$ from (\ref{qNahm_operators}).

On another hand, the physical point of view interprets quiver series as a partition function of a (0-dimensional) quiver supersymmetric quantum
mechanics \cite{KS2008,KS2011,EKL,EKL2}. Such supersymmetric quantum mechanics may arise as an
effective description of some (4-dimensional) SUSY theory, which can be
realized in brane systems. In this context, DT invariants captured by
the quiver generating series correspond to BPS invariants in such a
brane system. And if such BPS states can be encoded in a quiver
generating series, then it means that they can also be encoded in a
quantum quiver A-polynomial (\ref{quantum_quiver_poly}), which therefore must exist, and thus the
classical A-polynomial must be quantizable.
In some cases we know
explicit examples of such brane systems and effective descriptions in
terms o quivers (e.g. corresponding to systems of branes that encode
knots \cite{KRSS1,KRSS2,EKL,EKL2}, or for
strip geometries \cite{PS}). It is natural to expect
that other (all possible) quivers also provide effective description
of some brane systems, and thus corresponding A-polynomials should
also be quantizable.

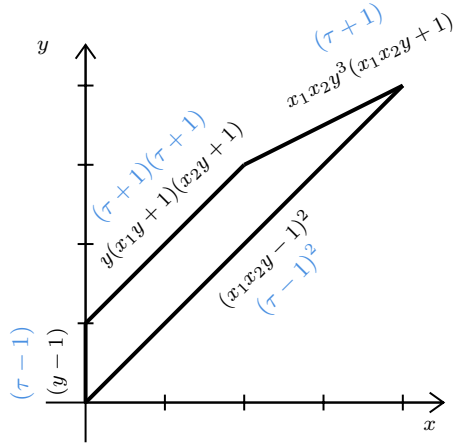
\begin{figure}[h!]
\centering
\begin{tikzpicture}[x=0.75pt,y=0.75pt,yscale=-1,xscale=1]

\draw  (138.94,940) -- (339.54,940)(159,759.46) -- (159,960.06) (332.54,935) -- (339.54,940) -- (332.54,945) (154,766.46) -- (159,759.46) -- (164,766.46)  ;
\draw [line width=1.5]    (159,900) -- (159,940) ;
\draw [fill={rgb, 255:red, 0; green, 0; blue, 0 }  ,fill opacity=1 ][line width=1.5]    (239,820) -- (159,900) ;
\draw [fill={rgb, 255:red, 0; green, 0; blue, 0 }  ,fill opacity=1 ][line width=1.5]    (319,780) -- (239,820) ;
\draw [fill={rgb, 255:red, 0; green, 0; blue, 0 }  ,fill opacity=1 ][line width=1.5]    (319,780) -- (159,940) ;
\draw [fill={rgb, 255:red, 0; green, 0; blue, 0 }  ,fill opacity=1 ]   (159,940) -- (159,760) (155,900) -- (163,900)(155,860) -- (163,860)(155,820) -- (163,820)(155,780) -- (163,780) ;

\draw    (159,940) -- (339,940) (199,936) -- (199,944)(239,936) -- (239,944)(279,936) -- (279,944)(319,936) -- (319,944) ;

\draw (328,948) node [anchor=north west][inner sep=0.75pt]  [font=\scriptsize]  {$x$};
\draw (133,756) node [anchor=north west][inner sep=0.75pt]  [font=\scriptsize]  {$y$};
\draw (221.41,886.24) node [anchor=north west][inner sep=0.75pt]  [font=\scriptsize,rotate=-315]  {$( x_{1} x_{2} y-1)^{2}$};
\draw (161.36,866) node [anchor=north west][inner sep=0.75pt]  [font=\scriptsize,rotate=-315]  {$y( x_{1} y+1)( x_{2} y+1)$};
\draw (138,938) node [anchor=north west][inner sep=0.75pt]  [font=\scriptsize,rotate=-270]  {$( y-1)$};
\draw (255.86,782.6) node [anchor=north west][inner sep=0.75pt]  [font=\scriptsize,rotate=-333]  {$x_{1} x_{2} y^{3}( x_{1} x_{2} y+1)$};
\draw (240.1,889.25) node [anchor=north west][inner sep=0.75pt]  [font=\footnotesize,color={rgb, 255:red, 74; green, 144; blue, 226 }  ,opacity=1 ,rotate=-315]  {$( \tau -1)^{2}$};
\draw (270.13,753.21) node [anchor=north west][inner sep=0.75pt]  [font=\footnotesize,color={rgb, 255:red, 74; green, 144; blue, 226 }  ,opacity=1 ,rotate=-333]  {$( \tau +1)$};
\draw (119.5,940.5) node [anchor=north west][inner sep=0.75pt]  [font=\footnotesize,color={rgb, 255:red, 74; green, 144; blue, 226 }  ,opacity=1 ,rotate=-270]  {$( \tau -1)$};
\draw (158.1,843.25) node [anchor=north west][inner sep=0.75pt]  [font=\footnotesize,color={rgb, 255:red, 74; green, 144; blue, 226 }  ,opacity=1 ,rotate=-315]  {$( \tau +1)( \tau +1)$};

\end{tikzpicture}
\caption{Newton polygon $N(A)$ and face polynomials for $\mathrm{diag}(2,2)$ quiver}
\label{fig:diag22_newton}
\end{figure}

Our main result is that for a diagonal quiver with $C = \mathrm{diag}(\underbrace{\alpha,\alpha,\dots,\alpha}_{m}),\ m\geq 2, \alpha\geq 2$, $A(x,y)$ is tempered (for the one-vertex quiver, the problem has been solved in \cite{GS}). Moreover,
all its face polynomials factorize into binomials, forming a remarkable combinatorial pattern.
This is the content of Section \ref{section_arbitrary_dimension}, and Theorem \ref{main_prop_diagonal} in particular. The beautiful combinatorial pattern is given in Proposition \ref{initial_forms_diag_arbirary}. It involves permutations of rows and columns of diagrams, representing the sub-resultants. One can think of it as a ``cellular automation'' acting on the faces of the Newton polytope.
To understand the mechanism better (and also for a nicer presentation), we study the low-dimensional cases $m = 2$ and $m = 3$ separately in Sections \ref{section_two_dimensional}  and \ref{section_three_dimensional}. E.g., for $\mathrm{diag}(2,2)$ there are four face polynomials: $\tau+1,\tau-1,(\tau+1)^2$ and $(\tau-1)^2$, and all their roots are equal to $\pm1$, as shown on Figure \ref{fig:diag22_newton}.

The key point is that we don't have to compute the resultant explicitly. Instead, we use the machinery of initial forms \cite{Stu,PetStu} and mixed polyhedral decompositions, developed in \cite{Stu} and \cite{KSZ}. These guys generalize extremal A-polynomials from knot theory \cite{GKS,FGS,KRSS1,KRSS2}, and, under certain assumption, are in bijection with the faces of $N(A)$. As a consequence, we obtain the ``extremalization'' of quiver A-polynomials, provided by a particular face of $N(A)$.

Lastly, the two appendices \ref{append_a} and \ref{append_b} are devoted to experimantal confirmations and computation of quiver resultants using Canny-Emiris matrix \cite{CE}.

\section*{Acknowledgement}

We thank Piotr Kucharski, H{\'e}lder Larragu{\'i}vel, Mi\l osz Panfil, Motohico Mulase, Markus Reineke and Piotr Su\l kowski for many fruitful discussions and comments; Ioannis Emiris and Matthias Franz for the provided software. This  work  has  been  supported  by the TEAM programme of the Foundation for Polish Science co-financed by the European Union under the European Regional Development Fund (POIR.04.04.00-00-5C55/17-00).

\section{Quiver representations and motivic DT series}\label{section_quiver_series}

We begin with the origins of the motivic DT series (\ref{quiver_series1}) from the quiver representation theory perspective \cite{KS,R,KS2008,B,S}.

\emph{Quiver} is a directed finite graph $Q = (Q_0,Q_1,h,t)$, where $Q_0,Q_1$ are the sets of vertices and arrows, and $h,t$ are the maps from $Q_1$ to $Q_0$, picking up a head or a tail vertex for a given arrow. For example, if there are two vertices $i,j$ connected by an arrow $a$ from $i$ to $j$, we write $ta=i,\ ha=j$.

The term ``quiver'' is used instead of ``graph'', since one also considers a \emph{quiver representation}: to every vertex $i \in Q_0$ it associates a finite-dimensional vector space $V(i)$ over a field $F$ (for example, one can take the field $\mathbb{C}$ of complex numbers), and to every arrow $a\in Q_1$ a linear map $f: V(ta)\rightarrow V(ha)$. Therefore, any representation is characterised by its \emph{dimension vector} $d = (d_1,\dots,d_m) \in \mathbb{Z}^m,\ d_i = \dim V(i)$. Without loss of generality, we take $V(i)=F^{d_i},\ \forall i\in Q_0$. Also, we will write $V(a)$ instead of $f(a)$, abusing the notation a bit.

We are not interested in particular representations, but rather in the \emph{representation space} of a fixed dimension vector $d$:
\begin{equation}\label{rep_space}
\mathrm{Rep}_d(Q):=\prod_{a\in Q_1}\mathrm{Mat}_{d(ha),d(ta)},
\end{equation}
where $\mathrm{Mat}_{m,n}$ is the space of all $m$ by $n$ matrices (with entries in $F$).
Since any element of $\mathrm{Mat}_{d(ha),d(ta)}$ is equal to $V(a)$ for some $V$, each point of (\ref{rep_space}) defines a representation of $Q$. Every representation $V$ has its own \emph{group of automorphisms} $\mathrm{Aut}(V)$, defined as the orbit of $G:= \prod_{i=1}^m \mathrm{GL}_{d_i}(F)$. The group $G$ acts on the points of (\ref{rep_space}) via conjugation:
\begin{equation}\label{rep_orbit}
    (g)(V(a)) := (g_j V(a) g_i^{-1})_{(a:i\rightarrow j)},\quad \forall\ a \in Q_1,\ g\in G
\end{equation}
By definition, the orbits of $G$ in $\mathrm{Rep}_d(Q)$ are precisely the \emph{isomorphism classes}
of quiver representations of $Q$ of dimension vector $d$ (two representation are said to be isomorphic if they are related by a change of bases of $F^{d_i},\ i=1\dots m$, which amounts to conjugation (\ref{rep_orbit})).

Now that we have defined all basic notions, it's time to count. Assume that our representations are  over a \emph{finite field} $\mathbb{F}_q$, where $q = p^r$ and $p$ is prime. These are non-negative integers modulo $q$, i.e. $\mathbb{F}_q=\{0,1,2,\dots,q-1\}$ with modular multiplication.
For a fixed $d$, denote 
\begin{equation}
    s_d = \sum_{[V],\ \dim V = d}\frac{1}{|\mathrm{Aut}(V)|},
\end{equation}
where the summation is over all isomorphism classes $[V]$ of representations with dimension vector $d$, and $|\mathrm{Aut}(V)|$ is the size of the corresponding automorphism group.
It is of course also finite, since we are dealing with a finite field.
Since $V(i)=\mathbb{F}_q^{d_i},\ \forall i\in Q_0,$ the total number of representations of dimension vector $d$ is $q^{\sum_{a\in Q_1}d_id_j} = q^{\sum_{i,j=1}^m C_{ij}d_id_j}$. On another hand, the number of points in the 
orbit of $V$ is $ \frac{|G|}{|\mathrm{Aut}(V)|}$. Therefore,
\begin{equation}
    q^{\sum_{i,j=1}^m C_{ij}d_id_j} = \sum_{[V],\ \dim V = d} \frac{|G|}{|\mathrm{Aut}(V)|},
\end{equation}
which gives
\begin{equation}\label{poincare_poly}
    s_d = \frac{q^{\sum_{i,j=1}^m C_{ij}d_id_j}}{\prod_{i=1}^m |\mathrm{GL}_{d_i}(\mathbb{F}_q)|},
\end{equation}
where $C$ is the adjacency matrix of $Q$, and $|\mathrm{GL}_n(\mathbb{F}_q)| = (q^n-1)(q^n-q)\dots (q^n-q^{n-1}) = (-q)^{\frac{n(n-1)}{2}}(q;q)_n$. The latter equality comes from counting of all admissible columns of an element in $\mathrm{GL}_n(\mathbb{F}_q)$. The first row can anything but zero vector, hence the factor $(q^n-1)$, the second row can be anything but the multiple of the first one, hence $(q^n-q)$, and so on.
Finally, we assemble the \emph{generating series}:
\begin{equation}\label{quiver_series}
    \sum_{(d_1,\dots,d_m)\geq 0}s_d\ x_1^{d_1}\dots x_m^{d_m}
\end{equation}
where $x_i$ are formal variables.
It is easy to see that (\ref{quiver_series}) coincides with (\ref{quiver_series1})
after transformation $x_i^{d_i} \mapsto (-q)^{\frac{d_i(d_i-1)}{2}}x_i^{d_i}$. This can be achieved by introducing the quantum torus variables \cite{KS2008,KS2011}. Therefore, the coefficients of $P_C$ can be interpreted as the Euler characteristics of (the ordinary cohomology of) $\mathrm{Rep}_d(Q)$, bearing the name of motivic DT generating series. If we take $q$ to be an arbitrary complex number, $s_d$ would have poles at the unit circle, due to the $q$-Pochhammer symbols in the denominator. Therefore, the perturbative expansion of (\ref{quiver_series}) at $q = 1$ will eventually lead to the Nahm equations (\ref{clNahm_eqs}).

\section{Algebraic K-theory and tempered polynomials}\label{section_k-theory}

Roughly speaking, the algebraic K-theory is about a study of the family of functors $K_n:\ \text{Rings} \rightarrow \text{Abelian groups}$ (it was invented to produce nice invariants of rings). $K_0,K_1$ and $K_2$ are classically known from the sixties. Higher $K$-groups, as well as those with the negative index, were defined in the following decades. However, our main character is the group $K_2(F)$, where $F$ is a field.
 The exposition here is mostly borrowed from Milnor's classical book \cite{Mil}.
We start with a rather informal definition:
\begin{equation}
\begin{aligned}
K_2(F) := &\ \text{a group of non-trivial relations satisfied by elementary matrices of any size} \\
&\ \text{with entries in $F$}
\end{aligned}
\end{equation}
Recall that elementary matrix is a matrix $e^{\lambda}_{ij}\in \mathrm{GL}_n(F)$, which differs from the identity matrix of size $n$ by a single element $\lambda$ in the $(i,j)$-th position, $i,j=1\dots n$, or a matrix obtained from such by elementary row operations. In other words, we can say that $e^{\lambda}_{ij}$ generate the subgroup of elementary matrices, sitting in $\mathrm{GL}_n(F)$.
If $e^{\lambda}_{ij},e^{\mu}_{kl}$ are elementary matrices, their commutator is
\begin{equation}
    [e^{\lambda}_{ij},e^{\mu}_{kl}] =
    \begin{cases}
               1;\ j\neq k,\ i\neq l \\
               e^{\lambda\mu}_{il};\ j = k,\ i\neq l \\
               e^{-\mu\lambda}_{kj};\ j\neq k,\ i = l
\end{cases}
\end{equation} 
We can forget for a moment about matrices, and consider an abstract group generated by the relations:
\begin{equation}
    \begin{aligned}
    x_{ij}^{\lambda}x_{ij}^{\mu} = &\ x_{ij}^{\lambda+\mu} \\
    [x_{ij}^{\lambda},x_{jl}^{\mu}] = &\ x_{il}^{\lambda\mu};\ i\neq l \\
    [x_{ij}^{\lambda},x_{kl}^{\mu}] = &\ 1;\ j\neq k,\ i\neq l
    \end{aligned}
\end{equation}
These relations define Steinberg group, denoted by $\mathrm{St}(n,F)$ for $n\geq 3$ (for $n<3$ the relations degenerate).

For each $n\geq 3$ we have a homomorphism of groups:
\begin{equation}
\psi:\    \mathrm{St}(n,F) \rightarrow \mathrm{GL}_n(F),
\end{equation}
which associates an elementary matrix of size $n$ to each element of $\mathrm{St}(n,F)$: $\psi(x^{\lambda}_{ij}) = e^{\lambda}_{ij}$.
Now we can pass through the direct limit of a sequence of groups when $n\rightarrow \infty$, denoting it $\mathrm{GL}(F)$, which is understood as follows:
\begin{equation}\label{glinf_seq}
    \mathrm{GL}_1(F) \subset \mathrm{GL}_2(F) \subset \mathrm{GL}_3(F) \subset \dots,
\end{equation}
and each $\mathrm{GL}_n(F)$ is injected into $\mathrm{GL}_{n+1}(F)$ by the map:
\begin{equation}
    * \mapsto \begin{pmatrix}
    * & 0 \\
    0 & 1
    \end{pmatrix},\quad \forall * \in \mathrm{GL}_n(F)
\end{equation}
Therefore, $\mathrm{GL}(F)$ is determined by taking the union of all elements in the infinite sequence (\ref{glinf_seq}). Analogously, one can define $\mathrm{St}(F)$.
In what follows is the formal definition of $K_2(F)$: 
\begin{equation}
K_2(F) := \text{Kernel of the map }\psi: \mathrm{St}(F) \rightarrow \mathrm{GL}(F),
\end{equation}
where the kernel elements are mapped to an identity matrix in $\mathrm{GL}(F)$. Let's show this by example: pick up a rotation by 90 degrees matrix, which is elementary:
\begin{equation}
e^1_{12}e^{-1}_{21}e^1_{12} = 
    \begin{pmatrix}
        0 & 1 \\
        -1 & 0
    \end{pmatrix}
\end{equation}
and is decomposed as a product of the generators $e^{\lambda}_{ij}$. This matrix has period 4:
\begin{equation}\label{the_relation_example}
    (e^1_{12}e^{-1}_{21}e^1_{12})^4 = 
        \begin{pmatrix}
        1 & 0 \\
        0 & 1
    \end{pmatrix}
\end{equation}
Therefore, the relation (\ref{the_relation_example}) is a non-trivial relation between elementary matrices, since the identity matrix is of course also elementary. If we associate to the left hand side of (\ref{the_relation_example}) the element in $\mathrm{St}$, that is, the preimage of $\psi$, it will belong to the kernel of $\psi$, and thus giving an element in $K_2(\mathbb{R})$:
\begin{equation}
    (x^1_{12}x^{-1}_{21}x^1_{12})^4 \in \ker \psi,\quad \psi \left( (x^1_{12}x^{-1}_{21}x^1_{12})^4 \right) = (e^1_{12}e^{-1}_{21}e^1_{12})^4 
\end{equation}
since it evaluates as an identity matrix, which means that ``the relation holds''. In general, such identities are of the form:
\begin{equation}
    e^{\lambda_1}_{i_1j_1}e^{\lambda_2}_{i_2j_2}\dots e^{\lambda_r}_{i_rj_r} = \mathrm{Id}\quad \longleftrightarrow \quad x^{\lambda_1}_{i_1j_1}x^{\lambda_2}_{i_2j_2}\dots x^{\lambda_r}_{i_rj_r}
\end{equation}
Following \cite{RV}, we restrict ourselves to $F = \mathbb{Q}(C)$ -- the field of rational functions on a compact Riemann surface $C$. Choose a pair $(x,y)$ of such functions. Since $C$ is compact, there is always a unique minimal irreducible polynomial $P(x,y)$ defining it.
For example, if $C$ is topologically a sphere, $x=x(t),y=y(t)$ give a rational parametrization of $P(x,y)$. For higher genus, however, we would need more parameters, in order to make a proper parametrization (see some examples in \cite{BRVD}).

Now take a pair of elementary matrices:
\begin{equation}
D_x = 
\begin{pmatrix}
x & 0 & 0 \\
0 & x^{-1} & 0 \\
0 & 0 & 1
\end{pmatrix},
\quad 
D_y' = 
\begin{pmatrix}
y & 0 & 0 \\
0 & 1 & 0 \\
0 & 0 & y^{-1}
\end{pmatrix},
\end{equation}
and define
\begin{equation}
\{x,y\} := uvu^{-1}v^{-1}
\end{equation}
with $u = \psi^{-1}(D_x),v = \psi^{-1}(D'_y)$. This bracket is called the universal symbol of $(x,y)$. The commutator is always identity matrix, therefore $\{x,y\}\in K_2(\mathbb{Q}(C))$. It turns out that $K_2(F)$ is \emph{generated} by the symbols $\{x,y\}$ (\cite{Mil}, Corollary 9.13 p. 78), and it holds exactly when $F$ is a field.

Now the K-theoretic condition for $P(x,y)$ would be stated as follows (\cite{RV}, also \cite{BRVD} and \cite{GS} give slightly different at the first sight, but in fact equivalent formulations):
\begin{equation}\label{K_criterion}
    \boxed{\{x,y\}^N \in K_{2,\emptyset}\ \text{for some $N\in\mathbb{N}$} \Longleftrightarrow P(x,y) \text{ is tempered}}
\end{equation}
where ``tempered'' means that the face polynomials of $P(x,y)$ have roots only on the unit circle (are products of cyclotomic polynomials), and $K_{2,\emptyset}$ is the set of ``trivial'' elements in $K_2(\mathbb{Q}(C))$:
\begin{equation}
    K_{2,\emptyset} := \bigcap_w \mathrm{ker}\lambda_w \quad \subset K_2(\mathbb{Q}(C)),
\end{equation}
where $w \in C$, and $\lambda_w: K_2 \rightarrow \mathbb{C}^*$ corresponds to the tame symbol:
\begin{equation}
    (x,y)_w := (-1)^{w(x)w(y)}\frac{x^{w(y)}}{y^{w(x)}}\biggr\rvert_{w}
\end{equation}
Here the point $w\in C$ induces a functional $w(\ )$ on $\mathbb{Q}(C)$, called the valuation. Such that $w(x(t))$ or $w(y(t))$ equals to the degree of a leading term of $x(t)$ (or $y(t)$) around $t = w$, where $x(t),y(t)$ are the Puiseax parametrizations of a local branch.

\textsc{Remark:} As the reader may notice, the tame symbol is a map $F^*\times F^* \rightarrow \mathbb{C}^*$, where $F^* := F\setminus \{0,1\}$. Where does then $\lambda_w$ come from? In fact, every symbol on $F$, that is, a map
\begin{equation}
  F^*\times F^* \rightarrow A,
\end{equation}
where $A$ is any abelian group, gives rise to a unique homomorphism $K_2(F) \rightarrow A$. This is the content of the theorem by Matsumoto \cite{RV}, which states that $K_2(F)$ is the universal target of all symbols on $F$. So in the case of the tame symbol, we simply denote this homomorphism by $\lambda_w$. Its kernel consists of all elements in $K_2(F)$, which are mapped to $1\in \mathbb{C}^*$. Rephrasing, we require that all tame symbols for any $w\in C$ are roots of unity.

It turns out that this criterion has many exciting implications: relation to modular forms and special values of Zeta function \cite{RV}, Chern-Simons geometric quantization \cite{GS}, knot theory \cite{CCGLS}, modularity properties of the Mahler measure \cite{RV,BRVD}, etc. The proof of (\ref{K_criterion}) is due to the fact that for each slope $\frac{p}{q}$ of $N(P)$, there is a valuation $v$ such that $\frac{p}{q} = -\frac{v(x)}{v(y)}$. 
Moreover, the value of the tame symbol $(x,y)_v$ equals to the root of the corresponding face polynomial with this slope (details in \cite{CCGLS}).

In other words, by choosing $(x,y)$, we have to evaluate tame symbols $(x,y)_w$ for each $w\in S$, where $S$ is the set of zeroes and poles of $x$ and $y$ on $C$, and thus must be sure to get the roots of unity. It holds if and only if the polynomial $P(x,y)$ is tempered.

In what follows, we will denote $N(P)$ the Newton polytope of a polynomial $P(x_1,\dots,x_n)$, i.e. the convex hull of its monomials as integer lattice points in $\mathbb{R}^n$, and $\mathrm{supp}(P)$ the support of $P$, i.e. all its monomials as integer lattice points. Therefore, $N(P) := \mathrm{conv}(\mathrm{supp}(P))$, where $\mathrm{conv}$ is the operation of taking the convex hull of a set of points.


\subsection*{(An) example}
\begin{figure}[h!]
    \centering
    \tikzset{every picture/.style={line width=0.75pt}} 

\begin{tikzpicture}[x=0.75pt,y=0.75pt,yscale=-1,xscale=1]

\draw [line width=2.25]    (290.88,4097.47) -- (365.64,4172.23) ;
\draw [line width=2.25]    (290.88,4172.23) -- (365.64,4172.23) ;
\draw [line width=2.25]    (290.88,4097.47) -- (290.88,4172.23) ;
\draw  (95,4172.85) -- (209.78,4172.85)(106.48,4069.55) -- (106.48,4184.33) (202.78,4167.85) -- (209.78,4172.85) -- (202.78,4177.85) (101.48,4076.55) -- (106.48,4069.55) -- (111.48,4076.55)  ;
\draw  [fill={rgb, 255:red, 0; green, 0; blue, 0 }  ,fill opacity=1 ] (141.43,4172.85) .. controls (141.43,4171.51) and (142.52,4170.43) .. (143.86,4170.43) .. controls (145.2,4170.43) and (146.28,4171.51) .. (146.28,4172.85) .. controls (146.28,4174.19) and (145.2,4175.27) .. (143.86,4175.27) .. controls (142.52,4175.27) and (141.43,4174.19) .. (141.43,4172.85) -- cycle ;
\draw  [fill={rgb, 255:red, 0; green, 0; blue, 0 }  ,fill opacity=1 ] (104.05,4135.47) .. controls (104.05,4134.13) and (105.14,4133.05) .. (106.48,4133.05) .. controls (107.82,4133.05) and (108.9,4134.13) .. (108.9,4135.47) .. controls (108.9,4136.81) and (107.82,4137.89) .. (106.48,4137.89) .. controls (105.14,4137.89) and (104.05,4136.81) .. (104.05,4135.47) -- cycle ;
\draw  [fill={rgb, 255:red, 0; green, 0; blue, 0 }  ,fill opacity=1 ] (104.05,4098.09) .. controls (104.05,4096.75) and (105.14,4095.67) .. (106.48,4095.67) .. controls (107.82,4095.67) and (108.9,4096.75) .. (108.9,4098.09) .. controls (108.9,4099.43) and (107.82,4100.51) .. (106.48,4100.51) .. controls (105.14,4100.51) and (104.05,4099.43) .. (104.05,4098.09) -- cycle ;
\draw  [fill={rgb, 255:red, 0; green, 0; blue, 0 }  ,fill opacity=1 ] (178.81,4172.85) .. controls (178.81,4171.51) and (179.9,4170.43) .. (181.24,4170.43) .. controls (182.57,4170.43) and (183.66,4171.51) .. (183.66,4172.85) .. controls (183.66,4174.19) and (182.57,4175.27) .. (181.24,4175.27) .. controls (179.9,4175.27) and (178.81,4174.19) .. (178.81,4172.85) -- cycle ;
\draw  [fill={rgb, 255:red, 0; green, 0; blue, 0 }  ,fill opacity=1 ] (104.05,4172.85) .. controls (104.05,4171.51) and (105.14,4170.43) .. (106.48,4170.43) .. controls (107.82,4170.43) and (108.9,4171.51) .. (108.9,4172.85) .. controls (108.9,4174.19) and (107.82,4175.27) .. (106.48,4175.27) .. controls (105.14,4175.27) and (104.05,4174.19) .. (104.05,4172.85) -- cycle ;
\draw  [fill={rgb, 255:red, 0; green, 0; blue, 0 }  ,fill opacity=1 ] (141.43,4135.47) .. controls (141.43,4134.13) and (142.52,4133.05) .. (143.86,4133.05) .. controls (145.2,4133.05) and (146.28,4134.13) .. (146.28,4135.47) .. controls (146.28,4136.81) and (145.2,4137.89) .. (143.86,4137.89) .. controls (142.52,4137.89) and (141.43,4136.81) .. (141.43,4135.47) -- cycle ;
\draw  (279.4,4172.23) -- (394.18,4172.23)(290.88,4068.93) -- (290.88,4183.7) (387.18,4167.23) -- (394.18,4172.23) -- (387.18,4177.23) (285.88,4075.93) -- (290.88,4068.93) -- (295.88,4075.93)  ;
\draw   (220.51,4120.07) -- (246.68,4120.07) -- (246.68,4113.84) -- (264.12,4126.3) -- (246.68,4138.76) -- (246.68,4132.53) -- (220.51,4132.53) -- cycle ;

\draw (203.49,4182.61) node [anchor=north west][inner sep=0.75pt]    {$x$};
\draw (115.4,4049.98) node [anchor=north west][inner sep=0.75pt]    {$y$};
\draw (386.89,4181.99) node [anchor=north west][inner sep=0.75pt]    {$x$};
\draw (300.43,4050.11) node [anchor=north west][inner sep=0.75pt]    {$y$};

\end{tikzpicture}
    \caption{$\mathrm{supp}(P)$ (left) and $N(P)$ (right) for $P(x,y) = x^2-2xy+y^2-2x-y+1$}
    \label{fig:newton_example}
\end{figure}
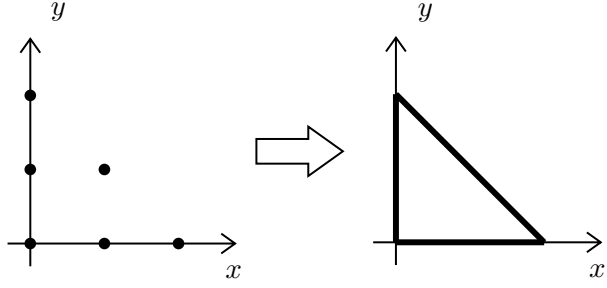

Take the genus zero curve:
\begin{equation}\label{tempered_example}
    P(x,y) = x^2-2xy+y^2-2x-y+1,
\end{equation}
also studied in \cite{BRVD}.
Its Newton polygon $N(P)$ is a triangle with vertices $(0,0),(0,2),(2,0)$ (Figure \ref{fig:newton_example}).
Notice that $P(x,y)$ is not self-reciprocal, since 
\begin{equation}
P(x^{-1},y^{-1})\neq \pm x^py^q P(x,y),\quad \text{for some integers $p,q$}
\end{equation}
which means it cannot be realized as a geometric A-polynomial for some hyperbolic 3-manifold \cite{CCGLS}.
The slopes are $0,\infty,-1$.
The face polynomials are $(\tau-1)^2,(\tau-1)^2,\tau^2-\tau+1$, where the variable $\tau$ decorates the monomials on a given edge of $N(P)$. Here's an explanation: write $P(x,y)=\sum c_{i,j}x^iy^j$.
To get a face polynomial $f_e$ for some edge $e$, label all the monomials on $e$ consequently from one vertex to another by $E = \{1,2,\dots,|e|\}$ where $|e|$ is the total number of monomials on $e$, and sum them up, replacing $x^iy^j$ by some power of $\tau$:
$
    f_{e} := \sum_{s\in E} c_{(i,j)(s)}\tau^s,\ E = \{1,2,\dots,|e|\}.
$
In this way, starting from a vertex and going through all edge monomials consequently, we end up in the opposite vertex, and get:
\begin{equation}
\begin{aligned}
x^2-2x+1 \mapsto &\ (\tau-1)^2 \\
y^2-y+1 \mapsto &\ \tau^2-\tau+1 \\
x^2-2xy+y^2 \mapsto &\ (\tau-1)^2
\end{aligned}
\end{equation}
All of them are cyclotomic. 
Choose the rational parametrization, e.g.:
\begin{equation}
    x(t) = \frac{t^2+t+1}{(t-1)^2},\ y(t) = \frac{3t^2}{(t-1)^2}
\end{equation}
Now compute the tame symbols at $w\in S$ for this parametrization.
In our case the set of zeroes and poles $S$ of $x(t)$ and $y(t)$ is
\begin{equation}
    S = \{0,1,\zeta_3^{(1)},\zeta_3^{(2)}\},
\end{equation}
where $\zeta_3^{(1)},\zeta_3^{(2)}$ are two complex-conjugated cubic roots of unity. We get:
\begin{equation}
    \text{horizontal : } (x,y)_0 = 1,\ \text{slope -1: } (x,y)_1 = 1,\ \text{vertical: } (x,y)_{\zeta_3^{(i)}} = \zeta_3^{(i)}
\end{equation}
For instance,
\begin{equation}
    (x,y)_0 = (-1)^{0\cdot 2}\frac{x(t)^2}{y(t)^0}\biggr\rvert_{t=0} = 1,
\end{equation}
since $x(t) = 1+ 3t+ 6t^2 + O(t^3)$, and $y(t) = 3t^2 + O(t^3)$ around $w = 0$, this gives $w(x) = 0$ and $w(y) = 2$.
As we see, each of the values $(x,y)_0,(x,y)_1,(x,y)_{\zeta_3^*}$ corresponds to a root of some face polynomial. All of them are roots of unity, which shows that $P(x,y)$ (\ref{tempered_example}) is tempered, i.e. the K-theoretic property holds for the underlying curve. Also, by computing the tame symbols we indeed see the surjection, but not the bijection between valuations and slopes (of course in this example one of the face polynomials has degree two and is irreducible, thus giving the two distinct roots with the same slope). 

\section{The main conjecture}\label{section_the_main_conjecture}

Here comes the main conjecture of the paper:
\begin{conjecture} Quiver A-polynomials are tempered, for every choice of the adjacency matrix.
\end{conjecture}
If true, it implies that all quiver A-polynomials are quantizable, according to \cite{GS}. Again, we would like to emphasise that this property is  a priori independent from the existence of quantized Nahm equations (\ref{qNahm_operators}), since no one guarantees that the two quantization techniques in \cite{GS} and \cite{LNPS} agree.
In our attempt to prove it, we are focusing on the diagonal case $C = \mathrm{diag}(\alpha_1,\dots,\alpha_m)$, because it involves a somewhat simplified combinatorics. In Appendix \ref{append_a_2} we provide a few confirmations for non-diagonal quivers.

We will study the Newton polytope of the resultant $\mathcal{R}$ from the Nahm equations, or simply the resultant polytope $N(\mathcal{R})$.
The strategy is:

\begin{itemize}
   \item find all monomials of $\mathcal{R}$, supported at the edges of $N(\mathcal{R})$
   \item study their projection onto $(x,y)$-plane and the polygon $N(A)$, given by the principal specialization
   \item binomiality of face polynomials of $A(x,y)$ would follow from factorization properties of the edge polynomials of $\mathcal{R}$, which project onto the edges of $N(A)$
   \item it would imply that $A(x,y)$ is tempered, since these binomials always have form $(\tau\pm 1)^k$ for some $k\in \mathbb{Z}_{+}$
\end{itemize}

We begin by defining the sparse mixed resultant \cite{GKZ,Stu,AJS}.

Fix a non-negative integer $m$ and a collection $\mathbf{A} = \{A_0,\dots,A_m\}$ of finite subsets $A_i\subset \mathbb{Z}^m,\ n_i = |A_i|$. Their convex hulls $Q_i=\mathrm{conv}(A_i)\subset \mathbb{R}^m$ are integral polytopes of dimension at most $m$. We are interested in Laurent polynomials, which are supported on $\mathbf{A}$.
Take a generic $(m+1)$-tuple $(f_0,\dots,f_m)$ of such polynomials:
\begin{equation}\label{f_i_polynomials}
    f_i(z_1,\dots,z_m) = \sum_{\mathbf{a}\in A_i} c_{i,\mathbf{a}}\mathbf{z^a},\ i=0,\dots,m
\end{equation}
Since $f_i$ are generic, the coefficients $c_{i,\mathbf{a}}\neq 0$ simultaneously for all $\mathbf{a}\in\mathcal{A}_i,\ i=0,\dots,m$. Therefore, we may think of the coefficient vector of (\ref{f_i_polynomials}) as a point in the product of complex projective spaces:
\begin{equation}
    (c_{0,A_0},\dots,c_{m,A_m}) \in \mathbb{P}^{n_0-1}\times \dots \times \mathbb{P}^{n_m-1}
\end{equation}
where $c_{i,A_i}$ encodes all the coefficients of $f_i$.
For example, if $m = 0$, $\mathbf{A} = \{A_0\}$ is just an integer, and $\mathrm{conv}(\mathbf{A}) = \mathbf{A}$ (zero-dimensional polytope). So the fist non-trivial case is $m = 1$. E.g., $A_0 = \{0, 1\},\ A_1 = \{0, 1,2\}$ gives: 
\begin{equation}\label{m=1_example}
                \begin{array}{ll}
                  f_0 = c_{0,0} + c_{0,1}z \\
                  f_1 = c_{1,0} + c_{1,1}z + c_{1,2}z^2
                \end{array}
\end{equation}
with $Q_0 = [0:1]$ and $Q_1 = [0:2]$ the two intervals, and the corresponding point would have projective coordinates $\{(c_{0,0} : c_{0,1}) , (c_{1,0} : c_{1,1} : c_{1,2})\}$: any of the two polynomials can be multiplied by a constant, which gives the same point in the projective space.

Consider now all $(m+1)$-tuples of the form (\ref{f_i_polynomials}), which have a common root $\mathbf{z}'\in (\mathbb{C}\setminus \{0\})^m:\ \{f_i(\mathbf{z}') = 0\}_{i=0\dots m}$. Since each such tuple corresponds to a single point in the projective product space (assuming that the coefficients are fixed numbers), all of them simultaneously will define a set of points, which closure we denote by $\overline{Z}$. It has a structure of projective variety.
In general, $\overline{Z}$ is an irreducible hypersurface in $\mathbb{P}^{n_0-1}\times \dots \times \mathbb{P}^{n_m-1}$ (see, for example, chapter 8 in \cite{GKZ}). However, sometimes degeneracies happen: for some ``bad'' choices of $\mathbf{A}$, $\overline{Z}$ may have codimension bigger than one.
\begin{definition}
Given a set $\mathbf{A}$, the sparse mixed resultant $\mathcal{R}_{\mathbf{A}}$ is the unique (up to an overall sign) irreducible polynomial in $c_{i,\mathbf{a}}$ with integral coefficients, which vanishes on $\overline{Z}$ if $\mathrm{codim}(\overline{Z}) = 1$, and $\mathcal{R}_{\mathbf{A}}:=1$ if $\mathrm{codim}(\overline{Z}) \geq 2$.
\end{definition}
Also, we will use the notion of a (sparse mixed) \emph{sub-resultant}, which is the sparse mixed resultant for a proper subset $\mathbf{A}' \subset \mathbf{A}$.

Returning to the example (\ref{m=1_example}), we get:
\begin{equation}\label{m=1_resultant}
    \mathcal{R}_{\{0,1\},\{0,1,2\}} = c_{1,0}c_{0,1}^2 - c_{1,1}c_{0,0}c_{0,1} + c_{1,2}c_{0,0}^2
\end{equation}
which agrees with the classical resultant (eliminant) $\mathrm{res}_z(f_0,f_1)$. The Newton polytope $N(\mathcal{R}_{\{0,1\},\{0,1,2\}})$ is a triangle in $\mathbb{R}^5$.
In this case, $\mathcal{R}_{\{0,1\},\{0,1,2\}} = 0$ is the defining equation for the hypersurface $\overline{Z}$ in $\mathbb{P}^{1}\times \mathbb{P}^{2}$. Indeed, in (\ref{m=1_resultant}) there are 5 parameters (only 3 of them are independent), but equating it to zero drops the dimension by 1, so that $\dim \overline{Z} = 2$ (in other words, its codimension is equal to 1). If we cross out any monomial in $f_0$, e.g.
\begin{equation}\label{example_sys2}
                \begin{array}{ll}
                  \tilde{f}_0 = \cancel{c_{0,0}} + c_{0,1}z \\
                  f_1 = c_{1,0} + c_{1,1}z + c_{1,2}z^2
                \end{array}
\end{equation}
the new (sub-)resultant would be $\mathcal{R}_{\{1\},\{0,1,2\}} = c_{0,1}$. The degenerate cases when $\mathrm{codim}(\overline{Z})\geq 2$, along with the conditions for $\mathbf{A}$ that guarantee $\mathcal{R}_{\mathbf{A}}$ to be non-trivial, are studied in \cite{Stu}.

Summing up, when dealing with a non-degenerate set of supports $\mathbf{A}$, $\mathcal{R}_{\mathbf{A}}$ agrees with the usual resultant, or eliminant with respect to $z_1,\dots,z_m$ from the system $\{f_i(z_1,\dots,z_m) = 0\}_{i=0\dots m}$, where the coefficients are taken as parameters.
Therefore, from a system of algebraic equations we obtain a single polynomial equation, which still encodes a lot of information about the original system. 
For example, in \cite{LNPS} the elimination has been performed for a large class of quivers, but things are getting messy fairly quickly.
There are several techniques to compute $\mathcal{R}_{\mathbf{A}}$ for a given $\mathbf{A}$. In most cases, any algorithm which performs elimination of variables from systems of equations (e.g. using Groebner bases) is able to compute $\mathcal{R}_{\mathbf{A}}$. One of the most powerful is the Canny-Emiris method \cite{CE}, an overview and computations for which we provide in Appendix \ref{append_b}.

That's being said; yet we need one more ingredient -- \emph{initial form} of a polynomial:
\begin{definition}\label{init_form_def}
    Given a polynomial $P(x_1,\dots,x_m)$ and an integer vector $\omega = (\omega_1,\dots,\omega_m)$, the initial form $init_{\omega}$ of $P$ with respect to $\omega$ is a polynomial, formed from all monomials of $P$ which have maximum weight with respect to $\omega$, in other words,
    \begin{equation}\label{init_deff}
        init_{\omega}(P) = \left.v^k P(v^{-\omega_1}x_1,\dots,v^{-\omega_m}x_m) \right|_{v = 0}
    \end{equation}
\end{definition}
(the exponent $k$ is chosen in order to get rid of the denominators, so that taking $v = 0$ does not give infinifies). Let's illustrate it again using the example (\ref{m=1_example}). Take $\omega = (0,1,1,0,2)$, then $\mathcal{R}_{\{0,1\},\{0,1,2\}}(c_{0,0},c_{0,1},c_{1,0},c_{1,1},c_{1,2})$ will have the initial form:
\begin{equation}
init_{\omega} = v^3\left.\mathcal{R}_{\{0,1\},\{0,1,2\}}(c_{0,0},v^{-1} c_{0,1},v^{-1} c_{1,0},c_{1,1},v^{-2}c_{1,2})\right|_{v = 0} = c_{1,0}c_{0,1}^2
\end{equation}
\textsc{Remark:} For any face of the Newton polytope $N(P)$ one can associate an initial form. Namely, if $\omega$ is the normal vector to some face, then $init_{\omega}$ is the restriction of $P$ to this face, meaning that we are left only with monomials belonging to the face. We will use this fact when dealing with the resultant polytope $N(\mathcal{R}_{\mathbf{A}})$. However, the vector $\omega$ itself will be not that important for us, since we will use another construction of initial forms, involving combinatorics.

It's time to get back to Nahm equations (\ref{clNahm_eqs}). We would like to treat them from the perspective of (\ref{f_i_polynomials}), therefore rewriting as
\begin{equation}\label{Nahm_eqs}
\begin{aligned}
    F_0 = &\ a_0 + a_1 z_1 \dots z_m \\
    F_i = &\ b_{i,0} + b_{i,1}z_i + b_{i,2}\prod_{j=1}^m z_j^{C_{i,j}}
\end{aligned}
\end{equation}
This gives: 
\begin{equation}
\mathbf{A}_{\mathrm{Nahm}} = \{\mathrm{supp}(F_i)\}_{i=0\dots m} = \{[\overline{0},\overline{1}],[\overline{0},\overline{e}_i,\dots],\dots,[\overline{0},\overline{e}_m,\dots]\},
\end{equation}
where $\overline{n} = (\overbrace{n,\dots,n}^{\text{$m$ times}})$, and $\overline{e}_i = (0,\dots,\underbrace{1}_{\text{$i$-th pos.}},\dots,0)$.
We will shorthand $\mathbf{b}:=\{b_{i,j}\},\ i=0\dots m,\ j=1,2$. 

\begin{definition}\label{quiver_resultant_def}
(Refined) quiver resultant $\mathcal{R}(a_0,a_1,\mathbf{b}) := \mathcal{R}_{\mathbf{A}_{\mathrm{Nahm}}}$ is the sparse mixed resultant from the supports $\mathbf{A}_{\mathrm{Nahm}}$ of $F_i,\ i=0\dots m$ (\ref{Nahm_eqs}).
\end{definition}
%
We have a chain of specializations:
\begin{equation}\label{quiver_A_poly_def}
\begin{aligned}
A(x_1,\dots,x_m,y)=&\ \mathcal{R}\left(y,-1\ \vert\ 1,-1,(-1)^{C_{1,1}}x_1\ \vert \ \dots \ \vert\ 1,-1,(-1)^{C_{m,m}}x_m \right) \\
A(x,y) =&\ A(\lambda_1x,\dots,\lambda_m x,y)
\end{aligned}
\end{equation}
There are several symbolical methods to compute sparse mixed resultants, one of them is discussed in Appendix \ref{append_b}. Also, there is a plenty of computer programs -- for example, standard elimination functions in Mathematica or Maple are capable to compute such resultants in cases when the set of supports is non-degenerate and not too complicated.

\begin{figure}[h!]
    \centering
    \tikzset{every picture/.style={line width=0.75pt}} 

\begin{tikzpicture}[x=0.75pt,y=0.75pt,yscale=-1,xscale=1]

\draw [color={rgb, 255:red, 144; green, 19; blue, 254 }  ,draw opacity=1 ][line width=1.5]    (90.4,184.6) -- (130.38,144.94) ;
\draw [color={rgb, 255:red, 0; green, 0; blue, 0 }  ,draw opacity=1 ][line width=0.75]  (72.84,184.6) -- (250.4,184.6)(90.6,59.14) -- (90.6,198.54) (243.4,179.6) -- (250.4,184.6) -- (243.4,189.6) (85.6,66.14) -- (90.6,59.14) -- (95.6,66.14)  ;
\draw [color={rgb, 255:red, 208; green, 2; blue, 27 }  ,draw opacity=1 ][fill={rgb, 255:red, 0; green, 0; blue, 0 }  ,fill opacity=1 ][line width=1.5]    (90.4,184.6) -- (210.4,184.6) ;
\draw [color={rgb, 255:red, 74; green, 144; blue, 226 }  ,draw opacity=1 ][fill={rgb, 255:red, 0; green, 0; blue, 0 }  ,fill opacity=1 ][line width=1.5]    (90.4,184.6) -- (90.4,104.6) ;
\draw [color={rgb, 255:red, 144; green, 19; blue, 254 }  ,draw opacity=1 ][fill={rgb, 255:red, 0; green, 0; blue, 0 }  ,fill opacity=1 ][line width=1.5]    (90.4,104.6) -- (130.4,64.6) ;
\draw [color={rgb, 255:red, 144; green, 19; blue, 254 }  ,draw opacity=1 ][fill={rgb, 255:red, 0; green, 0; blue, 0 }  ,fill opacity=1 ][line width=1.5]    (210.4,184.6) -- (250.4,144.6) ;
\draw [color={rgb, 255:red, 208; green, 2; blue, 27 }  ,draw opacity=1 ][fill={rgb, 255:red, 0; green, 0; blue, 0 }  ,fill opacity=1 ][line width=1.5]    (130.4,64.6) -- (250.4,64.6) ;
\draw [color={rgb, 255:red, 74; green, 144; blue, 226 }  ,draw opacity=1 ][fill={rgb, 255:red, 0; green, 0; blue, 0 }  ,fill opacity=1 ][line width=1.5]    (250.4,144.6) -- (250.4,64.6) ;
\draw  [fill={rgb, 255:red, 0; green, 0; blue, 0 }  ,fill opacity=1 ] (127.79,184.5) .. controls (127.79,182.91) and (129.08,181.62) .. (130.67,181.62) .. controls (132.26,181.62) and (133.54,182.91) .. (133.54,184.5) .. controls (133.54,186.09) and (132.26,187.38) .. (130.67,187.38) .. controls (129.08,187.38) and (127.79,186.09) .. (127.79,184.5) -- cycle ;
\draw  [fill={rgb, 255:red, 0; green, 0; blue, 0 }  ,fill opacity=1 ] (87.52,144.6) .. controls (87.52,143.01) and (88.81,141.72) .. (90.4,141.72) .. controls (91.99,141.72) and (93.28,143.01) .. (93.28,144.6) .. controls (93.28,146.19) and (91.99,147.48) .. (90.4,147.48) .. controls (88.81,147.48) and (87.52,146.19) .. (87.52,144.6) -- cycle ;
\draw  [fill={rgb, 255:red, 0; green, 0; blue, 0 }  ,fill opacity=1 ] (87.52,104.6) .. controls (87.52,103.01) and (88.81,101.72) .. (90.4,101.72) .. controls (91.99,101.72) and (93.28,103.01) .. (93.28,104.6) .. controls (93.28,106.19) and (91.99,107.48) .. (90.4,107.48) .. controls (88.81,107.48) and (87.52,106.19) .. (87.52,104.6) -- cycle ;
\draw  [fill={rgb, 255:red, 0; green, 0; blue, 0 }  ,fill opacity=1 ] (127.5,144.94) .. controls (127.5,143.35) and (128.79,142.06) .. (130.38,142.06) .. controls (131.97,142.06) and (133.25,143.35) .. (133.25,144.94) .. controls (133.25,146.53) and (131.97,147.82) .. (130.38,147.82) .. controls (128.79,147.82) and (127.5,146.53) .. (127.5,144.94) -- cycle ;
\draw  [fill={rgb, 255:red, 0; green, 0; blue, 0 }  ,fill opacity=1 ] (207.52,184.6) .. controls (207.52,183.01) and (208.81,181.72) .. (210.4,181.72) .. controls (211.99,181.72) and (213.28,183.01) .. (213.28,184.6) .. controls (213.28,186.19) and (211.99,187.48) .. (210.4,187.48) .. controls (208.81,187.48) and (207.52,186.19) .. (207.52,184.6) -- cycle ;
\draw  [fill={rgb, 255:red, 0; green, 0; blue, 0 }  ,fill opacity=1 ] (87.52,184.6) .. controls (87.52,183.01) and (88.81,181.72) .. (90.4,181.72) .. controls (91.99,181.72) and (93.28,183.01) .. (93.28,184.6) .. controls (93.28,186.19) and (91.99,187.48) .. (90.4,187.48) .. controls (88.81,187.48) and (87.52,186.19) .. (87.52,184.6) -- cycle ;

\draw (181.4,154.6) node [anchor=north west][inner sep=0.75pt]  [font=\footnotesize,color={rgb, 255:red, 208; green, 2; blue, 27 }  ,opacity=1 ]  {$Q_{1}$};
\draw (101.6,113.6) node [anchor=north west][inner sep=0.75pt]  [font=\footnotesize,color={rgb, 255:red, 74; green, 144; blue, 226 }  ,opacity=1 ]  {$Q_{2}$};
\draw (152,78) node [anchor=north west][inner sep=0.75pt]  [font=\footnotesize,color={rgb, 255:red, 0; green, 0; blue, 0 }  ,opacity=1 ]  {$Q_{0} +Q_{1} +Q_{2}$};
\draw (250.38,184.94) node [anchor=north west][inner sep=0.75pt]  [font=\small]  {$z_{1}$};
\draw (96.38,43.94) node [anchor=north west][inner sep=0.75pt]  [font=\small]  {$z_{2}$};
\draw (56.39,190.14) node [anchor=north west][inner sep=0.75pt]  [font=\scriptsize]  {$( 0,0)$};
\draw (56.42,93.33) node [anchor=north west][inner sep=0.75pt]  [font=\scriptsize]  {$( 0,\beta )$};
\draw (195.65,189.72) node [anchor=north west][inner sep=0.75pt]  [font=\scriptsize]  {$( \alpha ,0)$};
\draw (235.4,45.6) node [anchor=north west][inner sep=0.75pt]  [font=\scriptsize]  {$( \alpha+1 ,\beta+1 )$};
\draw (116,190) node [anchor=north west][inner sep=0.75pt]  [font=\scriptsize]  {$( 1,0)$};
\draw (56,134) node [anchor=north west][inner sep=0.75pt]  [font=\scriptsize]  {$( 0,1)$};
\draw (130.38,147.82) node [anchor=north west][inner sep=0.75pt]  [font=\footnotesize,color={rgb, 255:red, 144; green, 19; blue, 254 }  ,opacity=1 ]  {$Q_{0}$};

\end{tikzpicture}
    \caption{The Minkowski sum of three intervals (generators) $Q_0,Q_1,Q_2$ is a hexagon, each boundary face of which corresponds to one of its generators}
    \label{fig:Minkowski_sum}
\end{figure}
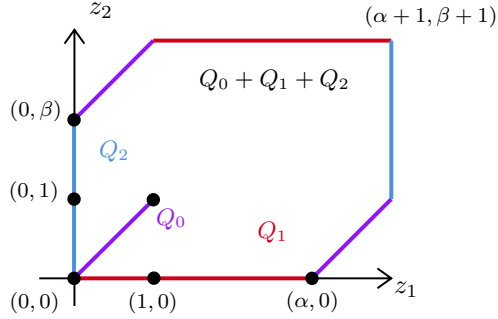

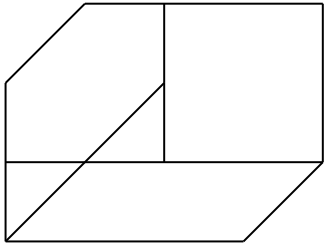
\begin{figure}[h!]
    \centering
    \tikzset{every picture/.style={line width=0.75pt}} 

\begin{tikzpicture}[x=0.75pt,y=0.75pt,yscale=-1,xscale=1]

\draw    (566,825) -- (686,825) ;
\draw    (686,825) -- (726,785) ;
\draw    (566,825) -- (566,745) ;
\draw    (726,785) -- (726,705) ;
\draw    (566,745) -- (606,705) ;
\draw    (606,705) -- (726,705) ;
\draw    (606,785) -- (566,825) ;
\draw    (606,785) -- (646,745) ;
\draw    (646,705) -- (646,745) ;
\draw    (566,785) -- (726,785) ;
\draw    (646,745) -- (646,785) ;

\end{tikzpicture}
    \caption{An example of non-tight decomposition for diagonal quiver with $m = 2$}
    \label{fig:TCMD_bad}
\end{figure}

Recall the notion of \emph{Minkowski sum} of subsets $Q_1,\dots,Q_m$ in $\mathbb{R}^n$ is simply the sum of all vectors in $Q_1,\dots,Q_m$. For example, the system (\ref{Nahm_eqs}) for $m = 2$ with $C = \mathrm{diag}(\alpha,\beta)$ produce 3 intervals $Q_i=\mathrm{conv}(F_i),\ i=0\dots 2$. Their Minkowski sum is a hexagon, shown on Figure \ref{fig:Minkowski_sum}. The hexagon, however, is not just a hexagon: it is a \emph{zonotope}, i.e. a projection of a 3-cube onto the plane. Its \emph{zones} (on the boundary) are given by three colours: red, magenta and blue (each one corresponding to its generator: $Q_1,Q_0$ and $Q_2$, correspondingly). It can be further generalized to higher dimensional zonotope, when the diagonal quiver will have more nodes. Therefore, the combinatorial simplicity of the diagonal quiver lies exactly here: its Newton polytope will inherit this combinatorial structure.
\\
Due to the results of \cite{Stu}, the dimension of $N(\mathcal{R})$ is equal to $(m-1)$, and the total degree is the mixed volume of the Minkowski sum $Q = Q_0 + \dots + Q_m$.
To work out higher dimensional resultant polytopes, we use the language of \emph{perfograms}. Each perfogram is just a pictorial presentation of a sub-resultant, for example:
\begin{equation}
\begin{tabular}{l l l}
  $b_0+b_1z_1+\cancel{b_2z_1^{C_{1,1}}\dots z_m^{C_{1,m}}}$ \\      
  $c_0+\cancel{c_1z_2}+c_2z_1^{C_{2,1}}\dots z_m^{C_{2,m}}$ \\      
  $d_0+\cancel{d_1z_3}+d_2z_1^{C_{3,1}}\dots z_m^{C_{3,m}}$
\end{tabular}
\Longleftrightarrow{}
\begin{tabular}{|c c c|}
  $\bullet$ & $\bullet$ & \\      
  $\bullet$ &  & $\bullet$ \\      
  $\bullet$ &  & $\bullet$
\end{tabular}
\end{equation}
Or, with $F_0$ included:
\begin{equation}
\begin{tabular}{l l l}
  $F_0$ = & $a_0+a_1z_1\dots z_m$ \\ \hline
  $F_1$ = & $b_0+b_1z_1+\cancel{b_2z_1^{C_{1,1}}\dots z_m^{C_{1,m}}}$ \\      
  $F_2$ = & $c_0+\cancel{c_1z_2}+c_2z_1^{C_{2,1}}\dots z_m^{C_{2,m}}$ \\      
  $F_3$ = & $d_0+\cancel{d_1z_3}+d_2z_1^{C_{3,1}}\dots z_m^{C_{3,m}}$
\end{tabular}
\Longleftrightarrow{}
\begin{tabular}{|c c c|}
  $\bullet$ & $\bullet$ & \\
  \hline
  $\bullet$ & $\bullet$ & \\      
  $\bullet$ &  & $\bullet$ \\      
  $\bullet$ &  & $\bullet$
\end{tabular}
\end{equation}
Now we shall review the combinatorics of a \emph{tight coherent mixed decomposition} (TCMD) of the Minkowski sum $Q$ of supports for some Laurent polynomial system.
Each cell of such decomposition is a Minkwoski sum of a sub-system, which means that it corresponds to some perfogram. For example, the decomposition on Figure \ref{fig:TCMD_bad} is not tight: not all its cells correspond to perfograms (triangles do not).
The word ``coherent'' is a bit more technical: it says that there is a convex piecewise linear function on $Q$, which domains of linearity are in 1:1 correspondence with the cells of our decomposition (the canonical examples are presented in Chapter 7.1 of \cite{GKS}; see also \cite{Stu}). 
In our study all mixed decompositions will be automatically tight and coherent by construction, so we will not refer to these properties henceforth.

\begin{figure}[h!]
    \centering
    \tikzset{every picture/.style={line width=0.75pt}} 

\begin{tikzpicture}[x=0.75pt,y=0.75pt,yscale=-1,xscale=1]

\draw    (38,827) -- (158,827) ;
\draw    (158,827) -- (198,787) ;
\draw    (38,827) -- (38,747) ;
\draw    (198,787) -- (198,707) ;
\draw    (38,747) -- (78,707) ;
\draw    (78,707) -- (198,707) ;
\draw    (78,787) -- (78,827) ;
\draw    (38,787) -- (78,787) ;
\draw    (78,787) -- (118,747) ;
\draw    (118,707) -- (118,747) ;
\draw    (118,747) -- (198,747) ;
\draw    (40,834) -- (156,834) ;
\draw [shift={(158,834)}, rotate = 180] [color={rgb, 255:red, 0; green, 0; blue, 0 }  ][line width=0.75]    (10.93,-3.29) .. controls (6.95,-1.4) and (3.31,-0.3) .. (0,0) .. controls (3.31,0.3) and (6.95,1.4) .. (10.93,3.29)   ;
\draw [shift={(38,834)}, rotate = 0] [color={rgb, 255:red, 0; green, 0; blue, 0 }  ][line width=0.75]    (10.93,-3.29) .. controls (6.95,-1.4) and (3.31,-0.3) .. (0,0) .. controls (3.31,0.3) and (6.95,1.4) .. (10.93,3.29)   ;
\draw    (30,825) -- (30,749) ;
\draw [shift={(30,747)}, rotate = 450] [color={rgb, 255:red, 0; green, 0; blue, 0 }  ][line width=0.75]    (10.93,-3.29) .. controls (6.95,-1.4) and (3.31,-0.3) .. (0,0) .. controls (3.31,0.3) and (6.95,1.4) .. (10.93,3.29)   ;
\draw [shift={(30,827)}, rotate = 270] [color={rgb, 255:red, 0; green, 0; blue, 0 }  ][line width=0.75]    (10.93,-3.29) .. controls (6.95,-1.4) and (3.31,-0.3) .. (0,0) .. controls (3.31,0.3) and (6.95,1.4) .. (10.93,3.29)   ;

\draw (66,733.33) node [anchor=north west][inner sep=0.75pt]    {$( a)$};
\draw (128,774.33) node [anchor=north west][inner sep=0.75pt]    {$( b)$};
\draw (47,793.33) node [anchor=north west][inner sep=0.75pt]    {$( c)$};
\draw (147,713.33) node [anchor=north west][inner sep=0.75pt]    {$( d)$};
\draw (93,838) node [anchor=north west][inner sep=0.75pt]    {$\alpha $};
\draw (9,775) node [anchor=north west][inner sep=0.75pt]    {$\beta $};

\end{tikzpicture}
    \caption{An example of TCMD for diagonal quiver with $m = 2$: the Minkowski sum $Q = Q_0 + Q_1 + Q_2$ is decomposed into the 4 non-overlapping cells}
    \label{fig:TCMD_demo}
\end{figure}
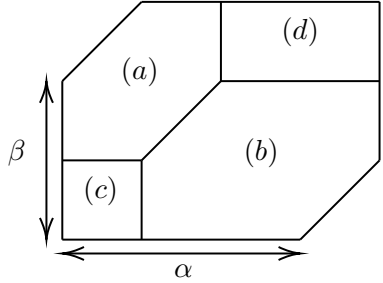

Given the Minkowski sum $Q = Q_1 + \dots + Q_m$ of convex hulls of supports of $F_i$, we may construct its TCMD as follows: each cell is a Minkowski sum of sub-supports, computed for some subsets
$A_0'\subset A_0,\dots,A_m' \subset A_m$. 
Then, another cell would be given by yet another subsets $A_0''\subset A_0,\dots,A_m''\subset A_m$, and so on, which yields a partition of $Q$ into non-overlapping cells (if the subsets are chosen properly).
Example of such mixed decomposition is on Figure \ref{fig:TCMD_demo}.
\\
Let's see which perfograms produce the cells on Figure \ref{fig:TCMD_demo}. For example:
\begin{equation}
\begin{aligned}
(a)\quad 
\begin{aligned}
    &\ a_0 + a_1 z_1 z_2 \\
    &\ b_{1,0} + b_{1,1}z_1 + \cancel{b_{1,2}z_1^2} \\
    &\ \cancel{b_{2,0}} + b_{2,1}z_2 + b_{2,2}z_2^2 \\
\end{aligned} \quad
\Longleftrightarrow{}
\quad
\begin{tabular}{|c c c|}
  $\bullet$ & $\bullet$ & \\
  \hline
  $\bullet$ & $\bullet$ & \\      
   & $\bullet$ & $\bullet$      
\end{tabular} \\
(b)\quad
\begin{aligned}
    &\ a_0 + a_1 z_1 z_2 \\
    &\ \cancel{b_{1,0}} + b_{1,1}z_1 + b_{1,2}z_1^2 \\
    &\ b_{2,0} + b_{2,1}z_2 + \cancel{b_{2,2}z_2^2} \\
\end{aligned} \quad
\Longleftrightarrow{}
\quad
\begin{tabular}{|c c c|}
  $\bullet$ & $\bullet$ & \\
  \hline
   & $\bullet$ & $\bullet$\\      
  $\bullet$ & $\bullet$ &      
\end{tabular}
\end{aligned}
\end{equation}
\begin{figure}[h!]
    \centering
    \tikzset{every picture/.style={line width=0.75pt}} 

\begin{tikzpicture}[x=0.75pt,y=0.75pt,yscale=-1,xscale=1]

\draw    (68.75,929.75) -- (90,908.5) ;
\draw    (90,908.5) -- (90,866) ;
\draw [color={rgb, 255:red, 245; green, 166; blue, 35 }  ,draw opacity=1 ][fill={rgb, 255:red, 255; green, 255; blue, 255 }  ,fill opacity=1 ][line width=1.5]    (5,887.25) -- (26.25,866) ;
\draw    (26.25,908.5) -- (26.25,929.75) ;
\draw [color={rgb, 255:red, 245; green, 166; blue, 35 }  ,draw opacity=1 ][fill={rgb, 255:red, 255; green, 255; blue, 255 }  ,fill opacity=1 ][line width=1.5]    (5,908.5) -- (26.25,908.5) ;
\draw [color={rgb, 255:red, 245; green, 166; blue, 35 }  ,draw opacity=1 ][fill={rgb, 255:red, 255; green, 255; blue, 255 }  ,fill opacity=1 ][line width=1.5]    (26.25,908.5) -- (47.5,887.25) ;
\draw [color={rgb, 255:red, 245; green, 166; blue, 35 }  ,draw opacity=1 ][fill={rgb, 255:red, 255; green, 255; blue, 255 }  ,fill opacity=1 ][line width=1.5]    (47.5,866) -- (47.5,887.25) ;
\draw    (47.5,887.25) -- (90,887.25) ;
\draw    (5,908.5) -- (5,929.75) ;
\draw [color={rgb, 255:red, 245; green, 166; blue, 35 }  ,draw opacity=1 ][fill={rgb, 255:red, 255; green, 255; blue, 255 }  ,fill opacity=1 ][line width=1.5]    (5,887.25) -- (5,908.5) ;
\draw    (5,929.75) -- (26.25,929.75) ;
\draw    (26.25,929.75) -- (68.75,929.75) ;
\draw    (47.5,866) -- (90,866) ;
\draw [color={rgb, 255:red, 245; green, 166; blue, 35 }  ,draw opacity=1 ][fill={rgb, 255:red, 255; green, 255; blue, 255 }  ,fill opacity=1 ][line width=1.5]    (26.25,866) -- (47.5,866) ;
\draw [color={rgb, 255:red, 245; green, 166; blue, 35 }  ,draw opacity=1 ][line width=1.5]    (187.75,929.75) -- (209,908.5) ;
\draw [color={rgb, 255:red, 0; green, 0; blue, 0 }  ,draw opacity=1 ][fill={rgb, 255:red, 255; green, 255; blue, 255 }  ,fill opacity=1 ][line width=0.75]    (124,887.25) -- (145.25,866) ;
\draw [color={rgb, 255:red, 245; green, 166; blue, 35 }  ,draw opacity=1 ][line width=1.5]    (145.25,908.5) -- (145.25,929.75) ;
\draw [color={rgb, 255:red, 0; green, 0; blue, 0 }  ,draw opacity=1 ][fill={rgb, 255:red, 255; green, 255; blue, 255 }  ,fill opacity=1 ][line width=0.75]    (124,908.5) -- (145.25,908.5) ;
\draw [color={rgb, 255:red, 245; green, 166; blue, 35 }  ,draw opacity=1 ][fill={rgb, 255:red, 255; green, 255; blue, 255 }  ,fill opacity=1 ][line width=1.5]    (145.25,908.5) -- (166.5,887.25) ;
\draw [color={rgb, 255:red, 0; green, 0; blue, 0 }  ,draw opacity=1 ][fill={rgb, 255:red, 255; green, 255; blue, 255 }  ,fill opacity=1 ][line width=0.75]    (166.5,866) -- (166.5,887.25) ;
\draw [color={rgb, 255:red, 245; green, 166; blue, 35 }  ,draw opacity=1 ][line width=1.5]    (166.5,887.25) -- (209,887.25) ;
\draw    (124,908.5) -- (124,929.75) ;
\draw [color={rgb, 255:red, 0; green, 0; blue, 0 }  ,draw opacity=1 ][fill={rgb, 255:red, 255; green, 255; blue, 255 }  ,fill opacity=1 ][line width=0.75]    (124,887.25) -- (124,908.5) ;
\draw    (124,929.75) -- (145.25,929.75) ;
\draw [color={rgb, 255:red, 245; green, 166; blue, 35 }  ,draw opacity=1 ][line width=1.5]    (145.25,929.75) -- (187.75,929.75) ;
\draw    (166.5,866) -- (209,866) ;
\draw [color={rgb, 255:red, 0; green, 0; blue, 0 }  ,draw opacity=1 ][fill={rgb, 255:red, 255; green, 255; blue, 255 }  ,fill opacity=1 ][line width=0.75]    (145.25,866) -- (166.5,866) ;
\draw    (209,887.25) -- (209,866) ;
\draw [color={rgb, 255:red, 245; green, 166; blue, 35 }  ,draw opacity=1 ][line width=1.5]    (209,908.5) -- (209,887.25) ;

\end{tikzpicture}
    \caption{Each cell is a Minkowski sum of sub-supports: the case (a), left and (b), right}
    \label{fig:minkowski_demo1}
\end{figure}
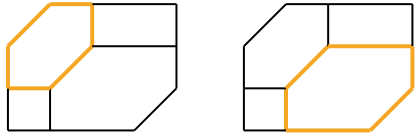
Now we see that the Minkowski sums of these two collections of sub-supports indeed produce the two small hexagons (Figure \ref{fig:minkowski_demo1}). E.g., $(a)$ gives $\tilde{Q}_0 = [(0,0):(1,1)],\ \tilde{Q}_1 = [(0,0):(1,0)],\ \tilde{Q}_2 = [(0,1):(0,2)]$, and $\tilde{Q}_0+\tilde{Q}_1+\tilde{Q}_2$ is indeed the left hexagon on Figure \ref{fig:minkowski_demo1}.
Analogously, the two rectangular cells are given by:
\begin{equation}
\begin{aligned}
(c)\quad 
\begin{aligned}
    &\ a_0 + \cancel{a_1 z_1 z_2} \\
    &\ b_{1,0} + b_{1,1}z_1 + \cancel{b_{1,2}z_1^2} \\
    &\ b_{2,0} + b_{2,1}z_2 + \cancel{b_{2,2}z_2^2} \\
\end{aligned} \quad
\Longleftrightarrow{}
\quad
\begin{tabular}{|c c c|}
  $\bullet$ &  & \\
  \hline
  $\bullet$ & $\bullet$ & \\      
  $\bullet$ & $\bullet$ &     
\end{tabular} \\
(d)\quad
\begin{aligned}
    &\ \cancel{a_0} + a_1 z_1 z_2 \\
    &\ \cancel{b_{1,0}} + b_{1,1}z_1 + b_{1,2}z_1^2 \\
    &\ \cancel{b_{2,0}} + b_{2,1}z_2 + b_{2,2}z_2^2 \\
\end{aligned} \quad
\Longleftrightarrow{}
\quad
\begin{tabular}{|c c c|}
  & $\bullet$ &  \\
  \hline
   & $\bullet$ & $\bullet$\\      
  & $\bullet$ & $\bullet$      
\end{tabular}
\end{aligned}
\end{equation}
\begin{figure}[h!]
    \centering
    \tikzset{every picture/.style={line width=0.75pt}} 

\begin{tikzpicture}[x=0.75pt,y=0.75pt,yscale=-1,xscale=1]

\draw    (68.75,929.75) -- (90,908.5) ;
\draw    (90,908.5) -- (90,866) ;
\draw [color={rgb, 255:red, 0; green, 0; blue, 0 }  ,draw opacity=1 ][fill={rgb, 255:red, 255; green, 255; blue, 255 }  ,fill opacity=1 ][line width=0.75]    (5,887.25) -- (26.25,866) ;
\draw [color={rgb, 255:red, 126; green, 211; blue, 33 }  ,draw opacity=1 ][line width=1.5]    (26.25,908.5) -- (26.25,929.75) ;
\draw [color={rgb, 255:red, 126; green, 211; blue, 33 }  ,draw opacity=1 ][fill={rgb, 255:red, 255; green, 255; blue, 255 }  ,fill opacity=1 ][line width=1.5]    (5,908.5) -- (26.25,908.5) ;
\draw [color={rgb, 255:red, 0; green, 0; blue, 0 }  ,draw opacity=1 ][fill={rgb, 255:red, 255; green, 255; blue, 255 }  ,fill opacity=1 ][line width=0.75]    (26.25,908.5) -- (47.5,887.25) ;
\draw [color={rgb, 255:red, 0; green, 0; blue, 0 }  ,draw opacity=1 ][fill={rgb, 255:red, 255; green, 255; blue, 255 }  ,fill opacity=1 ][line width=0.75]    (47.5,866) -- (47.5,887.25) ;
\draw    (47.5,887.25) -- (90,887.25) ;
\draw [color={rgb, 255:red, 126; green, 211; blue, 33 }  ,draw opacity=1 ][line width=1.5]    (5,908.5) -- (5,929.75) ;
\draw [color={rgb, 255:red, 0; green, 0; blue, 0 }  ,draw opacity=1 ][fill={rgb, 255:red, 255; green, 255; blue, 255 }  ,fill opacity=1 ][line width=0.75]    (5,887.25) -- (5,908.5) ;
\draw [color={rgb, 255:red, 126; green, 211; blue, 33 }  ,draw opacity=1 ][line width=1.5]    (5,929.75) -- (26.25,929.75) ;
\draw    (26.25,929.75) -- (68.75,929.75) ;
\draw    (47.5,866) -- (90,866) ;
\draw [color={rgb, 255:red, 0; green, 0; blue, 0 }  ,draw opacity=1 ][fill={rgb, 255:red, 255; green, 255; blue, 255 }  ,fill opacity=1 ][line width=0.75]    (26.25,866) -- (47.5,866) ;
\draw [color={rgb, 255:red, 0; green, 0; blue, 0 }  ,draw opacity=1 ][line width=0.75]    (187.75,929.75) -- (209,908.5) ;
\draw [color={rgb, 255:red, 0; green, 0; blue, 0 }  ,draw opacity=1 ][fill={rgb, 255:red, 255; green, 255; blue, 255 }  ,fill opacity=1 ][line width=0.75]    (124,887.25) -- (145.25,866) ;
\draw [color={rgb, 255:red, 0; green, 0; blue, 0 }  ,draw opacity=1 ][line width=0.75]    (145.25,908.5) -- (145.25,929.75) ;
\draw [color={rgb, 255:red, 0; green, 0; blue, 0 }  ,draw opacity=1 ][fill={rgb, 255:red, 255; green, 255; blue, 255 }  ,fill opacity=1 ][line width=0.75]    (124,908.5) -- (145.25,908.5) ;
\draw [color={rgb, 255:red, 0; green, 0; blue, 0 }  ,draw opacity=1 ][fill={rgb, 255:red, 255; green, 255; blue, 255 }  ,fill opacity=1 ][line width=0.75]    (145.25,908.5) -- (166.5,887.25) ;
\draw [color={rgb, 255:red, 126; green, 211; blue, 33 }  ,draw opacity=1 ][fill={rgb, 255:red, 255; green, 255; blue, 255 }  ,fill opacity=1 ][line width=1.5]    (166.5,866) -- (166.5,887.25) ;
\draw [color={rgb, 255:red, 126; green, 211; blue, 33 }  ,draw opacity=1 ][line width=1.5]    (166.5,887.25) -- (209,887.25) ;
\draw    (124,908.5) -- (124,929.75) ;
\draw [color={rgb, 255:red, 0; green, 0; blue, 0 }  ,draw opacity=1 ][fill={rgb, 255:red, 255; green, 255; blue, 255 }  ,fill opacity=1 ][line width=0.75]    (124,887.25) -- (124,908.5) ;
\draw    (124,929.75) -- (145.25,929.75) ;
\draw [color={rgb, 255:red, 0; green, 0; blue, 0 }  ,draw opacity=1 ][line width=0.75]    (145.25,929.75) -- (187.75,929.75) ;
\draw [color={rgb, 255:red, 126; green, 211; blue, 33 }  ,draw opacity=1 ][line width=1.5]    (166.5,866) -- (209,866) ;
\draw [color={rgb, 255:red, 0; green, 0; blue, 0 }  ,draw opacity=1 ][fill={rgb, 255:red, 255; green, 255; blue, 255 }  ,fill opacity=1 ][line width=0.75]    (145.25,866) -- (166.5,866) ;
\draw [color={rgb, 255:red, 126; green, 211; blue, 33 }  ,draw opacity=1 ][line width=1.5]    (209,887.25) -- (209,866) ;
\draw [color={rgb, 255:red, 0; green, 0; blue, 0 }  ,draw opacity=1 ][line width=0.75]    (209,908.5) -- (209,887.25) ;

\end{tikzpicture}
    \caption{Each cell is a Minkowski sum of sub-supports: the case (c), left and (d), right}
    \label{fig:minkowski_demo2}
\end{figure}
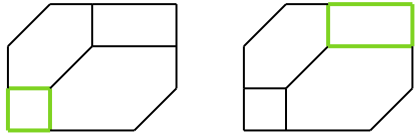

The fact that we treat hexagonal cells separately from rectangular ones, is not a coincidence. In fact, we can already see that rectangular cells give degenerate sets of supports, i.e. their sub-resultant would be equal identically to $1$. This is due to the fact that already in $F_0$ we are left with just a single monomial.
In what follows, we would like to associate an initial form to each TCMD, such that:
\begin{itemize}
 \item each hexagonal cell gives a binomial factor
 \item each rectangular cell gives a monomial factor
\end{itemize} 
Namely, given the following data:
\begin{itemize}
    \item a collection $\mathbf{A}=\{A_0,\dots,A_m\}$ of subsets in $\mathbb{Z}^m$ \item Minkowski sum of their convex hulls $Q = \sum_{i=1}^m \mathrm{conv}(A_i)$ in $\mathbb{R}^m$
    \item a coherent mixed decomposition $\mathrm{TCMD}$ of $Q$,
\end{itemize}
we can associate a polynomial to this $\mathrm{TCMD}$ by the formula:
\begin{equation}\label{initial_factorization}
    init_{\mathrm{TCMD}(Q)}(c_{0,0},c_{0,1},\dots) = \prod_{\text{``rectangles''}} \times \prod_{\text{``hexagons''}} =  \mu \prod_{\iota\in \mathrm{TCMD}(Q)} \tilde{\mathcal{R}}_{\iota}^{k_{\iota}},
\end{equation}
where $\mu = \prod_{\iota'}\mu_{\iota'}^{(k_{\iota'})}$ is a monomial, each letter $\mu_{\iota'}$ of which corresponds to a rectangle $\iota'$ in $\mathrm{TCMD}(Q)$.
On another hand, each $\tilde{\mathcal{R}}_{\iota}$ is a (sparse mixed) sub-resultant, which perfogram gives a hexagonal cell $\iota$. The latter product is taken over all hexagonal cells in $\mathrm{TCMD}(Q)$, and the exponents $k_{\iota}$ and $k_{\iota'}$ are chosen uniquely such that the volume of $\iota$ equals to the total degree of $\tilde{\mathcal{R}}_{\iota}^{k_{\iota}}$, for every cell $\iota$, and for $\iota'$ the volume of a rectangular cell $\iota'$ simply equals to $k_{\iota'}$.

The correspondence between TCMDs and initial forms of the sparse mixed resultant is due to the following
\begin{theorem} \label{init_def}
$init_{\mathrm{TCMD}(Q)}$ is the initial form for $\mathcal{R}_{\mathbf{A}}$.
\end{theorem}

This allows to associate a TCMD to each face of the resultant polytope, and then study their initial forms.
The theorem has been proven \cite{Stu} for $\mathbf{A}$ being ``good enough'', and quite recently \cite{AJS} in a full generality. 

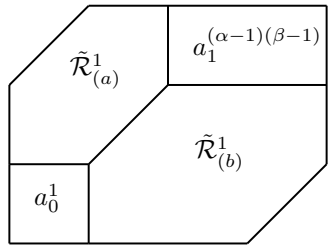
\begin{figure}[h!]
    \centering
    \tikzset{every picture/.style={line width=0.75pt}} 
\begin{tikzpicture}[x=0.75pt,y=0.75pt,yscale=-1,xscale=1]

\draw    (154,1592) -- (194,1552) ;
\draw    (34,1512) -- (74,1472) ;
\draw    (74,1552) -- (74,1592) ;
\draw    (34,1552) -- (74,1552) ;
\draw    (74,1552) -- (114,1512) ;
\draw    (114,1472) -- (114,1512) ;
\draw    (114,1512) -- (194,1512) ;
\draw    (194,1512) -- (194,1552) ;
\draw    (194,1472) -- (194,1512) ;
\draw    (114,1472) -- (194,1472) ;
\draw    (74,1472) -- (114,1472) ;
\draw    (34,1552) -- (34,1592) ;
\draw    (34,1512) -- (34,1552) ;
\draw    (34,1592) -- (74,1592) ;
\draw    (74,1592) -- (154,1592) ;

\draw (63,1493.33) node [anchor=north west][inner sep=0.75pt]  [font=\small,color={rgb, 255:red, 0; green, 0; blue, 0 }  ,opacity=1 ]  {$\tilde{\mathcal{R}}^{1}_{( a)}$};
\draw (45,1560.33) node [anchor=north west][inner sep=0.75pt]  [font=\small]  {$a^{1}_{0}$};
\draw (125,1481.33) node [anchor=north west][inner sep=0.75pt]  [font=\small]  {$a^{( \alpha -1)( \beta -1)}_{1}$};
\draw (126,1535.33) node [anchor=north west][inner sep=0.75pt]  [font=\small]  {$\tilde{\mathcal{R}}^{1}_{( b)}$};
\end{tikzpicture}
    \caption{This TCMD gives rise to the initial form (\ref{init_example}), computed as the product over all its cells}
    \label{fig:TCMD_example_init}
\end{figure}

Let us illustrate how it works using the example on Figure \ref{fig:TCMD_demo}. We claim that it gives the initial form:
\begin{equation}\label{init_example}
    a_0a_1^{(\alpha-1)(\beta-1)}(a_0^{\alpha-1}b_2c_1^{\alpha-1}+a_1^{\alpha-1}b_1c_0^{\alpha-1})(a_0^{\beta-1}c_2b_1^{\beta-1}+a_1^{\beta-1}c_1b_0^{\beta-1})
\end{equation}
The choice of $\omega$ is, however, not unique. E.g., we can take $\omega = ((0,0),(0,1,1),(1,1,1))+const$, where $const$ is an arbitrary constant vector. Once again, we want to emphasise that these vectors are not important for us, since we focus on the combinatorial structure given by the formula (\ref{initial_factorization}).
Here, hexagons are the two distinct binomial factors, while rectangles contribute to the monomial in (\ref{init_example}).
We have the following picture (Figure \ref{fig:TCMD_example_init}).
Indeed, the sparse mixed resultants from (a) and (b) are $\tilde{\mathcal{R}}_{(a)} = a_0^{\beta-1}c_2b_1^{\beta-1}+a_1^{\beta-1}c_1b_0^{\beta-1},\ \tilde{\mathcal{R}}_{(b)} = a_0^{\alpha-1}b_2c_1^{\alpha-1}+a_1^{\alpha-1}b_1c_0^{\alpha-1}$. The exponents $k_{(a)}$ and $k_{(b)}$ are equal to 1, since the total degrees of $\tilde{\mathcal{R}}_{(b)}$ and $\tilde{\mathcal{R}}_{(a)}$ are equal to $2(\alpha-1)+1$ and $2(\beta-1)+1$, correspondingly, which agrees with the areas of the two hexagons $(a)$ and $(b)$, Figures \ref{fig:TCMD_demo} and \ref{fig:minkowski_demo1}. On the monomial side, we have two degenerations: in the first one, $a_0$ survives, and the area of $(c)$ is equal to $k_{(c)} = 1$, so $a_0^1$. In the second one, $a_1$ survives and the area is $k_{(d)} = (\alpha-1)(\beta-1)$. Taking the product over all of them gives us the expression (\ref{init_example}).
\begin{definition}
Initial form is called simple, if it does not have ``interior'' monomials, i.e. if all its monomials lie on the 1-dimensional skeleton of the corresponding face of $N(\mathcal{R})$.
\end{definition}
In general, the correspondence between faces and initial forms is not 1:1. For each face, there may be many associated initial forms. It depends on whether we want to include the interior monomials or not, and which ones (by switching the intermediate bullets im each row of a perfogram). However, we may get better results with simple initial forms. Let $init_{\xi}$ be an initial form, such that $\mathrm{supp}(init_{\xi})\subset \xi$ and $\xi$ is a face of $N(\mathcal{R})$.
%
\begin{proposition} $init_{\xi}$ is simple if and only if all its perfograms, corresponding to each $\tilde{\mathcal{R}}_i$ in (\ref{initial_factorization}), do not have intermediate bullets in each of its row. Moreover, there is a bijection between the set of all simple initial forms and the set of faces of $N(\mathcal{R})$.
\end{proposition}
Here's an example:
\begin{equation}
\begin{tabular}{|c c c|}
  $\bullet$ & $\bullet$ & \\      \hline
  $\bullet$ &  & $\bullet$ \\      
  $\bullet$ &  & $\bullet$
\end{tabular}
\quad \text{is simple, whereas}\quad
\begin{tabular}{|c c c|}
  $\bullet$ & $\bullet$ & \\      \hline
  $\bullet$ & $\bullet$  & $\bullet$ \\      
  $\bullet$ &  & $\bullet$
\end{tabular}
\quad \text{is not.}
\end{equation}
\underline{Proof}. Start with an assumption that $init_{\xi}$ is simple, which means that all its monomials lie on the 1-dimensional skeleton of $\xi$. We can ignore the monomial prefactor $\mu_{\xi}$, since it simply rescales the lattice, which results in isomorphic polytopes. From the product formula (\ref{initial_factorization}) we deduce that the $N(\prod\tilde{\mathcal{R}}_i)$ decomposes as a Minkowski sum of $N(\tilde{\mathcal{R}}_i)$, for $i=1\dots |\mathrm{TCMD}_{\xi}|$. But since $\xi$ is simple, each $\tilde{\mathcal{R}}_i$ should be simple as well, i.e. not containing any interior monomials (otherwise it would hold also for their Minkowski sum). 
Another way around is immediate: since all $\tilde{\mathcal{R}}_i$ are simple, their Minkowksi sum does not have interior monomials, which implies $\xi$ is simple. Finally, the bijection is provided by:
\begin{equation}
    \mathrm{vertices}(\xi) = \mathrm{vertices}(\mathrm{conv}(\mathrm{supp}(init)_{\xi}))) \subseteq \mathrm{supp}(init_{\mathrm{\xi}}),
\end{equation}
and the set $\mathrm{supp}(init_{\mathrm{\xi}})\setminus \mathrm{vertices}(\xi)$ is fixed uniquely, by requiring all the perfograms to have no intermediate bullets ``$\bullet$'' in each of its row. $\square$

\begin{proposition}\label{phi_dimension_prop}
If $init_\xi$ is simple, then $\dim(\xi)$ is equal to the number of its distinct binomial factors.
\end{proposition}

\underline{Proof}. The case $\dim(\xi) = 1$ is trivial, since if $\xi$ is an edge and $init(\xi)$ is simple, then it cannot be anything but just a single binomial (one vertex + another vertex, and if there are intermediate monomials, it factorizes into a power of this binomial). When $\dim(\xi) = 2$, $init(\xi)$ would have two distinct binomial factors. Conversely, for any initial form with two binomial factors, these factors cannot belong to the same edge -- the initial form is said to be simple. The only monomials are vertices of its convex hull
(in the opposite situation we would encounter some monomials which are not the vertices -- a contradiction).

\begin{figure}[h!]
    \centering
    \includegraphics[scale=0.3]{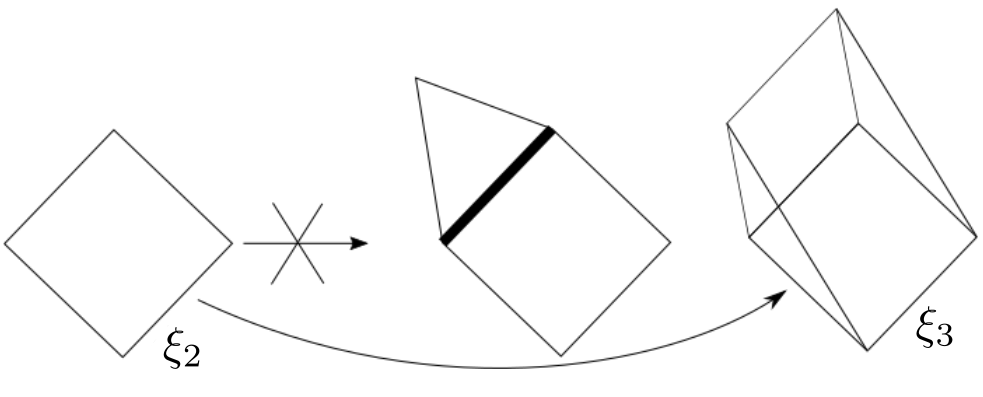}
    \caption{Simple initial form with two distinct binomial factors corresponds to a 2d face $\xi_2$. Joining an extra edge to $\xi_2$ will lead to $\xi_3$. Since the simplicity relation is preserved, it increases the dimension by one. The configuration in the middle does not preserve this relation, therefore is not simple. For the middle picture, the bold edge of $\xi_2$ is not an edge of the resulting convex hull.}
    \label{fig:phis_dimension}
\end{figure}

The same argument is applied by induction to any number of binomial factors.
Namely, assume we have a product of $n$ binomials, which defines a face of dimension $n$. If we join to them one more binomial, the dimension will increase to $n+1$, due to convexity and the fact that the faces $\xi_{n}$ and $\xi_{n+1}$ are both simple (so it will never happen that the extra face $\xi_{n+1}$ will be linearly dependent with any of sub-faces of $\xi_{n}$. Since if it would, then we will unavoidable loose some of its edges by taking the convex hull, which contradicts the simplicity property, and also the fact that $\xi_{n}$ is actually a face of $\xi_{n+1}$), Figure \ref{fig:phis_dimension}. $\square$

It's also important to mention that each initial form is a summand of $\mathcal{R}$ (this follows directly from Definition \ref{init_form_def}), and $\mathcal{R}$ itself corresponds to the ``filled'' diagram:
\begin{equation}
\mathcal{R} \simeq 
    \begin{tabular}{|c c c|}
  $\bullet$ & $\bullet$ & \\      \hline
  $\bullet$ & $\bullet$ & $\bullet$ \\
  \vdots & \vdots & \vdots \\
  $\bullet$ & $\bullet$  & $\bullet$
\end{tabular}
\end{equation}
In what follows, we will consider only simple initial forms. Also we will often write $\xi$ instead of $init_{\xi}$, since there is one-to-one correspondence between faces and simple initial forms.

\section{Two-dimensional case}\label{section_two_dimensional}

Our first result concerns the $m = 2$ case:
\begin{equation}\label{Nahm_2-vertex}
\begin{tabular}{l l l}
  $F_0$ = & $a_0+a_1z_1z_2$ \\
  \hline
  $F_1$ = & $b_0+b_1z_1+b_2z_1^{\alpha}$ \\      
  $F_2$ = & $c_0+c_1z_2+c_2z_2^{\beta}$
\end{tabular}
\end{equation}
Without loss of generality, we assume $\alpha,\beta\geq 2$ (for $\alpha,\beta\in \{0,1\}$ the resultant polytope degenerates; for the negative values, after multiplying each $F_i$ by a suitable monomial, we end up with an equivalent polytope).
We have subtracted the anti-diagonal, since it simply amounts to a choice of framing. Therefore, the case $C = \mathrm{diag}(\alpha,\beta)$ is of our main interest.

For any $\alpha,\beta$ distinct from\footnote{If $\alpha\in \{0,1\}$ or $\beta\in \{0,1\}$, the resultant polytope can be obtained by edge contraction from $N_{2,2}$.} 0 and 1, the polytope $N(\mathcal{R}_{\mathbf{A}})$ coincides with the Gelfand-Kapranov-Zelevinsky (GKZ) polytope $N_{2,2}$, depicted on Figure \ref{fig:N22_polytope} \cite{GKZ,GKZ2}. By definition, GKZ polytope $N_{m',n'}$ is the Newton polytope of classical resultant $\mathrm{res}_z(f_0,f_1)$:
\begin{equation}\label{classical_resultant}
    \left\lbrace\begin{aligned}
        f_0 = &\ \tilde{a}_0 + \tilde{a}_1 z + \dots \tilde{a}_{m'}z^{m'} \\
        f_1 = &\ \tilde{b}_0 + \tilde{b}_1 z + \dots \tilde{b}_{n'}z^{n'} \\
    \end{aligned}\right.
    \quad \longrightarrow \quad N_{m',n'} := N\left(\mathrm{res}_z(f_0,f_1)\right).
\end{equation}
When $(m',n')=(2,2)$, this system is equivalent to (\ref{Nahm_2-vertex}) with $(\alpha,\beta)=(2,2)$. Indeed, we can solve $F_0$ for any of $z_1,z_2$ and plug the result into $F_1$ or $F_2$, being left with an equivalent system of quadratic equations in one variable.
It implies that $N(\mathcal{R}_{\mathbf{A}})=N_{2,2}$ for $(\alpha,\beta)=(2,2)$. However, varying $\alpha,\beta\geq 2$ in (\ref{Nahm_2-vertex}) does not change the polytope, since it corresponds to dilation of the lattice of $\mathbf{A}$, which is an affine transformation. Therefore, $N(\mathcal{R}_{\mathbf{A}})=N_{2,2}$
for any $\alpha,\beta\geq 2$.

We begin with the simplest non-trivial example.

\subsection*{Warm-up example: $(\alpha,\beta)=(2,2)$}\label{section_examples}

\begin{figure}[h!]
    \centering
    \tikzset{every picture/.style={line width=0.75pt}} 

\begin{tikzpicture}[x=0.75pt,y=0.75pt,yscale=-1,xscale=1]

\draw  [fill={rgb, 255:red, 0; green, 0; blue, 0 }  ,fill opacity=1 ] (356.76,40.14) .. controls (356.78,38.64) and (358.01,37.45) .. (359.51,37.47) .. controls (361,37.49) and (362.2,38.72) .. (362.18,40.21) .. controls (362.16,41.71) and (360.93,42.9) .. (359.43,42.88) .. controls (357.94,42.86) and (356.74,41.63) .. (356.76,40.14) -- cycle ;
\draw  [fill={rgb, 255:red, 0; green, 0; blue, 0 }  ,fill opacity=1 ] (418.14,40.14) .. controls (418.16,38.64) and (419.39,37.45) .. (420.88,37.47) .. controls (422.38,37.49) and (423.57,38.72) .. (423.55,40.21) .. controls (423.53,41.71) and (422.3,42.9) .. (420.81,42.88) .. controls (419.31,42.86) and (418.11,41.63) .. (418.14,40.14) -- cycle ;
\draw   (305.77,40.18) .. controls (305.77,31.7) and (317.79,24.83) .. (332.62,24.83) .. controls (347.45,24.83) and (359.47,31.7) .. (359.47,40.18) .. controls (359.47,48.65) and (347.45,55.52) .. (332.62,55.52) .. controls (317.79,55.52) and (305.77,48.65) .. (305.77,40.18) -- cycle ;
\draw   (420.84,40.18) .. controls (420.84,31.7) and (432.87,24.83) .. (447.69,24.83) .. controls (462.52,24.83) and (474.55,31.7) .. (474.55,40.18) .. controls (474.55,48.65) and (462.52,55.52) .. (447.69,55.52) .. controls (432.87,55.52) and (420.84,48.65) .. (420.84,40.18) -- cycle ;
\draw   (326.86,40.18) .. controls (326.86,35.03) and (334.16,30.86) .. (343.17,30.86) .. controls (352.17,30.86) and (359.47,35.03) .. (359.47,40.18) .. controls (359.47,45.32) and (352.17,49.49) .. (343.17,49.49) .. controls (334.16,49.49) and (326.86,45.32) .. (326.86,40.18) -- cycle ;
\draw   (420.84,40.18) .. controls (420.84,35.03) and (428.14,30.86) .. (437.15,30.86) .. controls (446.15,30.86) and (453.45,35.03) .. (453.45,40.18) .. controls (453.45,45.32) and (446.15,49.49) .. (437.15,49.49) .. controls (428.14,49.49) and (420.84,45.32) .. (420.84,40.18) -- cycle ;

\draw (350.77,56.56) node [anchor=north west][inner sep=0.75pt]  [font=\normalsize]  {$x_{1}$};
\draw (412.15,56.56) node [anchor=north west][inner sep=0.75pt]  [font=\normalsize]  {$x_{2}$};

\end{tikzpicture}
    \caption{Quiver with adjacency matrix $C = \mathrm{diag}(2,2)$}
    \label{fig:quiver22}
\end{figure}
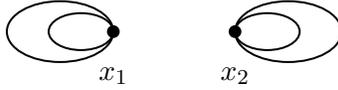

Take $C = \mathrm{diag}(2,2)$ (Figure \ref{fig:quiver22}).
Eliminating $z_1,z_2$ from (\ref{Nahm_2-vertex}) and specializing as in (\ref{quiver_A_poly_def}), we get 
\begin{equation}\label{diag22_resultant}
    A_{\mathrm{diag}(2,2)}(x_1,x_2,y) = x_1^2x_2^2y^4 + x_1x_2y^3 - 2x_1x_2y^2 + x_1y^2 + x_2y^2 + y + 1.
\end{equation}
\begin{figure}[h!]
    \centering
    \includegraphics[scale=0.3]{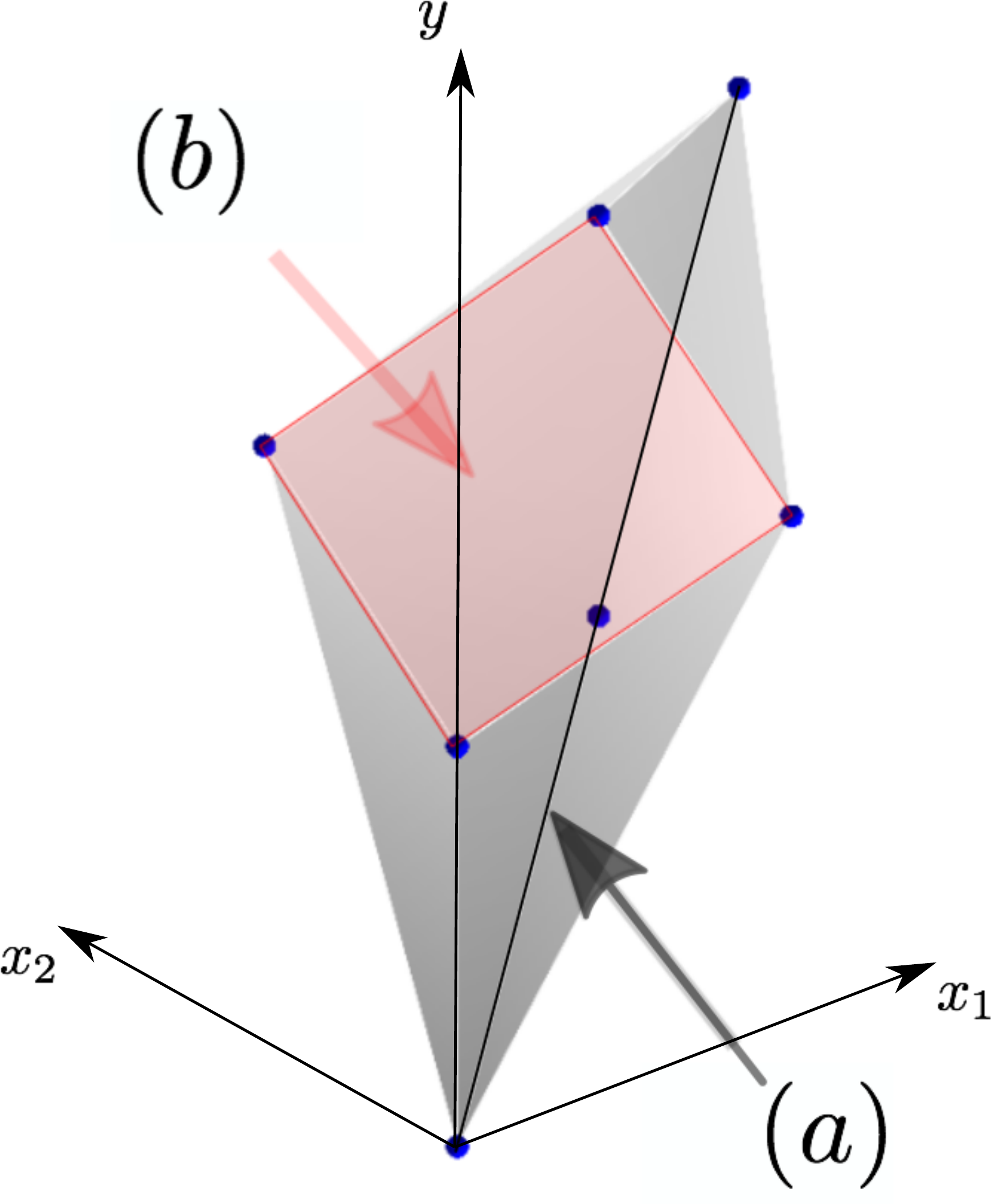}
    \caption{GKZ polytope $N_{2,2}$, along with the monomials (blue nodes) of $A_{\mathrm{diag}(2,2)}$, compare with the $(x,y)$-projection (\ref{fig:diag22_newton}).}
    \label{fig:N22_polytope}
\end{figure}
The vertices $(i,j,k)$ of $N_{2,2}$ encode the powers of monomials $x_1^ix_2^jy^k$ in (\ref{diag22_resultant}):
\begin{equation}
    (2,2,4),\ (1,1,3),\ (1,0,2),\ (0,1,2),\ (0,0,1),\ (0,0,0)
\end{equation}
Here only $-2x_1x_2y^2$ does not correspond to a vertex of $N_{2,2}$. Instead, it divides the bottom edge $(a)$ into two equal intervals (Figure \ref{fig:N22_polytope}). We can combine some monomials and re-write (\ref{diag22_resultant}) as
\begin{equation}\label{diag22_res_split}
    A_{\mathrm{diag}(2,2)}(x_1,x_2,y) = (x_1x_2y^2-1)^2 - y\,(x_1y+1)(x_2y+1)
\end{equation}
It turns out that the two binomial summands are supported on the two distinguished faces of $N_{2,2}$: the convex hull of $(x_1x_2y^2-1)^2$ gives the edge $(a)$, and $- y\,(x_1y+1)(x_2y+1)$ gives the 2-dimensional face $(b)$. The latter belongs to the plane defined by equation:
\begin{equation}
    1 + x_1 + x_2 - y = 0
\end{equation}
Its normal vector is $\omega = (1,1,-1)$, up to translation and multiplication by a scalar. Rescale the variables $x_1,x_2,y$ with respect to this vector:
\begin{equation}
    A_{\mathrm{diag}(2,2)}(c^{1}x_1,c^{1}x_2,c^{-1}y) = (x_1x_2y^2-1)^2 - c^{-1} y\,(x_1y+1)(x_2y+1),\quad c\in \mathbb{C}
\end{equation}
Therefore, the parameter $c$ separates the faces of $N_{2,2}$ as the two summands in (\ref{diag22_res_split}). So we get the two distinguished initial forms:
\begin{equation}\label{m22_inits}
\begin{aligned}
init_{a} = &\ \lim_{c\rightarrow \infty}\mathcal{R}_{\mathrm{diag}(2,2)}(c^{\omega_1}x_1,c^{\omega_2}x_2,c^{\omega_3}y) & = &\  (x_1x_2y^2-1)^2 \\
    init_{b} = &\ \lim_{c\rightarrow 0}(c\cdot \mathcal{R}_{\mathrm{diag}(2,2)}(c^{\omega_1}x_1,c^{\omega_2}x_2,c^{\omega_3}y)) & = &\ y\,(x_1y+1)(x_2y+1) \\
\end{aligned}   
\end{equation}
Let us move to the unspecialized case.
The (refined) quiver resultant from (\ref{Nahm_2-vertex}) reads:
\begin{equation}\label{diag22_resultant_unspec}
\mathcal{R}(a_0,a_1,\mathbf{b}) = (a_0^2b_2c_2-a_1^2b_0c_0)^2 + a_0a_1\,(a_0b_2c_1+a_1b_1c_0)(a_0b_1c_2+a_1b_0c_1).
\end{equation}
We have: $\mathcal{R}(y,-1,-1,1,x_1,-1,1,x_2) = A_{\mathrm{diag}(2,2)}(x_1,x_2,y)$.

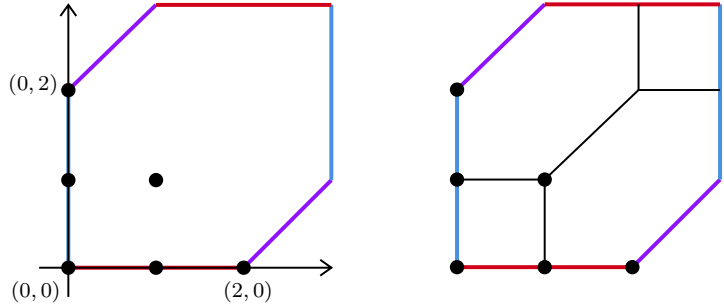
\begin{figure}[h!]
	\centering
    \begin{tikzpicture}[x=0.75pt,y=0.75pt,yscale=-0.8,xscale=0.8]

\draw [color={rgb, 255:red, 144; green, 19; blue, 254 }  ,draw opacity=1 ][fill={rgb, 255:red, 0; green, 0; blue, 0 }  ,fill opacity=1 ][line width=1.5]    (205.49,2403.72) -- (260.76,2348.45) ;
\draw [color={rgb, 255:red, 208; green, 2; blue, 27 }  ,draw opacity=1 ][fill={rgb, 255:red, 0; green, 0; blue, 0 }  ,fill opacity=1 ][line width=1.5]    (98.6,2403.72) -- (209.38,2403.72) ;
\draw [color={rgb, 255:red, 74; green, 144; blue, 226 }  ,draw opacity=1 ][fill={rgb, 255:red, 0; green, 0; blue, 0 }  ,fill opacity=1 ][line width=1.5]    (94.95,2399.83) -- (94.95,2291.72) ;
\draw [color={rgb, 255:red, 144; green, 19; blue, 254 }  ,draw opacity=1 ][fill={rgb, 255:red, 0; green, 0; blue, 0 }  ,fill opacity=1 ][line width=1.5]    (94.95,2291.72) -- (150.22,2237.91) ;
\draw  [fill={rgb, 255:red, 0; green, 0; blue, 0 }  ,fill opacity=1 ] (146.33,2403.72) .. controls (146.33,2401.58) and (148.07,2399.83) .. (150.22,2399.83) .. controls (152.37,2399.83) and (154.11,2401.58) .. (154.11,2403.72) .. controls (154.11,2405.87) and (152.37,2407.61) .. (150.22,2407.61) .. controls (148.07,2407.61) and (146.33,2405.87) .. (146.33,2403.72) -- cycle ;
\draw  [fill={rgb, 255:red, 0; green, 0; blue, 0 }  ,fill opacity=1 ] (91.06,2348.45) .. controls (91.06,2346.31) and (92.8,2344.56) .. (94.95,2344.56) .. controls (97.1,2344.56) and (98.84,2346.31) .. (98.84,2348.45) .. controls (98.84,2350.6) and (97.1,2352.34) .. (94.95,2352.34) .. controls (92.8,2352.34) and (91.06,2350.6) .. (91.06,2348.45) -- cycle ;
\draw  [fill={rgb, 255:red, 0; green, 0; blue, 0 }  ,fill opacity=1 ] (91.06,2291.72) .. controls (91.06,2289.57) and (92.8,2287.83) .. (94.95,2287.83) .. controls (97.1,2287.83) and (98.84,2289.57) .. (98.84,2291.72) .. controls (98.84,2293.87) and (97.1,2295.61) .. (94.95,2295.61) .. controls (92.8,2295.61) and (91.06,2293.87) .. (91.06,2291.72) -- cycle ;
\draw  [fill={rgb, 255:red, 0; green, 0; blue, 0 }  ,fill opacity=1 ] (201.6,2403.72) .. controls (201.6,2401.58) and (203.34,2399.83) .. (205.49,2399.83) .. controls (207.64,2399.83) and (209.38,2401.58) .. (209.38,2403.72) .. controls (209.38,2405.87) and (207.64,2407.61) .. (205.49,2407.61) .. controls (203.34,2407.61) and (201.6,2405.87) .. (201.6,2403.72) -- cycle ;
\draw  [fill={rgb, 255:red, 0; green, 0; blue, 0 }  ,fill opacity=1 ] (91.06,2403.72) .. controls (91.06,2401.58) and (92.8,2399.83) .. (94.95,2399.83) .. controls (97.1,2399.83) and (98.84,2401.58) .. (98.84,2403.72) .. controls (98.84,2405.87) and (97.1,2407.61) .. (94.95,2407.61) .. controls (92.8,2407.61) and (91.06,2405.87) .. (91.06,2403.72) -- cycle ;
\draw [color={rgb, 255:red, 74; green, 144; blue, 226 }  ,draw opacity=1 ][fill={rgb, 255:red, 0; green, 0; blue, 0 }  ,fill opacity=1 ][line width=1.5]    (260.76,2348.45) -- (260.76,2237.91) ;
\draw [color={rgb, 255:red, 208; green, 2; blue, 27 }  ,draw opacity=1 ][fill={rgb, 255:red, 0; green, 0; blue, 0 }  ,fill opacity=1 ][line width=1.5]    (150.22,2237.91) -- (260.76,2237.91) ;
\draw  [fill={rgb, 255:red, 0; green, 0; blue, 0 }  ,fill opacity=1 ] (146.33,2348.45) .. controls (146.33,2346.31) and (148.07,2344.56) .. (150.22,2344.56) .. controls (152.37,2344.56) and (154.11,2346.31) .. (154.11,2348.45) .. controls (154.11,2350.6) and (152.37,2352.34) .. (150.22,2352.34) .. controls (148.07,2352.34) and (146.33,2350.6) .. (146.33,2348.45) -- cycle ;
\draw [color={rgb, 255:red, 144; green, 19; blue, 254 }  ,draw opacity=1 ][fill={rgb, 255:red, 0; green, 0; blue, 0 }  ,fill opacity=1 ][line width=1.5]    (450.53,2403.37) -- (505.8,2348.1) ;
\draw [color={rgb, 255:red, 208; green, 2; blue, 27 }  ,draw opacity=1 ][fill={rgb, 255:red, 0; green, 0; blue, 0 }  ,fill opacity=1 ][line width=1.5]    (343.63,2403.37) -- (454.41,2403.37) ;
\draw [color={rgb, 255:red, 74; green, 144; blue, 226 }  ,draw opacity=1 ][fill={rgb, 255:red, 0; green, 0; blue, 0 }  ,fill opacity=1 ][line width=1.5]    (339.98,2399.48) -- (339.98,2291.37) ;
\draw [color={rgb, 255:red, 144; green, 19; blue, 254 }  ,draw opacity=1 ][fill={rgb, 255:red, 0; green, 0; blue, 0 }  ,fill opacity=1 ][line width=1.5]    (339.98,2291.37) -- (395.25,2237.56) ;
\draw  [fill={rgb, 255:red, 0; green, 0; blue, 0 }  ,fill opacity=1 ] (391.37,2403.37) .. controls (391.37,2401.22) and (393.11,2399.48) .. (395.25,2399.48) .. controls (397.4,2399.48) and (399.14,2401.22) .. (399.14,2403.37) .. controls (399.14,2405.51) and (397.4,2407.26) .. (395.25,2407.26) .. controls (393.11,2407.26) and (391.37,2405.51) .. (391.37,2403.37) -- cycle ;
\draw  [fill={rgb, 255:red, 0; green, 0; blue, 0 }  ,fill opacity=1 ] (336.09,2348.1) .. controls (336.09,2345.95) and (337.84,2344.21) .. (339.98,2344.21) .. controls (342.13,2344.21) and (343.87,2345.95) .. (343.87,2348.1) .. controls (343.87,2350.24) and (342.13,2351.99) .. (339.98,2351.99) .. controls (337.84,2351.99) and (336.09,2350.24) .. (336.09,2348.1) -- cycle ;
\draw  [fill={rgb, 255:red, 0; green, 0; blue, 0 }  ,fill opacity=1 ] (336.09,2291.37) .. controls (336.09,2289.22) and (337.84,2287.48) .. (339.98,2287.48) .. controls (342.13,2287.48) and (343.87,2289.22) .. (343.87,2291.37) .. controls (343.87,2293.51) and (342.13,2295.25) .. (339.98,2295.25) .. controls (337.84,2295.25) and (336.09,2293.51) .. (336.09,2291.37) -- cycle ;
\draw  [fill={rgb, 255:red, 0; green, 0; blue, 0 }  ,fill opacity=1 ] (446.64,2403.37) .. controls (446.64,2401.22) and (448.38,2399.48) .. (450.53,2399.48) .. controls (452.67,2399.48) and (454.41,2401.22) .. (454.41,2403.37) .. controls (454.41,2405.51) and (452.67,2407.26) .. (450.53,2407.26) .. controls (448.38,2407.26) and (446.64,2405.51) .. (446.64,2403.37) -- cycle ;
\draw  [fill={rgb, 255:red, 0; green, 0; blue, 0 }  ,fill opacity=1 ] (336.09,2403.37) .. controls (336.09,2401.22) and (337.84,2399.48) .. (339.98,2399.48) .. controls (342.13,2399.48) and (343.87,2401.22) .. (343.87,2403.37) .. controls (343.87,2405.51) and (342.13,2407.26) .. (339.98,2407.26) .. controls (337.84,2407.26) and (336.09,2405.51) .. (336.09,2403.37) -- cycle ;
\draw [color={rgb, 255:red, 74; green, 144; blue, 226 }  ,draw opacity=1 ][fill={rgb, 255:red, 0; green, 0; blue, 0 }  ,fill opacity=1 ][line width=1.5]    (505.8,2348.1) -- (505.8,2237.56) ;
\draw [color={rgb, 255:red, 208; green, 2; blue, 27 }  ,draw opacity=1 ][fill={rgb, 255:red, 0; green, 0; blue, 0 }  ,fill opacity=1 ][line width=1.5]    (395.25,2237.56) -- (505.8,2237.56) ;
\draw  [fill={rgb, 255:red, 0; green, 0; blue, 0 }  ,fill opacity=1 ] (391.37,2348.1) .. controls (391.37,2345.95) and (393.11,2344.21) .. (395.25,2344.21) .. controls (397.4,2344.21) and (399.14,2345.95) .. (399.14,2348.1) .. controls (399.14,2350.24) and (397.4,2351.99) .. (395.25,2351.99) .. controls (393.11,2351.99) and (391.37,2350.24) .. (391.37,2348.1) -- cycle ;
\draw    (395.25,2348.1) -- (343.87,2348.1) ;
\draw    (395.25,2407.26) -- (395.25,2349.31) ;
\draw    (505.8,2291.61) -- (454.41,2291.61) ;
\draw    (454.41,2291.61) -- (454.41,2237.56) ;
\draw    (454.41,2291.61) -- (395.25,2348.1) ;
\draw  (76.53,2403.72) -- (260.76,2403.72)(94.95,2237.91) -- (94.95,2422.15) (253.76,2398.72) -- (260.76,2403.72) -- (253.76,2408.72) (89.95,2244.91) -- (94.95,2237.91) -- (99.95,2244.91)  ;

\draw (56.74,2410.66) node [anchor=north west][inner sep=0.75pt]  [font=\scriptsize]  {$( 0,0)$};
\draw (190.6,2410.5) node [anchor=north west][inner sep=0.75pt]  [font=\scriptsize]  {$( 2,0)$};
\draw (55.21,2280.01) node [anchor=north west][inner sep=0.75pt]  [font=\scriptsize]  {$( 0,2)$};

\end{tikzpicture}
	\caption{Mixed decompositions in $(z_1,z_2)$-plane: for $init_a$ (left) and $init_b$ (right)}
	\label{fig:TCMD_m2}
\end{figure}

The advantage of the refined quiver resultant is that we can study combinatorics of the Minkowski sum $Q = Q_0 + Q_1 + Q_2$, where $Q_i = \mathrm{conv}(F_i)$. This would be impossible when dealing with the specialized case (\ref{diag22_resultant}).
Consider the first initial form: $(a_0^2b_2c_2-a_1^2b_0c_0)^2$. 
It is attached to the bottom edge $(a)$ of $N_{2,2}$. The square comes from the areal factor of $Q$, which is the largest hexagon (Figure \ref{fig:TCMD_m2}, left).
The total degree of a given binomial equals to the euclidean volume of the corresponding cell of a mixed decomposition, which is equal to 8 in our case. Notice that the binomial $a_0^2b_2c_2-a_1^2b_0c_0$ is the sub-resultant for $b_1 = c_1 = 0$. 

At last, consider the second initial form $a_0a_1\,(a_0b_2c_1+a_1b_1c_0)(a_0b_1c_2+a_1b_0c_1)$. It splits into the product of four distinct sub-resultants, which represent four distinct cells of our mixed decomposition on Figure \ref{fig:TCMD_m2}, right:
\begin{equation}
\begin{aligned}
&\ b_0 = c_2 = 0, &\ a_0b_2c_1 + a_1b_1c_0 \\
&\ b_2 = c_0 = 0, &\ a_0b_1c_2 + a_1b_0c_1 \\
&\ a_1 = b_2 = c_2 = 0, &\ a_0 \\
&\ a_0 = b_0 = c_0 = 0, &\ a_1
\end{aligned}
\end{equation}

Therefore, one may search for all possible mixed decompositions of $Q$ and compute the associated initial forms from the cell arrangement in each decomposition. This is the meaning of the formula (\ref{initial_factorization}).

\subsection*{General $(\alpha,\beta)$}

Moving to the general case $\alpha,\beta\geq 2$, we have to introduce an operator which implements the rule for computing the exponents $k_{\iota}$ in (\ref{initial_factorization}).

\begin{figure}[h!]
    \centering
    \input{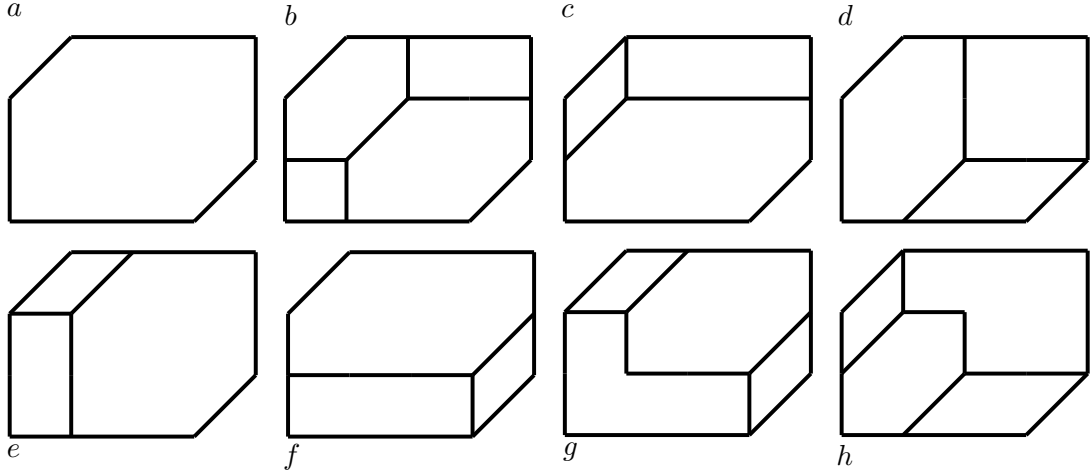}
    \caption{TCMDs of $Q$, associated to the faces of $N_{2,2}$}
    \label{fig:n22TCMD_all}
\end{figure}

\begin{definition}\label{gcd_map}
Given a binomial $\eta^{\mathbf{p}} + \theta^{\mathbf{q}}, \mathbf{p}=(p_1,\dots,p_k),\mathbf{q}=(q_1,\dots,q_k)$, define
\begin{equation}
    \mathrm{GCD}\left(\eta^{\mathbf{p}} + \theta^{\mathbf{q}}\right) := \left(\eta^{\frac{\mathbf{p}}{\mathrm{gcd}(\mathbf{p},\mathbf{q})}} + \theta^{\frac{\mathbf{q}}{\mathrm{gcd}(\mathbf{p},\mathbf{q})}}\right)^{\mathrm{gcd}(\mathbf{p},\mathbf{q})},
\end{equation}
where $\mathrm{gcd}(\mathbf{p},\mathbf{q})$ acts on the two vectors component-wise. Also, for any integer $s\geq 1$
\begin{equation}\label{GCD_product_rule}
    \mathrm{GCD}\left(\prod_{i=1\dots s}(\eta^{\mathbf{p}_i}+\theta^{\mathbf{q}_i})\right) = \prod_{i=1\dots s} \mathrm{GCD}(\eta^{\mathbf{p}_i}+\theta^{\mathbf{q}_i})
\end{equation}
\end{definition}
For example:
\begin{equation}
\begin{aligned}
    \mathrm{GCD}\left(a_0^2b_2^2c_2 + a_1^2b_0^2c_0  \right) = &\ a_0^2b_2^2c_2 + a_1^2b_0^2c_0, \\
        \mathrm{GCD}\left(a_0^4b_2^2c_2^2 + a_1^4b_0^2c_0^2  \right) = &\ (a_0^2b_2c_2 + a_1^2b_0c_0)^2. 
\end{aligned}
\end{equation}
\begin{proposition}\label{m=2_proposition_initial} The Newton polytope $N(\mathcal{R})$ for the system (\ref{Nahm_2-vertex}) supports the following simple initial forms:
\\
\scalebox{0.8}{\parbox{.5\linewidth}{%
\begin{align*}
init_a =&\ \mathrm{GCD}\left(a_0^{\alpha\beta}b_2^{\beta}c_2^{\alpha}+(-1)^{\alpha\beta+\alpha+\beta}a_1^{\alpha\beta}b_0^{\beta}c_0^{\alpha}\right), 
\begin{tabular}{|c c c|}
  $\bullet$ & $\bullet$ & \\      \hline
  $\bullet$ &  & $\bullet$ \\      
  $\bullet$ &  & $\bullet$ \\
\end{tabular} 
\\
init_b =&\ a_0a_1^{(\alpha-1)(\beta-1)}(a_0^{\alpha-1}b_2c_1^{\alpha-1}+a_1^{\alpha-1}b_1c_0^{\alpha-1})(a_0^{\beta-1}c_2b_1^{\beta-1}+a_1^{\beta-1}c_1b_0^{\beta-1}),  
\begin{tabular}{|c c c|}
  $\bullet$ & $\bullet$ & \\      \hline
  $\bullet$ & $\bullet$ & \\      
  & $\bullet$ & $\bullet$ \\
\end{tabular} 
\times
\begin{tabular}{|c c c|}
  $\bullet$ & $\bullet$ & \\      \hline
  & $\bullet$ & $\bullet$ \\      
  $\bullet$ & $\bullet$ & \\
\end{tabular} 
\\
init_c =&\ a_1^{\alpha(\beta-1)}b_0^{\beta-1}(a_0^{\alpha}b_2c_1^{\alpha}+a_1^{\alpha}b_0c_2^{\alpha}), 
\begin{tabular}{|c c c|}
  $\bullet$ & $\bullet$ & \\      \hline
  $\bullet$ & & $\bullet$ \\      
  $\bullet$ & $\bullet$ & \\
\end{tabular}
\\
init_d =&\ a_1^{(\alpha-1)\beta}c_0^{\alpha-1}(a_0^{\beta}b_1^{\beta}c_2+a_1^{\beta}b_2^{\beta}c_0), 
\begin{tabular}{|c c c|}
  $\bullet$ & $\bullet$ & \\      \hline
  $\bullet$ & $\bullet$ & \\      
  $\bullet$ & & $\bullet$ \\
\end{tabular}
\\
init_e =&\ a_0^{\beta}c_2\cdot\mathrm{GCD}\left(a_0^{(\alpha-1)\beta}b_2^{\beta}c_2^{\alpha-1}+(-1)^{(\alpha-1)\beta+(\alpha-1)+\beta}a_1^{(\alpha-1)\beta}b_1^{\beta}c_0^{\alpha-1}\right), 
\begin{tabular}{|c c c|}
  $\bullet$ & $\bullet$ & \\      \hline
  & $\bullet$ & $\bullet$ \\      
  $\bullet$ & & $\bullet$ \\
\end{tabular}
\\
init_f =&\ a_0^{\alpha}b_2\cdot\mathrm{GCD}\left(a_0^{\alpha(\beta-1)}b_2^{\beta-1}c_2^{\alpha}+(-1)^{(\alpha-1)\beta+(\alpha-1)+\beta}a_1^{\alpha(\beta-1)}b_0^{\beta-1}c_1^{\alpha}\right), 
\begin{tabular}{|c c c|}
  $\bullet$ & $\bullet$ & \\      \hline
  $\bullet$ & & $\bullet$ \\      
  & $\bullet$ & $\bullet$ \\
\end{tabular}
\\
init_g =&\ a_0^{\alpha+\beta-1}b_2c_2\cdot\mathrm{GCD}\left(a_0^{(\alpha-1)(\beta-1)}b_2^{\beta-1}c_2^{\alpha-1}+(-1)^{(\alpha-1)(\beta-1)+(\alpha-1)+(\beta-1))}a_1^{(\alpha-1)(\beta-1)}b_1^{\beta-1}c_1^{\alpha-1}\right),
\begin{tabular}{|c c c|}
  $\bullet$ & $\bullet$ & \\      \hline
  & $\bullet$ & $\bullet$ \\      
  & $\bullet$ & $\bullet$ \\
\end{tabular}
\\
init_h =&\ a_1^{\alpha\beta-1}b_0^{\beta-1}c_0^{\alpha-1}(a_0b_1c_1+a_1b_0c_0), 
\begin{tabular}{|c c c|}
  $\bullet$ & $\bullet$ & \\      \hline
  $\bullet$ & $\bullet$ & \\      
  $\bullet$ & $\bullet$ & \\
\end{tabular}
\end{align*}   
}}
\\
where $a,b,c,d,e,f,g,h$ are the faces of $N(\mathcal{R})$ (Figure \ref{fig:N22}), and perfograms on the right correspond to distinct binomial factors.
\begin{figure}[h!]
    \centering
    \includegraphics[scale=0.6]{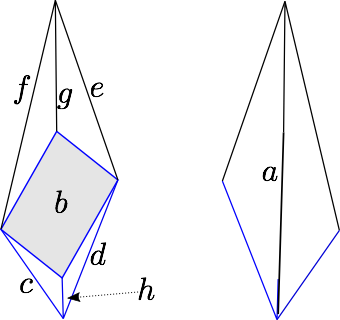}
    \caption{The Newton polytope $N_{2,2}$ with the initial forms (\ref{Nahm_2-vertex}): top (left) and bottom (right). Blue faces are those, which do not have other points rather than the vertices}
    \label{fig:N22}
\end{figure}
\end{proposition}
\underline{Proof}. Every binomial factor in $init_*$ correspond to a sub-resultant, which perfogram is given on the right side of each expression in Proposition \ref{m=2_proposition_initial}. Let's associate mixed decompositions to these initial forms, as shown on Figure \ref{fig:n22TCMD_all}.
This provides a desired combinatorial interpretation of the faces. Each hexagon in a mixed decomposition gives the distinct binomial factor in the corresponding initial form, and all rectangles together determine the monomial prefactor. The GCD operator has the following interpretation: 
each $k_{\iota}\geq 1$ in (\ref{initial_factorization}) is uniquely fixed when $(\alpha,\beta)$ are fixed, so that the total degree of $\tilde{\mathcal{R}}_{\iota}^{k_i}$ equals to the area of the $\iota$-th cell of a mixed decomposition.

E.g., for $init_a$ there is only a single hexagon (the top-left in Figure \ref{fig:n22TCMD_all}), which is $Q$ itself -- so there is no monomial prefactor. This hexagon gives the sub-resultant $\tilde{\mathcal{R}} = a_0^{\alpha\beta}b_2^{\beta}c_2^{\alpha}+(-1)^{\alpha\beta+\alpha+\beta}a_1^{\alpha\beta}b_0^{\beta}c_0^{\alpha}$. We see that the area of $Q$ is $\alpha\beta+\alpha+\beta$, so if $\alpha$ and $\beta$ are not co-prime, it would give $k > 1$, hence $$init_a = \mathrm{GCD}\left(a_0^{\alpha\beta}b_2^{\beta}c_2^{\alpha}+(-1)^{\alpha\beta+\alpha+\beta}a_1^{\alpha\beta}b_0^{\beta}c_0^{\alpha}\right).$$

In the case of $init_b$ we have two hexagons (giving the two distinct binomial factors) and two quadrangles for the monomial: the bottom square is $a_0$, and the top quadrangle is $a_1^{(\alpha-1)(\beta-1)}$ (compare with Figure \ref{fig:TCMD_demo}). The rest is carried out analogously.

It turns out that using the mixed decompositions $a,b,c,d,e,f,g,h$ we completely described the bijection between the faces and simple initial forms. $\square$

\begin{corollary}
Quiver A-polynomial for any two-vertex quiver is tempered, with its face polynomials all being binomials.
\end{corollary}
It follows directly from factorization formulas for the initial forms $a,b,c,d,e,f,g,h$.
The polytope $N(\mathcal{R})$ projects onto $N(A)$ in such a way that the faces of $N(\mathcal{R})$ do not overlap each other (colliding the axes $x_1$ and $x_2$ on Figure \ref{fig:N22_polytope} to obtain the projection shown on Figure \ref{fig:diag22_newton}).
Binomiality of the initial forms of Proposition \ref{m=2_proposition_initial} implies binomiality of the face polynomials, which means that $A(x,y)$ is tempered.
Notice that the non-diagonal case is simply a framing transformation $x\mapsto xy^f$, which amounts to equivalence of the polytopes, therefore not bringing any substantial changes. This is clear on the level of quiver A-polynomials even for generic $m$: framing transformation maps $A(x_1,\dots,x_m,y)$ into $A'=A(x_1y^f,\dots,x_my^f,y)$, so the two polytopes $N(A)$ and $N(A')$ are equivalent up to dilation of the axes $x_1,\dots,x_m$, and the shape of the initial forms is preserved.

\section{Three-dimensional case}\label{section_three_dimensional}
The Nahm equations for $C = \mathrm{diag}(\alpha,\beta,\gamma)$ take form:
\begin{equation}\label{Nahm_m3}
\begin{tabular}{l l l}
  $F_0$ = & $a_0+a_1z_1z_2z_3$ \\
  \hline
  $F_1$ = & $b_0+b_1z_1+b_2z_1^{\alpha}$ \\      
  $F_2$ = & $c_0+c_1z_2+c_2z_2^{\beta}$ \\
  $F_3$ = & $d_0+d_1z_3+d_2z_3^{\gamma}$
\end{tabular}
\end{equation}
We assume $\alpha,\beta,\gamma\geq 2$ and introduce the initial forms $init_{\phi_{p,q}}$ (which we will shortly write as $\phi_{p,q}$, at the same time referring to the corresponding face of $N(\mathcal{R})$), labelled by the two non-negative integers.
These initial forms are given by products over all permutations of perfograms with $p$ rows of the form $[\bullet \bullet\ \ ]$
and $q$ rows of the form $[\textcolor{red}{\ \ \bullet \bullet}]$, such that $p+q = m$ (the red color is just for a better visuals). 
The only exception is $\phi_{0,0}$, which rows are of the form $[\bullet \ \ \bullet]$.
For $m = 3$, they are given by
\begin{equation}\label{m=3_phis}
\begin{aligned}
\phi_{0,0}: &\
\begin{tabular}{|c c c|}
  $\bullet$ &  $\bullet$ & \\
  \hline
  $\bullet$ &  & $\bullet$ \\      
  $\bullet$ &  & $\bullet$ \\      
  $\bullet$ &  & $\bullet$ \\
\end{tabular}
\qquad
\phi_{3,0}: 
\begin{tabular}{|c c c|}
  $\bullet$ &  $\bullet$ & \\
  \hline
  $\bullet$ &   $\bullet$ & \\      
  $\bullet$ &   $\bullet$ & \\      
  $\bullet$ &   $\bullet$ & \\
\end{tabular}
\qquad
\phi_{0,3}: 
\begin{tabular}{|c c c|}
  $\bullet$ &  $\bullet$ & \\
  \hline
  & \textcolor{red}{$\bullet$} & \textcolor{red}{$\bullet$} \\      
  & \textcolor{red}{$\bullet$} & \textcolor{red}{$\bullet$} \\          
  & \textcolor{red}{$\bullet$} & \textcolor{red}{$\bullet$} \\    
\end{tabular}
\\ \ \\
\phi_{2,1}: &\ 
\begin{tabular}{|c c c|}
  $\bullet$ &  $\bullet$ & \\ 
  \hline
  & \textcolor{red}{$\bullet$} & \textcolor{red}{$\bullet$} \\    
  $\bullet$ & $\bullet$ &  \\      
  $\bullet$ & $\bullet$ &       
\end{tabular}
\ \times\ 
\begin{tabular}{|c c c|}
  $\bullet$ &  $\bullet$ & \\
  \hline
  $\bullet$ & $\bullet$ &  \\      
  & \textcolor{red}{$\bullet$} & \textcolor{red}{$\bullet$} \\    
  $\bullet$ & $\bullet$ &       
\end{tabular}
\ \times\ 
\begin{tabular}{|c c c|}
  $\bullet$ &  $\bullet$ & \\
  \hline
  $\bullet$ & $\bullet$ &  \\      
  $\bullet$ & $\bullet$ &   \\    
  & \textcolor{red}{$\bullet$} & \textcolor{red}{$\bullet$} \\    
\end{tabular}
\\ \ \\
\phi_{1,2}: &\ 
\begin{tabular}{|c c c|}
  $\bullet$ &  $\bullet$ & \\
  \hline
  $\bullet$ & $\bullet$ & \\
  & \textcolor{red}{$\bullet$} & \textcolor{red}{$\bullet$} \\         
  & \textcolor{red}{$\bullet$} & \textcolor{red}{$\bullet$} \\          
\end{tabular}
\ \times\ 
\begin{tabular}{|c c c|}
  $\bullet$ &  $\bullet$ & \\
  \hline
  & \textcolor{red}{$\bullet$} & \textcolor{red}{$\bullet$} \\          
  $\bullet$ & $\bullet$ & \\
  & \textcolor{red}{$\bullet$} & \textcolor{red}{$\bullet$} \\           
\end{tabular}
\ \times\ 
\begin{tabular}{|c c c|}
  $\bullet$ &  $\bullet$ & \\
  \hline
  & \textcolor{red}{$\bullet$} & \textcolor{red}{$\bullet$} \\          
  & \textcolor{red}{$\bullet$} & \textcolor{red}{$\bullet$} \\        
  $\bullet$ & $\bullet$ &
\end{tabular}
\end{aligned}
\end{equation}
\begin{equation}\label{m=3_phi_formulas}
\begin{aligned}
\phi_{0,0} = &\ \mathrm{GCD}\left({a_{{0}}}^{\alpha\,\beta\,\gamma}{b_{{2}}}^{\beta\,\gamma}{c_{{2}}}^{
\alpha\,\gamma}{d_{{2}}}^{\alpha\,\beta}+ \left( -1 \right) ^{\sigma+1}{a_{{1}}}^{
\alpha\,\beta\,\gamma}{b_{{0}}}^{\beta\,\gamma}{c_{{0}}}^{\alpha\,
\gamma}{d_{{0}}}^{\alpha\,\beta}
\right), \\
\phi_{3,0} = &\ \mu_{3,0}\cdot(a_0b_1c_1d_1 - a_1b_0c_0d_0),
\\
\phi_{2,1} = &\
\mu_{2,1}\cdot \mathrm{GCD}\left(\left( {a_{{0}}}^{\alpha-1}b_{{2}}{c_{{1}}}^{\alpha-1}{d_{{1}}}^{
\alpha-1}+ \left( -1 \right) ^{\alpha+1}{a_{{1}}}^{\alpha-1}b_{{1}}{c_{{0}}}^{\alpha-1}{d_{{0}}}^{\alpha-1} \right)\right.\times \\
&\ \left( {a_{{0}}}^{\beta
-1}c_{{2}}{b_{{1}}}^{\beta-1}{d_{{1}}}^{\beta-1}+ \left( -1 \right) ^{
\beta+1}{a_{{1}}}^{\beta-1}c_{{1}}{b_{{0}}}^{\beta-1}{d_{{0}}}^{\beta-
1} \right)\times  \\
&\ \left.\left( {a_{{0}}}^{\gamma-1}d_{{2}}{b_{{1}}}^{\gamma-1}{c_{
{1}}}^{\gamma-1}+ \left( -1 \right) ^{\gamma+1}{a_{{1}}}^{\gamma-1}d_{
{1}}{b_{{0}}}^{\gamma-1}{c_{{0}}}^{\gamma-1} \right) \right), \\
\phi_{1,2} = &\ 
\mu_{1,2}\cdot \mathrm{GCD}\left(\left({a_{{0}}}^{ \left( \beta-1 \right)  \left( \gamma-1 \right) }{b_{{1}}}
^{ \left( \beta-1 \right)  \left( \gamma-1 \right) }{c_{{2}}}^{\gamma-
1}{d_{{2}}}^{\beta-1}+\right.\right. \\
&\
\left.\left.\left( -1 \right) ^{\beta+\gamma+1}{a_{{1}}}^{
 \left( \beta-1 \right)  \left( \gamma-1 \right) }{b_{{0}}}^{ \left( 
\beta-1 \right)  \left( \gamma-1 \right) }{c_{{1}}}^{\gamma-1}{d_{{1}}
}^{\beta-1}\right)\right. \times \\
&\ \left({a_{{0}}}^{ \left( \alpha-1 \right)  \left( \gamma-1 \right) }{c_{{1}}}^{ \left( \alpha-1 \right)  \left( \gamma-1 \right) }{b_{{2}}}^{
\gamma-1}{d_{{2}}}^{\alpha-1}+\right. \\
&\
\left.
\left( -1 \right) ^{\alpha+\gamma+1}{a_
{{1}}}^{ \left( \alpha-1 \right)  \left( \gamma-1 \right) }{c_{{0}}}^{
 \left( \alpha-1 \right)  \left( \gamma-1 \right) }{b_{{1}}}^{\gamma-1
}{d_{{1}}}^{\alpha-1}\right) \times \\
&\ \left. \left({a_{{0}}}^{ \left( \alpha-1 \right)  \left( \beta-1 \right) }{d_{{1}}}
^{ \left( \alpha-1 \right)  \left( \beta-1 \right) }{b_{{2}}}^{\beta-1
}{c_{{2}}}^{\alpha-1}+\right.\right. \\
&\
\left.\left.
\left( -1 \right) ^{\alpha+\beta+1}{a_{{1}}}^{
 \left( \alpha-1 \right)  \left( \beta-1 \right) }{d_{{0}}}^{ \left( 
\alpha-1 \right)  \left( \beta-1 \right) }{b_{{1}}}^{\beta-1}{c_{{1}}}
^{\alpha-1}\right) \right), \\
\phi_{0,3} = &\ \mu_{0,3}\cdot\mathrm{GCD}\left(a_0^{(\alpha-1)(\beta-1)(\gamma-1)}b_2^{(\beta-1)(\gamma-1)}c_2^{(\alpha-1)(\gamma-1)}d_2^{(\alpha-1)(\beta-1)}+\right. \\
&\ \left. (-1)^{(\alpha-1)+(\beta-1)+(\gamma-1)+1}a_1^{(\alpha-1)(\beta-1)(\gamma-1)}b_1^{(\beta-1)(\gamma-1)}c_1^{(\alpha-1)(\gamma-1)}d_1^{(\alpha-1)(\beta-1)} \right),
\end{aligned}
\end{equation}
where $\sigma = \alpha\,
\beta\,\gamma+\alpha\,\beta+\alpha\,\gamma+\beta\,\gamma$, and the monomials are:
\begin{equation}\label{m=3_phi_monomials}
\begin{aligned}
\mu_{3,0} = &\ a_1^{\alpha\beta\gamma-1}b_0^{\beta\gamma-1}c_0^{\alpha\gamma-1}d_0^{\alpha\beta-1}, \\
\mu_{2,1} = &\ a_{{0}}{a_{{1}}}^{\alpha\,\beta\,\gamma-\alpha-\beta-\gamma+2}{b_{{0}}
}^{ \left( \beta-1 \right)  \left( \gamma-1 \right) }{c_{{0}}}^{
 \left( \alpha-1 \right)  \left( \gamma-1 \right) }{d_{{0}}}^{ \left( 
\alpha-1 \right)  \left( \beta-1 \right) }, \\
\mu_{1,2} = &\ {a_{{0}}}^{\alpha+\beta+\gamma-2}{a_{{1}}}^{ \left( \alpha-1 \right) 
 \left( \beta-1 \right)  \left( \gamma-1 \right) }b_{{2}}c_{{2}}d_{{2}
}, \\
\mu_{0,3} = &\ a_0^{\alpha\beta+\alpha\gamma+\beta\gamma-\alpha-\beta-\gamma+1}b_2^{\beta+\gamma-1}c_2^{\alpha+\gamma-1}d_2^{\alpha+\beta-1}.
\end{aligned}
\end{equation}
One can check the dimensions of the corresponding faces of $N(\mathcal{R})$: $\dim \phi_{0,0} = \dim \phi_{3,0} = \dim \phi_{0,3} = 1$, whereas $\dim \phi_{2,1} = \dim \phi_{1,2} = 3$.
%
%
The initial form $\phi_{0,0}$ corresponds to the bottom edge of $N(\mathcal{R})$. Its vertices are the monomials of minimal and maximal weight in $\mathcal{R}$.

\begin{figure}[h!]
    \centering
    \includegraphics[scale=0.5]{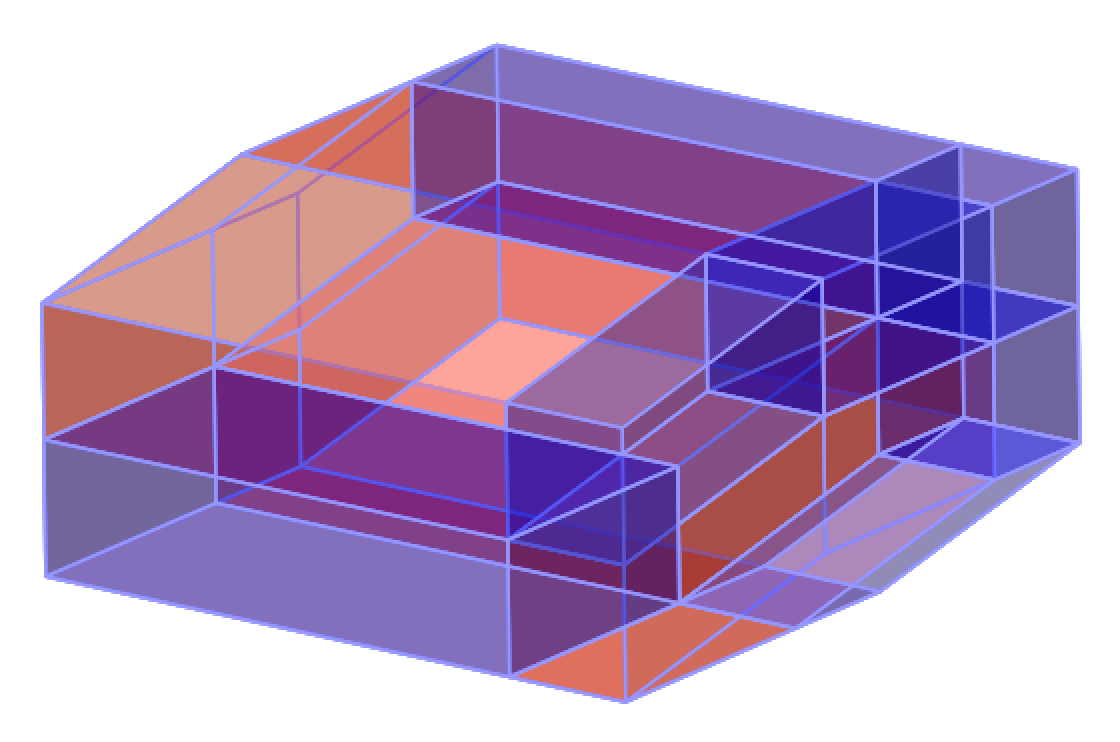}
    \caption{The mixed decomposition induced by $\phi_{1,2}$: there are three red cells, which correspond to three binomials in (\ref{m=3_phi_formulas}), while the union of blue cells give the monomial $\mu_{1,2}$ (\ref{m=3_phi_monomials}). This picture is a generalization of Figure \ref{fig:TCMD_demo} to 3d}
    \label{fig:TCMD_phi12}
\end{figure}

For example, let's calculate $\phi_{0,3}$: by looking at the corresponding perfogram in (\ref{m=3_phis}), we know which monomials to cross-out in (\ref{Nahm_m3}), to get:
\begin{equation}\label{m3_phi03}
    \begin{aligned}
        \left\{a_0 = -a_1z_1z_2z_3,\ z_1^{\alpha-1}=-\frac{b_1}{b_2},\ z_2^{\beta-1}=-\frac{c_1}{c_2},\ z_3^{\gamma-1}=-\frac{d_1}{d_2} \right\}
    \end{aligned}
\end{equation}
To compute the binomial factor, we have to compute the sparse mixed resultant from this system (\ref{m3_phi03}). This is quite easy: rasing the first equation to $(\alpha-1)$ immediately eliminates $z_1$:
\begin{equation}
    a_0^{(\alpha-1)}=(-a_1z_2z_3)^{(\alpha-1)}\left(-\frac{b_1}{b_2} \right)
\end{equation}
Consequently, we raise it to $(\beta-1)$ and $(\gamma-1)$ and getting rid of numerators, to obtain
\begin{equation}\label{m3_phi03_step2}
\begin{aligned}
&\ \left(a_0^{(\alpha-1)(\beta-1)(\gamma-1)}b_2^{(\beta-1)(\gamma-1)}c_2^{(\alpha-1)(\gamma-1)}d_2^{(\alpha-1)(\beta-1)}+\right. \\
&\ \left. (-1)^{(\alpha-1)+(\beta-1)+(\gamma-1)+1}a_1^{(\alpha-1)(\beta-1)(\gamma-1)}b_1^{(\beta-1)(\gamma-1)}c_1^{(\alpha-1)(\gamma-1)}d_1^{(\alpha-1)(\beta-1)} \right)
\end{aligned}
\end{equation}
Now the binomial part of $\phi_{0,3}$ will be equal to the GCD applied to (\ref{m3_phi03_step2}).
The monomial part (\ref{m=3_phi_monomials}) is a bit more subtle, since we have to play with ``Lego boxes'' to form a proper subdivision of $Q$. Only when all the boxes are aligned properly, we get a mixed decomposition, which amounts to the expressions for $\mu_{p,q}$ (see Figure \ref{fig:TCMD_phi12} as an example).

Now we focus on the case $\beta=\gamma=\alpha$. It turns out that this is well-behaved: the data of (\ref{m=3_phis}) is sufficient to describe all the edges of $N(A)$. In particular:
\begin{itemize}
\item every vertex of $N(A)$ has a unique preimage, which is a vertex of $N(\mathcal{R})$
\item every edge of $N(A)$ is an image of a unique simple face from the set $\{\phi_{0,0}$, $\phi_{3,0}$, $\phi_{2,1}$, $\phi_{1,2},\phi_{0,3}\}$
\end{itemize}
First, let's put $\alpha=\beta=\gamma$ into the formulas (\ref{m=3_phi_formulas}).
For any $\phi_{p,q}$, denote its extremal monomials (having minimal/maximal powers of $a_0$) by $\phi_{p,q}^{\mathrm{min}}$ and $\phi_{p,q}^{\mathrm{max}}$.
We get for $\phi_{0,0}$, $\phi_{2,1}$ and $\phi_{1,2}$, respectively:
\begin{equation}\label{phi_3d_minmax}
\begin{aligned}
& \left[{a_{{1}}}^{{\alpha}^{3}}{b_{{0}}}^{{\alpha}^{2}}{c_{{0}}
}^{{\alpha}^{2}}{d_{{0}}}^{{\alpha}^{2}},{a_{{0}}}^{{\alpha}^{3}}{b_{{2}}}^{{\alpha}^{2}}{c_{{2}}}^{{\alpha}^{
2}}{d_{{2}}}^{{\alpha}^{2}}\right] \\
& \left[a_{{0}}{a_{{1}}}^{{\alpha}^{3}-1}{b_{{0
}}}^{{\alpha}^{2}-1}b_{{1}}{c_{{0}}}^{{\alpha}^{2}-1}c_{{1}}{d_{{0}}}^
{{\alpha}^{2}-1}d_{{1}},\right. \\
&\left. {a_{{0}}}^{3\,\alpha-2}{a_{{1}}}^{{\alpha}^{3}-3\,\alpha+2}{b_{{0}}}^{
{\alpha}^{2}-2\,\alpha+1}{b_{{1}}}^{2\,\alpha-2}b_{{2}}{c_{{0}}}^{{
\alpha}^{2}-2\,\alpha+1}{c_{{1}}}^{2\,\alpha-2}c_{{2}}{d_{{0}}}^{{
\alpha}^{2}-2\,\alpha+1}{d_{{1}}}^{2\,\alpha-2}d_{{2}}\right] \\
& \left[{a_{{0}}}^{3\,\alpha-2}{a_{{1}}}^{{
\alpha}^{3}-3\,\alpha+2}{b_{{0}}}^{{\alpha}^{2}-2\,\alpha+1}{b_{{1}}}^
{2\,\alpha-2}b_{{2}}{c_{{0}}}^{{\alpha}^{2}-2\,\alpha+1}{c_{{1}}}^{2\,
\alpha-2}c_{{2}}{d_{{0}}}^{{\alpha}^{2}-2\,\alpha+1}{d_{{1}}}^{2\,
\alpha-2}d_{{2}}, \right.\\
&\left. {a_{{0}}}^{3\,{\alpha}^{2}-3\,\alpha+1}{a_{{1}}}^{{\alpha}^{3}-3\,{
\alpha}^{2}+3\,\alpha-1}{b_{{1}}}^{{\alpha}^{2}-2\,\alpha+1}{b_{{2}}}^
{-1+2\,\alpha}{c_{{1}}}^{{\alpha}^{2}-2\,\alpha+1}{c_{{2}}}^{-1+2\,
\alpha}{d_{{1}}}^{{\alpha}^{2}-2\,\alpha+1}{d_{{2}}}^{-1+2\,\alpha}\right]
\end{aligned}
\end{equation}
Now coming back to $\phi_{3,0}$ and $\phi_{0,3}$: they are simply given by
\begin{equation}
    \phi_{3,0} = \phi_{0,0}^{\mathrm{min}}+\phi_{2,1}^{\mathrm{min}},\ \phi_{0,3} = \phi_{1,2}^{\mathrm{max}} + \phi_{0,0}^{\mathrm{max}}.
\end{equation}
So (\ref{phi_3d_minmax}) are the monomials, which project onto the vertices $N(A)$. Each face corresponding to $\phi_{p,q}$ projects onto one of the edges of $N(A)$, and from (\ref{phi_3d_minmax}) we can write down all the vertices of $N(A)$, starting from the origin $(x,y)=(0,0)$ and going clockwise around the polygon:
\begin{equation}\label{NA_diag_aaa}
    \mathrm{diag}(\alpha,\alpha,\alpha): \ (0,0),\ (0,1),\ (3,3\alpha-2),\ (6\alpha-3,3\alpha^2-3\alpha+1),\ (3\alpha^2,\alpha^3).
\end{equation}
Hence binomiality of the face polynomials for $\mathrm{diag}(\alpha,\alpha,\alpha)$ follows from binomiality of its preimages in $\mathcal{R}$: the initial forms (\ref{m=3_phis}).
Unfortunately, the same does not work for generic $\alpha,\beta,\gamma$. In fact, we have to take into account some extra edges $r,r',r'',\dots$ lying between $\phi_{1,2}$ and $\phi_{2,1}$. Consider the following initial forms:
\begin{equation}\label{3vertex_init_r}
\begin{aligned}
init_{r} = &\ \mu_r\cdot \mathrm{GCD}\left({a_{{0}}}^{ \left( \beta-1 \right)  \left( \gamma-1 \right) }{b_{{1}}}
^{ \left( \beta-1 \right)  \left( \gamma-1 \right) }{c_{{2}}}^{\gamma-
1}{d_{{2}}}^{\beta-1}+\right. \\
&\ \left.\left( -1 \right) ^{\beta+\gamma+1}{a_{{1}}}^{
 \left( \beta-1 \right)  \left( \gamma-1 \right) }{b_{{0}}}^{ \left( 
\beta-1 \right)  \left( \gamma-1 \right) }{c_{{1}}}^{\gamma-1}{d_{{1}}
}^{\beta-1}\right), \\
\mu_{r} = &\ {a_{{0}}}^{-1+\beta+\gamma}{a_{{1}}}^{\sigma}{b_{{1}}}^{-1+\beta+\gamma}{c_{{0}}}^{\gamma
-2+\alpha}c_{{2}}{d_{{0}}}^{ \left( \alpha-1 \right)  \left( \beta-1
 \right) +\alpha-1}d_2
\end{aligned}
\end{equation}
where $\sigma =  \left( \alpha-1 \right) 
 \left( \beta-1 \right)  \left( \gamma-1 \right) +\alpha-1+ \left( 
\alpha-1 \right)  \left( \beta-1 \right) + \left( \alpha-1 \right) 
 \left( \gamma-1 \right) $, and
\begin{equation}\label{3vertex_init_rr}
\begin{aligned}
init_{r'} = &\ \mu_{r'} \cdot \mathrm{GCD}\left(\left( {a_{{0}}}^{ \left( \beta-1 \right)  \left( \gamma-1 \right) }{
b_{{1}}}^{ \left( \beta-1 \right)  \left( \gamma-1 \right) }{c_{{2}}}^
{\gamma-1}{d_{{2}}}^{\beta-1}+\right.\right.
\\
&\
\left.\left.\left( -1 \right) ^{\beta+\gamma+1}{a_{
{1}}}^{ \left( \beta-1 \right)  \left( \gamma-1 \right) }{b_{{0}}}^{
 \left( \beta-1 \right)  \left( \gamma-1 \right) }{c_{{1}}}^{\gamma-1}
{d_{{1}}}^{\beta-1} \right) \right. \times \\
&\
\left. \left( {a_{{0}}}^{\alpha-1}b_{{2}}{c_{{1}
}}^{\alpha-1}{d_{{1}}}^{\alpha-1}+ \left( -1 \right) ^{\alpha+1}{a_{{1
}}}^{\alpha-1}b_{{1}}{c_{{0}}}^{\alpha-1}{d_{{0}}}^{\alpha-1} \right) \right), \\
\mu_{r'} = &\ {a_{{0}}}^{-1+\beta+\gamma}{a_1}^{\sigma'}{b_{{1}}}^{\beta-2+\gamma}{c_{{0}}}^{ \left( \alpha-1
 \right)  \left( \gamma-1 \right) }c_{{2}}{d_{{0}}}^{ \left( \alpha-1
 \right)  \left( \beta-1 \right) }d_{{2}}
\end{aligned}
\end{equation}
where $\sigma' =  \left( \alpha-1 \right) 
 \left( \beta-1 \right)  \left( \gamma-1 \right) + \left( \alpha-1
 \right)  \left( \gamma-1 \right) + \left( \alpha-1 \right)  \left( 
\beta-1 \right) $.
In what follows, we assume $2\leq \alpha < \beta < \gamma$.
\begin{conjecture}\label{3vertex_conj}
For the diagonal quiver with 3 vertices, the only contribution to the edges of $N(A)$ is due to the initial forms $\{\phi_{i,j}\}_{i+j=3}$, completed with $init_r$ and $init_{r'}$:
\begin{enumerate}
    \item $(\alpha,\alpha,\alpha)$: $\phi_{0,0},\phi_{3,0},\phi_{2,1},\phi_{1,2},\phi_{0,3}$ project onto the edges of $N(A)$, and are in bijection with the edges
    \item $(\alpha,\alpha,\beta)$: $\{\phi_{i,j}\}_{i+j=3}$ are still OK, although some of the monomials of $\phi_{1,2}$ and $\phi_{2,1}$ will project onto the interior of $N(A)$, unlike in the case above. Nevertheless, this does not affect binomiality of the edges of $N(A)$
    \item $(\alpha,\beta,\beta)$: we need to include the extra initial form $init_{r'}$, since $\phi_{1,2}$ would have an issue: it will capture the vertices, but not all the intermediate points of the edge. Therefore, $init_{r'}$ will fully cover this problematic edge, and since it is a binomial, so is true for the projection.
    \item $(\alpha,\beta,\gamma)$ all distinct: instead of 
    $init_{r'}$, we have to take $init_r$
\end{enumerate}
\end{conjecture}
We verify this conjecture using the computer program (see Appendix \ref{append_a_1} for the examples).
To sum up, the four cases of Conjecture \ref{3vertex_conj} produce non-equivalent projections $N(A)$. Proposed initial forms (\ref{m=3_phi_formulas}), (\ref{3vertex_init_r}) and (\ref{3vertex_init_rr}) contribute to the edges of $N(A)$. Their binomiality would imply that the quiver A-polynomial for $C = \mathrm{diag}(\alpha,\beta,\gamma)$ is tempered, providing that the contribution happens in the same way as for the examples studied. However, a
general proof for $\mathrm{diag}(\alpha,\beta,\gamma)$ is still missing.


\section{Arbitrary dimension}\label{section_arbitrary_dimension}

This section contains the main result of the paper.
Consider the quiver 
\begin{equation}
C = \mathrm{diag}(\alpha_1,\dots,\alpha_m),
\quad \alpha_i\geq 2,\ m\geq 2.
\end{equation}
Let $\mathcal{R}$ be the refined quiver resultant (Definition \ref{quiver_resultant_def}) and $N(\mathcal{R})$ its Newton polytope -- a $(m+1)$-dimensional polytope in $\mathbb{R}^{2+3m}$.

We reveal the combinatorial structure of the 1-dimensional skeleton of $N(\mathcal{R})$, captured by its initial forms\footnote{Note that non-diagonal quivers can be build upon the same skeleton, but with additional assumptions, which are outside of the scope of this paper.}. When $\alpha_i=\alpha,\ i=1\dots m$, binomiality of these forms implies the K-theoretic property (\ref{K_criterion}) for quiver A-polynomial $A(x,y)$. Therefore the latter is quantizable.

Let $m = p + q$. Our main actors are simple initial forms $\{\phi_{p,q}\}$, where $p$ is the number of $[\bullet \bullet \ ]$-type rows -- we indicate them by the index subset $I=\{i_1,\dots,i_p\}$, and $q$ is the number of $[\textcolor{red}{\ \bullet \bullet}]$-type rows $K=\{k_1,\dots,k_q\}$, in each of the perfogram contained in $\phi_{p,q}$. 
Dimension of the face on which $\phi_{p,q}$ is supported, is equal to the number of its distinct binomial factors (due to Proposition \ref{phi_dimension_prop}).
In what follows, we give a full description of $\{\phi_{p,q}\}$ for the diagonal quiver.

\begin{proposition}\label{initial_forms_diag_arbirary}
Let $I = \{i_1,\dots,i_p\},\ K = \{k_1,\dots,k_q\},\ p+q = m$. Define
\begin{equation}\label{diag_general_binomials}
\begin{aligned}
\varphi_{I,K} := & \left(a_0\prod_{i\in I}b_{i,1}\right)^{\prod_{k\in K}(\alpha_k-1)}\prod_{k\in K}b_{k,2}^{\prod'_{\substack{k'\in K \\ k' \neq k}}(\alpha_{k'}-1)}
+ \\ 
& (-1)^{1+\sum_{k\in K}\prod_{\substack{k' \in K \\ k' \neq k}}(\alpha_{k'}-1)}\left(a_1\prod_{i\in I}b_{i,0}\right)^{\prod_{k\in K}(\alpha_k-1)}\prod_{k\in K}b_{k,1}^{\prod'_{\substack{k'\in K \\ k' \neq k}}(\alpha_{k'}-1)}
\end{aligned}
\end{equation}
where the product $\prod'$ equals to $1$ if $q=1$. Then 
\begin{equation}
\phi_{p,q} := \mu_{p,q} \cdot \prod_{\substack{I,K\subset \{1,\dots,m\} \\ |I|=p, |K|=q}}\mathrm{GCD}(\varphi_{I,K}) 
\end{equation}
are well-defined initial forms,
where the product is taken over all $\frac{m!}{p!q!}$ choices of the subsets $I,K$, and the monomial $\mu_{p,q}$ is given by
\begin{equation}\label{diag_general_monomial}
\begin{aligned}
\mu_{p,q} = & a_0^{1+\sum_{|K'| = 1 \dots q-1}\prod_{k' \in K'}(\alpha_{k'}-1)}a_1^{\sum_{|K'| = q+1 \dots m}\prod_{k' \in K'}(\alpha_{k'}-1)} \times \\
& 
\prod_{i=1\dots m}b_{i,0}^{\sum_{\substack{|K'| = q-2 \dots m-1 \\ i \notin K'}}\prod_{k' \in K'}(\alpha_{k'}-1)}b_{i,2}^{\delta(q)+\sum_{\substack{|K'| = 1 \dots q-2 \\ i \notin K'}}\prod_{k' \in K'}(\alpha_{k'}-1)}
\end{aligned}
\end{equation}
where $\delta(q) = 0$ if $q\leq m-1$, and $\delta(q) = 1$ otherwise. 
\end{proposition}
\underline{Proof}. The expression for $\varphi_{I,K}$ is rather easy. Recall that we are dealing with polynomials (\ref{Nahm_eqs}) and $C = \mathrm{diag}(\alpha_1,\dots,\alpha_m)$. Write the first equation $F_0 = 0$ as
\begin{equation}
a_0 = -a_1 z_1 \dots z_m,
\end{equation}
and then raise it consequently in powers $(\alpha_k-1)$ where $k\in K$ corresponds to $[\ \bullet \bullet]$-type rows, each time plugging $z_k^{\alpha_k-1} = -\frac{b_{k,1}}{b_{k,2}}$ (repeating for all permuted perfograms).

The monomial part is more involved.  We will use the short-hand notation $(\pi)$ for the permutation class of a perfogram, e.g.:
\begin{equation}\label{pi_definition}
\begin{aligned}
\substack{m+1\\ \text{rows}} \left\lbrace   \begin{tabular}{|c c c|}
  \textcolor{blue}{$\bullet$} &  & \\ \hline 
  & \textcolor{red}{$\bullet$} & \textcolor{red}{$\bullet$} \\
  $\bullet$ & $\bullet$ & \\      
  $\bullet$ & $\bullet$ & \\
  $\bullet$ & $\bullet$ & \\
  \vdots & \vdots & \vdots \\
  $\bullet$ & $\bullet$ &      
\end{tabular}_{(\pi)}\right.
:= &\ \quad
    \begin{tabular}{|c c c|}
    \textcolor{blue}{$\bullet$} &  & \\ \hline 
  & \textcolor{red}{$\bullet$} & \textcolor{red}{$\bullet$} \\
  $\bullet$ & $\bullet$ & \\      
  $\bullet$ & $\bullet$ & \\
  $\bullet$ & $\bullet$ & \\
  \vdots & \vdots & \vdots \\
  $\bullet$ & $\bullet$ &      
\end{tabular}
\times
 \begin{tabular}{|c c c|}
 \textcolor{blue}{$\bullet$} &  & \\ \hline 
  $\bullet$ & $\bullet$ & \\      
  & \textcolor{red}{$\bullet$} & \textcolor{red}{$\bullet$} \\
  $\bullet$ & $\bullet$ & \\
  $\bullet$ & $\bullet$ & \\
  \vdots & \vdots & \vdots \\
  $\bullet$ & $\bullet$ &      
\end{tabular}
\times
 \begin{tabular}{|c c c|}
 \textcolor{blue}{$\bullet$} &  & \\ \hline 
  $\bullet$ & $\bullet$ & \\      
  $\bullet$ & $\bullet$ & \\
  & \textcolor{red}{$\bullet$} & \textcolor{red}{$\bullet$} \\
  $\bullet$ & $\bullet$ & \\
  \vdots & \vdots & \vdots \\
  $\bullet$ & $\bullet$ &      
\end{tabular}
\times \dots \times
 \begin{tabular}{|c c c|}
 \textcolor{blue}{$\bullet$} &  & \\ \hline 
  $\bullet$ & $\bullet$ & \\      
  $\bullet$ & $\bullet$ & \\
  $\bullet$ & $\bullet$ & \\
  $\bullet$ & $\bullet$ & \\
  \vdots & \vdots & \vdots \\
  & \textcolor{red}{$\bullet$} & \textcolor{red}{$\bullet$} \\      
\end{tabular}
\\ \ \\
\begin{tabular}{|c c c|}
\textcolor{blue}{$\bullet$} &  & \\ \hline 
  & \textcolor{red}{$\bullet$} & \textcolor{red}{$\bullet$} \\
  & \textcolor{red}{$\bullet$} & \textcolor{red}{$\bullet$} \\      
  $\bullet$ & $\bullet$ & \\
  $\bullet$ & $\bullet$ & \\
  \vdots & \vdots & \vdots \\
  $\bullet$ & $\bullet$ &      
\end{tabular}_{(\pi)}
:= &\ \quad
\begin{tabular}{|c c c|}
\textcolor{blue}{$\bullet$} &  & \\ \hline 
  & \textcolor{red}{$\bullet$} & \textcolor{red}{$\bullet$} \\
  & \textcolor{red}{$\bullet$} & \textcolor{red}{$\bullet$} \\      
  $\bullet$ & $\bullet$ & \\
  $\bullet$ & $\bullet$ & \\
  \vdots & \vdots & \vdots \\
  $\bullet$ & $\bullet$ &      
\end{tabular}
\times
\begin{tabular}{|c c c|}
\textcolor{blue}{$\bullet$} &  & \\ \hline 
  & \textcolor{red}{$\bullet$} & \textcolor{red}{$\bullet$} \\
  $\bullet$ & $\bullet$ & \\      
  & \textcolor{red}{$\bullet$} & \textcolor{red}{$\bullet$} \\
  $\bullet$ & $\bullet$ & \\
  \vdots & \vdots & \vdots \\
  $\bullet$ & $\bullet$ &      
\end{tabular}
\times
\begin{tabular}{|c c c|}
\textcolor{blue}{$\bullet$} &  & \\ \hline 
  $\bullet$ & $\bullet$ & \\
  & \textcolor{red}{$\bullet$} & \textcolor{red}{$\bullet$}\\      
  & \textcolor{red}{$\bullet$} & \textcolor{red}{$\bullet$} \\
  $\bullet$ & $\bullet$ & \\
  \vdots & \vdots & \vdots \\
  $\bullet$ & $\bullet$ &      
 \end{tabular}
 \times \dots \times
\begin{tabular}{|c c c|}
\textcolor{blue}{$\bullet$} &  & \\ \hline 
  $\bullet$ & $\bullet$ & \\     
  $\bullet$ & $\bullet$ & \\
  $\bullet$ & $\bullet$ & \\
  \vdots & \vdots & \vdots \\
  & \textcolor{red}{$\bullet$} & \textcolor{red}{$\bullet$}\\
  & \textcolor{red}{$\bullet$} & \textcolor{red}{$\bullet$}     
 \end{tabular}
\end{aligned}
\end{equation}
and similarly for other types, which consists of all possible perfograms with fixed number of black and red rows. Therefore, the number of such permutations is equal to 
\begin{equation}
|(\pi)| = \frac{m!}{(\# \text{black rows})! \ \cdot \ (\# \text{\textcolor{red}{red rows}})!}
\end{equation}
Note that the product ``$\times$'' is commutative, since it corresponds to taking unions of Minkowski sums (the ordering in (\ref{pi_definition}) is chosen just to illustrate the idea).
Here we used black (red) color for bullets in order to easily distinguish $[\bullet \bullet \ ]$- ($[\ \bullet \bullet]$-) type of rows, correspondingly -- it does not carry any additional structure. Next, the blue bullet indicate the equation, which remains fixed under permutation (it is always the one with just a single bullet).
In (\ref{pi_definition}) it is $\tilde{F}_0 = \textcolor{blue}{a_0} + \cancel{a_1 z_1\dots z_m}$, and its monomial $\textcolor{blue}{a_0}$ is the only guy which survived in its row. Therefore, all such perfograms contribute to $a_0$ in the monomial $\mu_{p,q}$ (\ref{diag_general_monomial}). On another hand, all perfograms for $a_1$ will start with $\tilde{F}_0 = \cancel{a_0} + \textcolor{blue}{a_1 z_1\dots z_m}$, and so forth. 
Therefore, we claim that the following perfograms generate all letters in the monomial $\mu_{p,q}$:

\scalebox{0.8}{\parbox{.5\linewidth}{%
\begin{align}\label{phi_themonomial}
a_0: & \quad 
\begin{tabular}{|c c c|}
  \textcolor{blue}{$\bullet$} &  & \\ \hline 
  $\bullet$ & $\bullet$ & \\
  $\bullet$ & $\bullet$ & \\      
  $\bullet$ & $\bullet$ & \\
  $\bullet$ & $\bullet$ & \\
  \vdots & \vdots & \vdots \\
  $\bullet$ & $\bullet$ &      
\end{tabular}
\times
\begin{tabular}{|c c c|}
  \textcolor{blue}{$\bullet$} &  & \\ \hline 
 & \textcolor{red}{$\bullet$} & \textcolor{red}{$\bullet$} \\
  $\bullet$ & $\bullet$ & \\      
  $\bullet$ & $\bullet$ & \\
  $\bullet$ & $\bullet$ & \\
  \vdots & \vdots & \vdots \\
  $\bullet$ & $\bullet$ &      
\end{tabular}_{(\pi)}
\times
\begin{tabular}{|c c c|}
  \textcolor{blue}{$\bullet$} &  & \\ \hline 
 & \textcolor{red}{$\bullet$} & \textcolor{red}{$\bullet$} \\
 & \textcolor{red}{$\bullet$} & \textcolor{red}{$\bullet$} \\      
  $\bullet$ & $\bullet$ & \\
  $\bullet$ & $\bullet$ & \\
  \vdots & \vdots & \vdots \\
  $\bullet$ & $\bullet$ &      
\end{tabular}_{(\pi)}
\times \dots \times
\begin{tabular}{|c c c|}
  \textcolor{blue}{$\bullet$} &  & \\ \hline 
 & \textcolor{red}{$\bullet$} & \textcolor{red}{$\bullet$} \\
  \vdots & \vdots & \vdots \\
 & \textcolor{red}{$\bullet$} & \textcolor{red}{$\bullet$} \\ 
  $\bullet$ & $\bullet$ & \\
  \vdots & \vdots & \vdots \\
  $\bullet$ & $\bullet$ &      
\end{tabular}_{(\pi)},
\quad \text{unless } \#\textcolor{red}{\text{red rows}}\leq q-1
\\
a_1: & \quad 
\begin{tabular}{|c c c|}
   & \textcolor{blue}{$\bullet$} & \\ \hline 
 & \textcolor{red}{$\bullet$} & \textcolor{red}{$\bullet$} \\
 & \textcolor{red}{$\bullet$} & \textcolor{red}{$\bullet$} \\ 
  & \textcolor{red}{$\bullet$} & \textcolor{red}{$\bullet$} \\
  & \textcolor{red}{$\bullet$} & \textcolor{red}{$\bullet$} \\
  \vdots & \vdots & \vdots \\
 & \textcolor{red}{$\bullet$} & \textcolor{red}{$\bullet$} \\     
\end{tabular}
\times
\begin{tabular}{|c c c|}
   & \textcolor{blue}{$\bullet$} & \\ \hline 
  $\bullet$ & $\bullet$ & \\
   & \textcolor{red}{$\bullet$} & \textcolor{red}{$\bullet$} \\      
   & \textcolor{red}{$\bullet$} & \textcolor{red}{$\bullet$} \\
   & \textcolor{red}{$\bullet$} & \textcolor{red}{$\bullet$} \\
  \vdots & \vdots & \vdots \\
   & \textcolor{red}{$\bullet$} & \textcolor{red}{$\bullet$} \\      
\end{tabular}_{(\pi)}
\times
\begin{tabular}{|c c c|}
   & \textcolor{blue}{$\bullet$} & \\ \hline 
  $\bullet$ & $\bullet$ & \\
  $\bullet$ & $\bullet$ & \\      
   & \textcolor{red}{$\bullet$} & \textcolor{red}{$\bullet$} \\
   & \textcolor{red}{$\bullet$} & \textcolor{red}{$\bullet$} \\
  \vdots & \vdots & \vdots \\
   & \textcolor{red}{$\bullet$} & \textcolor{red}{$\bullet$} \\      
\end{tabular}_{(\pi)}
\times \dots \times
\begin{tabular}{|c c c|}
   & \textcolor{blue}{$\bullet$} & \\ \hline 
  $\bullet$ & $\bullet$ & \\
  \vdots & \vdots & \vdots \\
  $\bullet$ & $\bullet$ & \\ 
   & \textcolor{red}{$\bullet$} & \textcolor{red}{$\bullet$} \\
  \vdots & \vdots & \vdots \\
   & \textcolor{red}{$\bullet$} & \textcolor{red}{$\bullet$} \\      
\end{tabular}_{(\pi)},
\quad \text{unless } \#\textcolor{red}{\text{red rows}}\geq q+1
\\
b_{*,0}: & \quad 
\begin{tabular}{|c c c|}
   $\bullet$ & $\bullet$ & \\ \hline 
 \textcolor{blue}{$\bullet$} &  & \\
 & \textcolor{red}{$\bullet$} & \textcolor{red}{$\bullet$} \\
  & \textcolor{red}{$\bullet$} & \textcolor{red}{$\bullet$} \\      
  & \textcolor{red}{$\bullet$} & \textcolor{red}{$\bullet$} \\
  & \textcolor{red}{$\bullet$} & \textcolor{red}{$\bullet$} \\
  \vdots & \vdots & \vdots \\
 & \textcolor{red}{$\bullet$} & \textcolor{red}{$\bullet$} \\      
\end{tabular}
\times
\begin{tabular}{|c c c|}
   $\bullet$ & $\bullet$ & \\ \hline 
  \textcolor{blue}{$\bullet$} &  & \\
  $\bullet$ & $\bullet$ & \\      
 & \textcolor{red}{$\bullet$} & \textcolor{red}{$\bullet$} \\
 & \textcolor{red}{$\bullet$} & \textcolor{red}{$\bullet$} \\
 & \textcolor{red}{$\bullet$} & \textcolor{red}{$\bullet$} \\
  \vdots & \vdots & \vdots \\
  & \textcolor{red}{$\bullet$} & \textcolor{red}{$\bullet$} \\     
\end{tabular}_{(\pi)}
\times
\begin{tabular}{|c c c|}
   $\bullet$ & $\bullet$ & \\ \hline 
  \textcolor{blue}{$\bullet$} &  & \\
  $\bullet$ & $\bullet$ & \\      
   $\bullet$ & $\bullet$ & \\
 & \textcolor{red}{$\bullet$} & \textcolor{red}{$\bullet$} \\
  & \textcolor{red}{$\bullet$} & \textcolor{red}{$\bullet$} \\
  \vdots & \vdots & \vdots \\
  & \textcolor{red}{$\bullet$} & \textcolor{red}{$\bullet$} \\      
\end{tabular}_{(\pi)}
\times \dots \times
\begin{tabular}{|c c c|}
   $\bullet$ & $\bullet$ & \\ \hline 
  \textcolor{blue}{$\bullet$} &  & \\
  $\bullet$ & $\bullet$ & \\ 
  \vdots & \vdots & \vdots \\
  $\bullet$ & $\bullet$ & \\ 
 & \textcolor{red}{$\bullet$} & \textcolor{red}{$\bullet$} \\
  \vdots & \vdots & \vdots \\
 & \textcolor{red}{$\bullet$} & \textcolor{red}{$\bullet$} \\      
\end{tabular}_{(\pi)},
\quad \text{unless } \#\textcolor{red}{\text{red rows}}\geq q+1
\\
\label{phi_themonomial_b2}
b_{*,2}: & \quad 
\begin{tabular}{|c c c|}
   $\bullet$ & $\bullet$ & \\ \hline 
  &  & \textcolor{blue}{$\bullet$} \\
  $\bullet$ & $\bullet$ & \\
  $\bullet$ & $\bullet$ & \\      
  $\bullet$ & $\bullet$ & \\
  $\bullet$ & $\bullet$ & \\
  \vdots & \vdots & \vdots \\
  $\bullet$ & $\bullet$ &      
\end{tabular}
\times
\begin{tabular}{|c c c|}
   $\bullet$ & $\bullet$ & \\ \hline 
 &  & \textcolor{blue}{$\bullet$} \\
 & \textcolor{red}{$\bullet$} & \textcolor{red}{$\bullet$} \\      
  $\bullet$ & $\bullet$ & \\
  $\bullet$ & $\bullet$ & \\
  $\bullet$ & $\bullet$ & \\
  \vdots & \vdots & \vdots \\
  $\bullet$ & $\bullet$ &      
\end{tabular}_{(\pi)}
\times
\begin{tabular}{|c c c|}
   $\bullet$ & $\bullet$ & \\ \hline 
 &  & \textcolor{blue}{$\bullet$} \\
 & \textcolor{red}{$\bullet$} & \textcolor{red}{$\bullet$} \\      
  & \textcolor{red}{$\bullet$} & \textcolor{red}{$\bullet$}   \\
  $\bullet$ & $\bullet$ & \\
  $\bullet$ & $\bullet$ & \\
  \vdots & \vdots & \vdots \\
  $\bullet$ & $\bullet$ &      
\end{tabular}_{(\pi)}
\times \dots \times
\begin{tabular}{|c c c|}
   $\bullet$ & $\bullet$ & \\ \hline 
 &  & \textcolor{blue}{$\bullet$} \\
 & \textcolor{red}{$\bullet$} & \textcolor{red}{$\bullet$} \\ 
  \vdots & \vdots & \vdots \\
  & \textcolor{red}{$\bullet$} & \textcolor{red}{$\bullet$} \\
  $\bullet$ & $\bullet$ & \\
  \vdots & \vdots & \vdots \\
  $\bullet$ & $\bullet$ &      
\end{tabular}_{(\pi)},
\quad \text{unless } \#\textcolor{red}{\text{red rows}}\leq q-1
\end{align}
}}
\\
where the number of red rows corresponds to $|K'|$ in (\ref{diag_general_monomial}). Also, for $b_{*,0}$ and $b_{*,2}$ ``*'' means that we can choose any row $i'$ for the blue bullet, moving it around from the first to the $m$-th row of each perfogram (the zeroth is of course not, since $F_0$ does not depend on $b_{i,j}$).
Somewhat surprisingly, $\mu_{p,q}$ does not depend on $b_{*,1}$ at all. This is due to the fact that there is simply no more space for such extra cells. Recall that each $\phi_{p,q}$ corresponds to a TCMD of $Q$. Its total volume equals to $\mathrm{vol}(Q)$, and does not depend on $p,q$, but only on $m$.
Therefore, the sum over all cells (perfograms) entering the monomial plus all the perfograms entering the binomials, gives the total volume. 
For example, let's verify that the exponent of $a_0$ in $\mu_{p,q}$ (\ref{diag_general_monomial}) is indeed equal to $1+\sum_{|K'| = 1 \dots q-1}\prod_{k' \in K'}(\alpha_{k'}-1)$. This is simply the sum over all volumes of Minkowski sums for its perfograms:
\begin{equation}\label{perfograms_volumes}
\begin{tabular}{c}
$\tilde{F}_0$ \\
$\tilde{F}_1$ \\
$\tilde{F}_2$ \\
$\tilde{F}_3$ \\ 
$\tilde{F}_4$ \\
\vdots \\
$\tilde{F}_m$
\end{tabular}
\underbrace{\begin{tabular}{|c c c|}
  \textcolor{blue}{$\bullet$} &  & \\ \hline 
  $\bullet$ & $\bullet$ & \\
  $\bullet$ & $\bullet$ & \\      
  $\bullet$ & $\bullet$ & \\
  $\bullet$ & $\bullet$ & \\
  \vdots & \vdots & \vdots \\
  $\bullet$ & $\bullet$ &      
\end{tabular}}_{\mathrm{vol = 1}}    
\qquad\qquad
\underbrace{
\begin{tabular}{|c c c|}
  \textcolor{blue}{$\bullet$} &  & \\ \hline 
 & \textcolor{red}{$\bullet$} & \textcolor{red}{$\bullet$} \\
  $\bullet$ & $\bullet$ & \\      
  $\bullet$ & $\bullet$ & \\
  $\bullet$ & $\bullet$ & \\
  \vdots & \vdots & \vdots \\
  $\bullet$ & $\bullet$ &      
\end{tabular}}_{\mathrm{vol = \alpha_1 - 1}}  
\qquad\qquad
\underbrace{
\begin{tabular}{|c c c|}
  \textcolor{blue}{$\bullet$} &  & \\ \hline 
 & \textcolor{red}{$\bullet$} & \textcolor{red}{$\bullet$} \\
 & \textcolor{red}{$\bullet$} & \textcolor{red}{$\bullet$} \\      
  $\bullet$ & $\bullet$ & \\
  $\bullet$ & $\bullet$ & \\
  \vdots & \vdots & \vdots \\
  $\bullet$ & $\bullet$ &      
\end{tabular}}_{\mathrm{vol = (\alpha_1 - 1)(\alpha_2 - 1)}}  
\qquad\qquad
\dots
\end{equation}
In the first case we are left with $\tilde{F}_0=a_0$ and $\tilde{F}_i = b_{i,0} + b_{i,2}z,\ i=1\dots m$, which corresponds to $\tilde{Q}$ being a unit $m$-cube with $\mathrm{vol}(\tilde{Q}) = 1$. Similarly, each $j$-th row switching to red, gives the factor $(\alpha_j-1)$ to the volume of $\tilde{Q}$. Calculating the volumes for each perfogram of the variables $a_0,a_1,b_{*,0},b_{*,2}$ and then summing them up, we obtain the exponents in (\ref{diag_general_monomial}). What about the total volume? One can verify that the total degree of $\phi_{p,q}$ is equal to the total volume $\mathrm{vol}(Q)$. First of all, we have:
\begin{equation}\label{mixed_volume_of_Q}
    \mathrm{vol}(Q) = \alpha_1\dots \alpha_m + \sum_{j=1}^m\alpha_1\dots \cancel{\alpha_j}\dots \alpha_{m}. 
\end{equation}
Take the identical $\mathrm{TCMD}(Q) = Q$, which corresponds to  the ``bottom'' 1-dimensional face of $N(\mathcal{R})$, with initial form
\begin{equation}
\phi_{0,0} := \mathrm{GCD}\left(a_0^{\prod \alpha_j}\prod b_{j,2}^{\prod_{j'\neq j}\alpha_{j'}}+
(-1)^{\prod\alpha_j + \sum\prod \alpha_{j'}}a_1^{\prod \alpha_j}\prod b_{j,0}^{\prod_{j'\neq j}\alpha_{j'}}
\right)
\simeq 
\begin{tabular}{|c c c|}
 $\bullet$ & $\bullet$  & \\ \hline
  $\bullet$ & & $\bullet$ \\
  $\bullet$ & &  $\bullet$  \\      
  $\vdots$ & $\vdots$ & $\vdots$ \\
  $\bullet$ & & $\bullet$       
\end{tabular}
\end{equation}
The GCD operator does not change the total degree of a polynomial, therefore 
the degree of $\phi_{0,0}$ is equal to the right hand side of (\ref{mixed_volume_of_Q}). On another hand, since the cell is unique and equals to $Q$ itself, its total degree is equal to the volume of $Q$. For $m = 2$ we have the complete set of $\phi_{p,q}$'s:
\begin{equation}
    \{\phi_{p,q}\} = \{\phi_{0,0}\} \cup \{\phi_{2,0},\ \phi_{0,2},\ \phi_{1,1}\}.
\end{equation}
E.g.,
\begin{equation}
    \phi_{1,1} = a_0a_1^{(\alpha_1-1)(\alpha_2-1)}(a_0^{\alpha_1-1}b_2c_1^{\alpha_1-1}+a_1^{\alpha_1-1}b_1c_0^{\alpha_1-1})(a_0^{\alpha_2-1}c_2b_1^{\alpha_2-1}+a_1^{\alpha_2-1}c_1b_0^{\alpha_2-1})
\end{equation}
(recall (\ref{init_example})). The volume $\mathrm{vol}(Q) = \alpha_1\alpha_2 + \alpha_1 + \alpha_2$. We see that the condition $\deg(\phi_{1,1}) = \mathrm{vol}(Q)$ is satisfied. Analogously, for $m = 3$ it can be checked from the formulas (\ref{m=3_phi_formulas}).
For arbitrary $m$ it then follows by induction. Indeed, increasing $m$ by one amounts to adding one extra row to all perfograms we have, and also introducing some new perfograms. Since the volume of a cell given by a perfogram is the product over all its rows, the individual volumes are modified as
\begin{equation}
\begin{aligned}
    \tilde{\alpha}_1\dots\tilde{\alpha}_m \mapsto &\ \tilde{\alpha}_1\dots\tilde{\alpha}_m\tilde{\alpha}_{m+1} \\
    \tilde{\alpha}_1\dots\cancel{\tilde{\alpha}_j}\dots\tilde{\alpha}_m \mapsto &\ \tilde{\alpha}_1\dots\cancel{\tilde{\alpha}_j}\dots\tilde{\alpha}_m\tilde{\alpha}_{m+1}, \ j\neq m+1 \\
    \tilde{\alpha}_1\dots\cancel{\tilde{\alpha}_j,\tilde{\alpha}_{j'}}\dots\tilde{\alpha}_m \mapsto &\ \tilde{\alpha}_1\dots\cancel{\tilde{\alpha}_j,\tilde{\alpha}_{j'}}\dots\tilde{\alpha}_m\tilde{\alpha}_{m+1},\ j,j'\neq m+1 \\
    \vdots
\end{aligned}
\end{equation}
where $\tilde{\alpha}_i := \alpha_i - 1$. But this picture is not yet symmetric, since $\tilde{\alpha}_{m+1}$ is never crossed out. To make it fully symmetric, we have to take the permutation classes $(\pi)$ in both binomial (\ref{diag_general_binomials}) and monomial (\ref{diag_general_monomial}) parts.
When we sum up the volumes of all cells in $\phi_{p,q}$ with $p+q=m$, some cancellations occur (i.e. in the total volume all monomials $\alpha_{i_1}\dots\alpha_{i_k}$ with $k< m-1$ are cancelled), which results in
(\ref{mixed_volume_of_Q}). The same type of cancellation happens if we add one extra dimension, since
the formula is symmetrized (thanks to the permutations involved), and only the number of factors is increased by one. Therefore, $\{\phi_{p,q}\}$ are well-defined initial forms for any $m\geq 2$. $\square$
\\ \ \\
 The following result is the most important for us, since it implies that $A(x,y)$ is tempered for any quiver with $C=\mathrm{diag}(\alpha,\alpha,\dots,\alpha)$, of size $m$:
\begin{theorem}\label{main_prop_diagonal}
The edges of $N(A)$ are in 1:1 correspondence with $\phi_{p,q}$'s, if and only if $(\alpha_1,\dots,\alpha_m) = (\alpha,\dots,\alpha)$ (Figure \ref{fig:newton_projection_phi}). Moreover, the vertices of $N(A)$ are then given by $\phi_{p,q}^{\mathrm{min}}$ and $\phi_{p,q}^{\mathrm{max}}$, for all $p,q$ such that $p+q=m$ and $(p,q)=(0,0)$ for the bottom edge.
\end{theorem}
\underline{Proof}. 
%
%
\begin{figure}[h!]\label{NewtonA_aaa}
    \centering
    \includegraphics[scale=0.5]{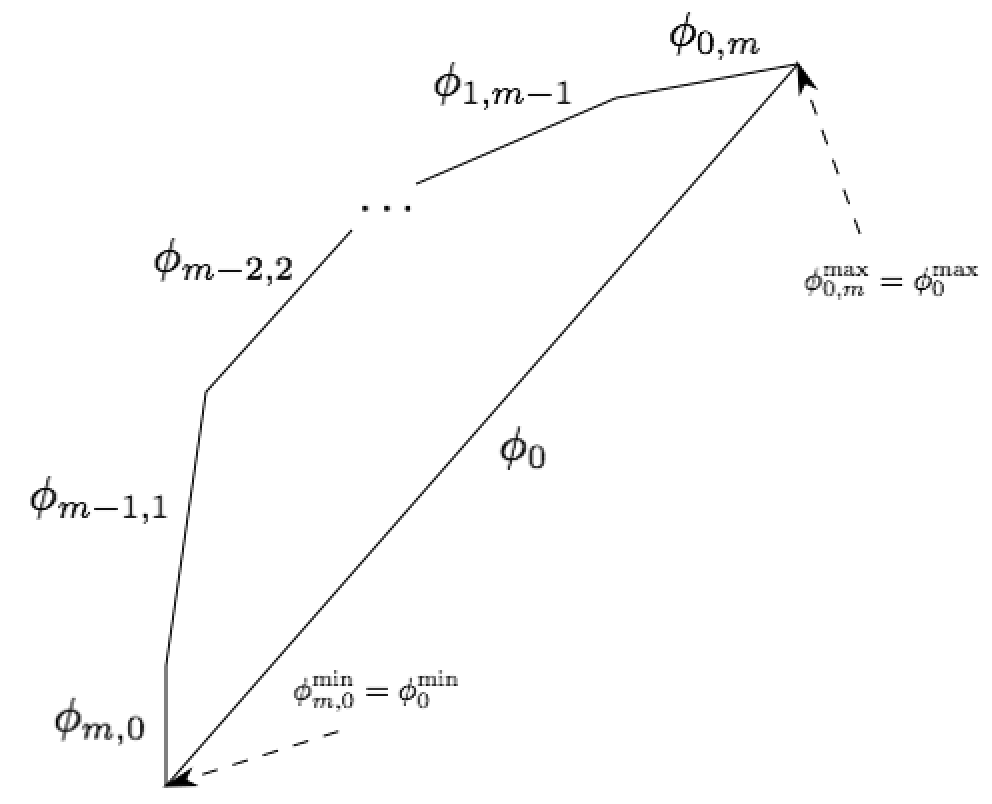}
    \caption{$N(A)$ for $\mathrm{diag}(\alpha,\dots,\alpha)$: each $\phi_{i,j}$ projects onto the corresponding edge, such that the min/max monomials of $\phi_{i,j}$ are in bijection with the vertices of $N(A)$; $\phi_0 := \phi_{0,0}$.}
    \label{fig:newton_projection_phi}
\end{figure}
We have:
\begin{equation}
\begin{aligned}
\phi_{0,0}^{\mathrm{min}} = &\ (-1)^{\prod\alpha_j + \sum\prod \alpha_{j'}}a_1^{\prod \alpha_j}\prod b_{j,0}^{\prod_{j'\neq j}\alpha_{j'}}, & \quad \phi_{0,0}^{\mathrm{max}} = &\  a_0^{\prod \alpha_j}\prod b_{j,2}^{\prod_{j'\neq j}\alpha_{j'}}
\end{aligned}
\end{equation}
which gives the lowest and highest powers of $y$ in the A-polynomial -- the points $(0,0)$ and $(\sum\prod_{j'\neq j}\alpha_{j'},\prod\alpha_j)$ on the $(x,y)$-plane, are thus the vertices of $N(A)$. 
Let's write min/max monomials for $\phi_{p,q}$:
\begin{equation}
\begin{aligned}
\phi_{p,q}^{\mathrm{min}} = &\ \mu_{p,q}\cdot \prod_{\pi(I,K)}  (-1)^{1+\sum_{k\in K}\prod_{\substack{k' \in K \\ k' \neq k}}(\alpha_{k'}-1)}\left(a_1\prod_{i\in I}b_{i,0}\right)^{\prod_{k\in K}(\alpha_k-1)}\prod_{k\in K}b_{k,1}^{\prod_{\substack{k'\in K \\ k' \neq k}}(\alpha_{k'}-1)}  \\
\phi_{p,q}^{\mathrm{max}} = &\ \mu_{p,q}\cdot \prod_{\pi(I,K)}  \left(a_0\prod_{i\in I}b_{i,1}\right)^{\prod_{k\in K}(\alpha_k-1)}\prod_{k\in K}b_{k,2}^{\prod_{\substack{k'\in K \\ k' \neq k}}(\alpha_{k'}-1)}
\end{aligned}
\end{equation}
For example, the first few nodes project onto $(x,y)$-plane with coordinates:
\begin{equation}
\begin{aligned}
\phi_{m,0}^{\mathrm{min}}:\ (0,0), & \quad
\phi_{m,0}^{\mathrm{max}}:\ (0,1), \\
\phi_{m-1,1}^{\mathrm{min}}:\ (0,1), & \quad
\phi_{m-1,1}^{\mathrm{max}}:\ \left(m,1+\sum_{i=1\dots m}(\alpha_i-1)\right)
\end{aligned}
\end{equation}
where $(x_i,y_i) = (\mathrm{deg}(\phi^{\mathrm{min/max}}_*,x),\mathrm{deg}(\phi_*^{\mathrm{min/max}},y))$.
The vertical edge given by $\phi_{m,0}$ is always presented in $A(x,y)$, since it encodes the analytic branch of $y$ as a function of $x$ (when the leading coefficient in the Puiseaux expansion has non-negative degree), see \cite{LNPS}.

Uniqueness of the preimage of each vertex of $N(A)$ follows from uniqueness of the corresponding mixed decomposition, where the $a_0$-type and $\left(\bigcup_i b_{i,2}\right)$-type cells (represented by the perfograms (\ref{phi_themonomial}) and (\ref{phi_themonomial_b2}), respectively) are fixed. There is no space to vary the other cells, as they would be fixed rigidly by their perfograms, and therefore produce a unique extremal monomial. Being projected, each of them gives a unique vertex of $N(A)$.

\begin{figure}[h!]
    \centering
    \includegraphics[scale=0.4]{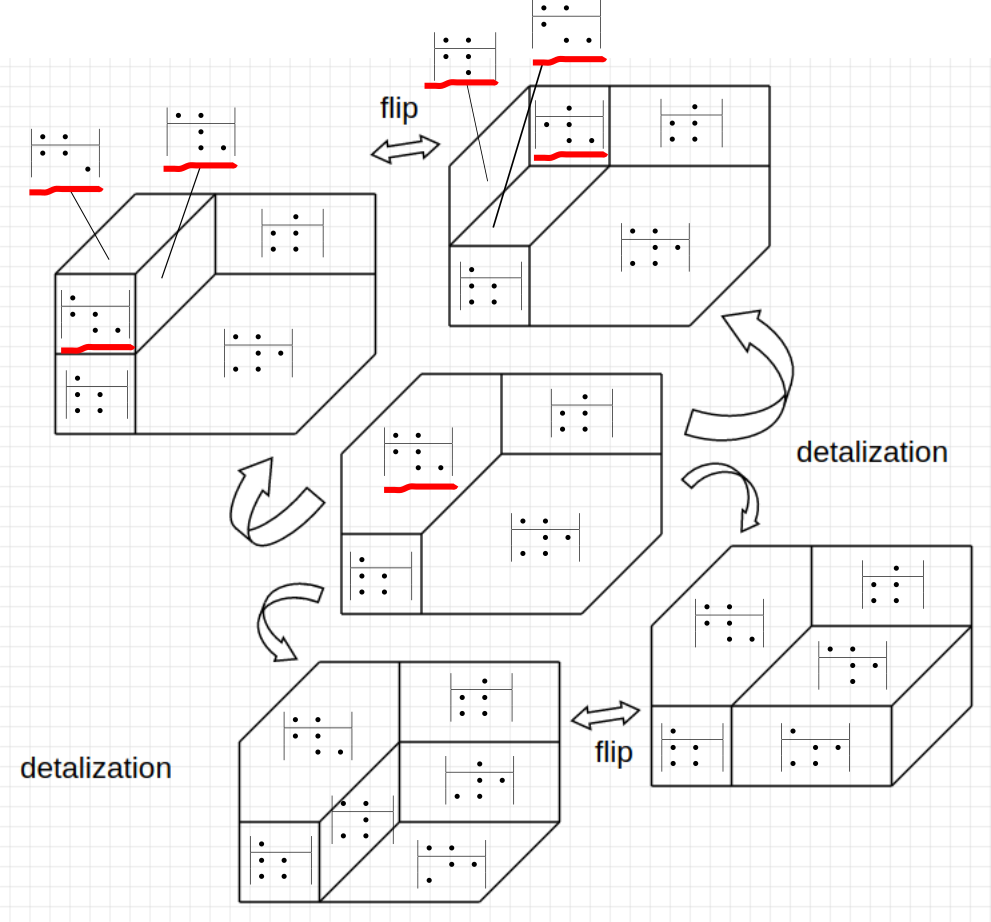}
    \caption{Detalization of $\phi_{1,1}$, applied once to each of its smaller hexagons}
    \label{fig:detalization_detailed}
\end{figure}

\begin{figure}[h!]
    \centering
    \input{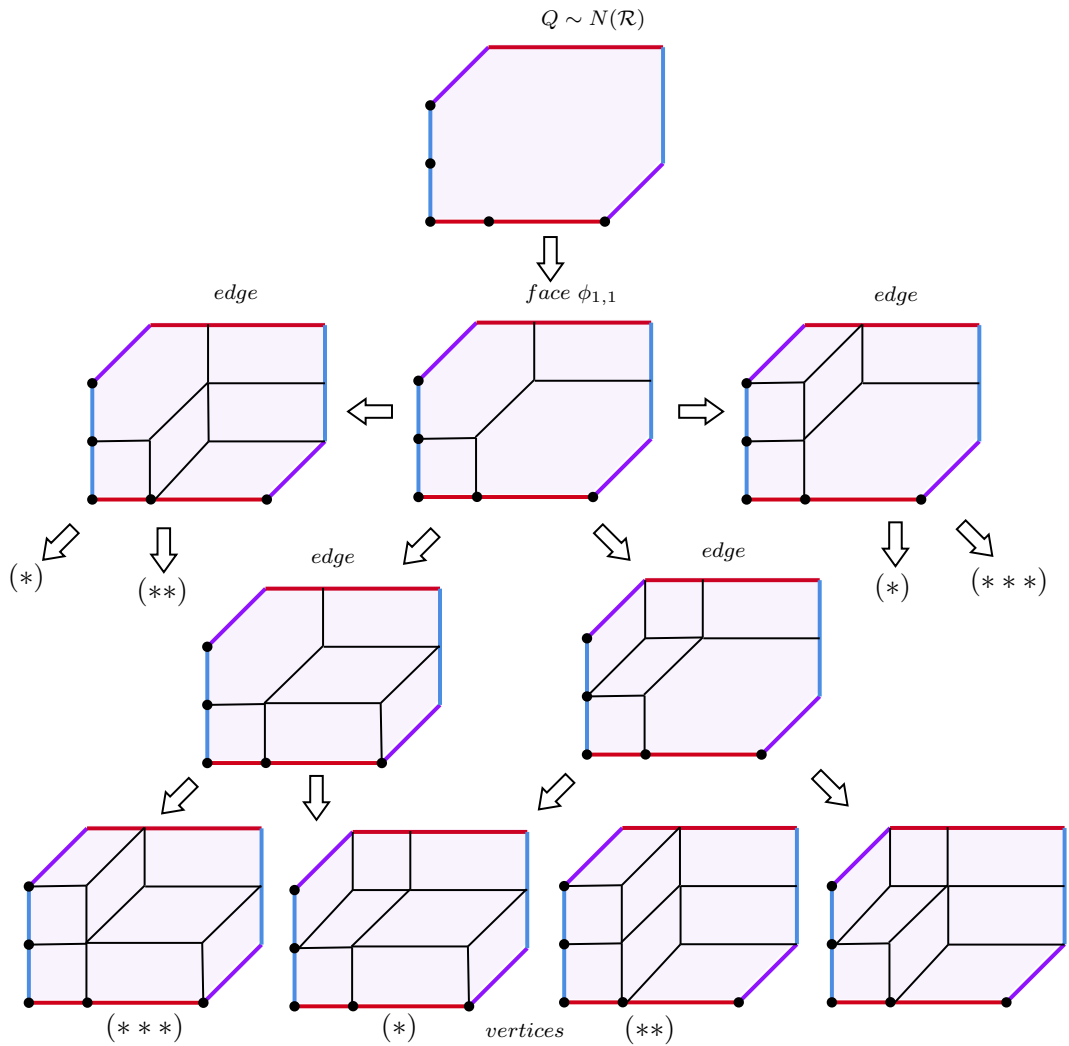}
    \caption{Detalization of $\phi_{1,1}$: the complete picture}
    \label{fig:delatization_map}
\end{figure}

We have to introduce one extra notion: the \emph{detalization map}, which subdivides a mixed decomposition, refining its cell structure by dividing cells into smaller cells. For a given face, it corresponds to picking up a particular sub-face. We can think of it as acting on the two mutually dual levels: 1) the level of TCMDs, and 2) the level of perfograms.
If we have a simple initial form supported on a face, given by a collection of perfograms, we can apply the detalization map to each of its perfograms as follows: assume we have a perfogram which corresponds to some $\varphi_{I,K}$ in (\ref{diag_general_binomials}):
\begin{equation}
 \varphi_{I,K} \quad \sim \quad 
 \begin{tabular}{|c c c|}
    \textcolor{black}{$\bullet$} & ${\bullet}$  & \\ \hline 
    & \textcolor{red}{${\bullet}$} & \textcolor{red}{$\bullet$} \\
    \vdots & \vdots & \vdots \\
    & \textcolor{red}{${\bullet}$} & \textcolor{red}{$\bullet$} \\
    ${\bullet}$ & $\bullet$ & \\
    \vdots & \vdots & \vdots \\
    ${\bullet}$ & $\bullet$ &      
    \end{tabular}
\end{equation}
(the positions of red and black rows are not so important, they can be arbitrary, as their numbers, so everything can be considered up to permutation of the rows). We proceed with the following steps:
\begin{enumerate}
    \item in the first equation $F_0$, highlight the leftmost (rightmost) bullet, by putting it into the ``box''
    \item in the rest of equations $F_1,\dots,F_m$, do the same for the rightmost (leftmost) bullets:
    \begin{equation}\label{detal_step1}
    \begin{tabular}{|c c c|}
    \textcolor{black}{$\bullet$} & $\boxed{\bullet}$  & \\ \hline 
    & \textcolor{red}{$\boxed{\bullet}$} & \textcolor{red}{$\bullet$} \\
    \vdots & \vdots & \vdots \\
    & \textcolor{red}{$\boxed{\bullet}$} & \textcolor{red}{$\bullet$} \\
    $\boxed{\bullet}$ & $\bullet$ & \\
    \vdots & \vdots & \vdots \\
    $\boxed{\bullet}$ & $\bullet$ &      
    \end{tabular}\quad \text{or} \quad
    \begin{tabular}{|c c c|}
    \textcolor{black}{$\boxed{\bullet}$} & $\bullet$  & \\ \hline 
    & \textcolor{red}{$\bullet$} & \textcolor{red}{$\boxed{\bullet}$} \\
    \vdots & \vdots & \vdots \\
    & \textcolor{red}{$\bullet$} & \textcolor{red}{$\boxed{\bullet}$} \\
    $\bullet$ & $\boxed{\bullet}$ & \\
    \vdots & \vdots & \vdots \\
    $\bullet$ & $\boxed{\bullet}$ &      
    \end{tabular}
    \end{equation}
    \item copy this perfogram as many times as the number of its rows, every time removing one of the bullets which are not in the box:
    
    \scalebox{0.65}{\parbox{.5\linewidth}{
$$\label{detal_step2}
    \begin{aligned}
    \begin{tabular}{|c c c|}
    \textcolor{black}{$\bullet$} & $\boxed{\bullet}$  & \\ \hline 
    & \textcolor{red}{$\boxed{\bullet}$} & \textcolor{red}{$\bullet$} \\
    \vdots & \vdots & \vdots \\
    & \textcolor{red}{$\boxed{\bullet}$} & \textcolor{red}{$\bullet$} \\
    $\boxed{\bullet}$ & $\bullet$ & \\
    \vdots & \vdots & \vdots \\
    $\boxed{\bullet}$ & $\bullet$ &      
    \end{tabular}
    \qquad \xrightarrow{\text{detalization}} \qquad \left\{ \quad
      \begin{tabular}{|c c c|}
     & $\boxed{\bullet}$  & \\ \hline 
    & \textcolor{red}{$\boxed{\bullet}$} & \textcolor{red}{$\bullet$} \\
    \vdots & \vdots & \vdots \\
    & \textcolor{red}{$\boxed{\bullet}$} & \textcolor{red}{$\bullet$} \\
    $\boxed{\bullet}$ & $\bullet$ & \\
    \vdots & \vdots & \vdots \\
    $\boxed{\bullet}$ & $\bullet$ &      
    \end{tabular}
    \qquad 
      \begin{tabular}{|c c c|}
    \textcolor{black}{$\bullet$} & $\boxed{\bullet}$  & \\ \hline 
    & \textcolor{red}{$\boxed{\bullet}$} &  \\
    \vdots & \vdots & \vdots \\
    & \textcolor{red}{$\boxed{\bullet}$} & \textcolor{red}{$\bullet$} \\
    $\boxed{\bullet}$ & $\bullet$ & \\
    \vdots & \vdots & \vdots \\
    $\boxed{\bullet}$ & $\bullet$ &      
    \end{tabular}
    \qquad
      \begin{tabular}{|c c c|}
    \textcolor{black}{$\bullet$} & $\boxed{\bullet}$  & \\ \hline 
    & \textcolor{red}{$\boxed{\bullet}$} & \textcolor{red}{$\bullet$} \\
    \vdots & \vdots & \vdots \\
    & \textcolor{red}{$\boxed{\bullet}$} &  \\
    $\boxed{\bullet}$ & $\bullet$ & \\
    \vdots & \vdots & \vdots \\
    $\boxed{\bullet}$ & $\bullet$ &      
    \end{tabular}
    \qquad
     \begin{tabular}{|c c c|}
    \textcolor{black}{$\bullet$} & $\boxed{\bullet}$  & \\ \hline 
    & \textcolor{red}{$\boxed{\bullet}$} & \textcolor{red}{$\bullet$} \\
    \vdots & \vdots & \vdots \\
    & \textcolor{red}{$\boxed{\bullet}$} & \textcolor{red}{$\bullet$} \\
    $\boxed{\bullet}$ &  & \\
    \vdots & \vdots & \vdots \\
    $\boxed{\bullet}$ & $\bullet$ &      
    \end{tabular}
    \qquad 
      \begin{tabular}{|c c c|}
    \textcolor{black}{$\bullet$} & $\boxed{\bullet}$  & \\ \hline 
    & \textcolor{red}{$\boxed{\bullet}$} & \textcolor{red}{$\bullet$} \\
    \vdots & \vdots & \vdots \\
    & \textcolor{red}{$\boxed{\bullet}$} & \textcolor{red}{$\bullet$} \\
    $\boxed{\bullet}$ & $\bullet$ & \\
    \vdots & \vdots & \vdots \\
    $\boxed{\bullet}$ &  &      
    \end{tabular}\quad \right\}
    \end{aligned}
$$
     }}
     
    which defines the map
    $$
        \text{edge} \qquad \longrightarrow \qquad \text{vertex}
    $$
    \item the whole set of such reduced perfograms corresponds to a vertex of $N(\mathcal{R})$: a tail or a head of an edge given by $\varphi_{I,K}$ (if the leftmost or rightmost configuration is chosen, respectively)
\end{enumerate}
Therefore, if we have just a single perfogram (remember that it always corresponds to an edge), there are only two options: detalization gives either its head or tail vertex (a kind of ``morsification'', since we assume that the head is always above the tail). On another hand, if the face has a bigger dimension, we can apply detalization to each of its perfograms independently, each time choosing either a head or a tail. In this way, all possible choices generate the complete set of sub-faces.
Here is an example of how it looks on the level of TCMDs, borrowed from Section \ref{section_two_dimensional} (Figure \ref{fig:detalization_detailed}).
\\
Now we see how the ``abstract'' steps 1-4 work. For example, the underlined perfograms align into the pattern of step 3. The two possibilities -- the choice of the rightmost or leftmost boxes -- implement the \emph{cubical flip} inside of each hexagon, which is being detailed.
Returning to Figure \ref{fig:N22}, we see that the middle TCMD corresponds to the 2-dimensional face $b$, whereas each of the four detalizations give one of its edges.  
On another hand, we can iterate the procedure to obtain the complete set of sub-faces of $\phi_{1,1}$ (Figure \ref{fig:delatization_map}).

Let's prove an important intermediate statement, which clarifies the incidence relations for $\{\phi_{p,q}\}$.
\begin{proposition} The incidence relation for min/max:
\begin{equation}
\phi_{m-i,i}^{\mathrm{max}} = \phi_{m-i-1,i+1}^{\mathrm{min}},\ \forall i=0 \dots m; \ m \geq 2    
\end{equation}
\end{proposition}
\underline{Proof}. To begin with, let's see how the detalization map acts on each perfogram, representing a binomial in $\phi_{p,q}$. We study the pattern for $\mathrm{diag}(\alpha_1,\alpha_2)$: $\phi_{2,1}^{\mathrm{max}} = \phi_{1,2}^{\mathrm{min}}$. Then it will follow for any $\phi_{p,q}$ by induction on $m$.
First, for $\phi_{2,1}$:
\begin{equation}
\phi_{2,1} \sim  \qquad \begin{tabular}{|c c c|}
  $\boxed{\bullet}$ & $\bullet$ & \\ \hline
  & $\bullet$ & $\boxed{\bullet}$ \\
  $\bullet$ & $\boxed{\bullet}$ &  \\      
  $\bullet$ & $\boxed{\bullet}$ & 
\end{tabular}
\ \times\ 
\begin{tabular}{|c c c|}
  $\boxed{\bullet}$ & $\bullet$ & \\ \hline
  $\bullet$ & $\boxed{\bullet}$ &  \\      
  & $\bullet$ & $\boxed{\bullet}$ \\
  $\bullet$ & $\boxed{\bullet}$ & 
\end{tabular}
\ \times \
\begin{tabular}{|c c c|}
  $\boxed{\bullet}$ & $\bullet$ & \\ \hline
  $\bullet$ & $\boxed{\bullet}$ &  \\      
  $\bullet$ & $\boxed{\bullet}$ &   \\    
  & $\bullet$ & $\boxed{\bullet}$
\end{tabular}
\end{equation}
We take its detalization which gives the maximal weight:
\begin{eqnarray}\label{minmax_phi21}
&\ \begin{tabular}{|c c c|}
  $\boxed{\bullet}$ & & \\ \hline
  & $\bullet$ & $\boxed{\bullet}$ \\
  $\bullet$ & $\boxed{\bullet}$ &  \\      
  $\bullet$ & $\boxed{\bullet}$ & 
\end{tabular}
\ \times\ 
\begin{tabular}{|c c c|}
  $\boxed{\bullet}$ & & \\ \hline
  $\bullet$ & $\boxed{\bullet}$ & \\      
  & $\bullet$ & $\boxed{\bullet}$ \\
  $\bullet$ & $\boxed{\bullet}$ & 
\end{tabular}
\ \times \
\begin{tabular}{|c c c|}
  $\boxed{\bullet}$ & & \\ \hline
  $\bullet$ & $\boxed{\bullet}$ &  \\      
  $\bullet$ & $\boxed{\bullet}$ &   \\    
  & $\bullet$ & $\boxed{\bullet}$
\end{tabular}
\\ \ \\ 
&\ \begin{tabular}{|c c c|}
  $\boxed{\bullet}$ & $\bullet$ & \\ \hline
  & & $\boxed{\bullet}$ \\
  $\bullet$ & $\boxed{\bullet}$ &  \\      
  $\bullet$ & $\boxed{\bullet}$ & 
\end{tabular}
\ \times\ 
\begin{tabular}{|c c c|}
  $\boxed{\bullet}$ & $\bullet$ & \\ \hline
  & $\boxed{\bullet}$ &  \\      
  & $\bullet$ & $\boxed{\bullet}$ \\
  $\bullet$ & $\boxed{\bullet}$ & 
\end{tabular}
\ \times \
\begin{tabular}{|c c c|}
  $\boxed{\bullet}$ & $\bullet$ & \\ \hline
  & $\boxed{\bullet}$ &  \\      
  $\bullet$ & $\boxed{\bullet}$ &   \\    
  & $\bullet$ & $\boxed{\bullet}$
\end{tabular}
\\ \ \\ 
&\ \begin{tabular}{|c c c|}
  $\boxed{\bullet}$ & $\bullet$ & \\ \hline
  & $\bullet$ & $\boxed{\bullet}$ \\
  & $\boxed{\bullet}$ &  \\      
  $\bullet$ & $\boxed{\bullet}$ & 
\end{tabular}
\ \times\ 
\begin{tabular}{|c c c|}
  $\boxed{\bullet}$ & $\bullet$ & \\ \hline
  $\bullet$ & $\boxed{\bullet}$ &  \\      
  & & $\boxed{\bullet}$ \\
  $\bullet$ & $\boxed{\bullet}$ & 
\end{tabular}
\ \times \
\begin{tabular}{|c c c|}
  $\boxed{\bullet}$ & $\bullet$ & \\ \hline
  $\bullet$ & $\boxed{\bullet}$ &  \\      
  & $\boxed{\bullet}$ &   \\    
  & $\bullet$ & $\boxed{\bullet}$
\end{tabular}
\\ \ \\ 
&\ \begin{tabular}{|c c c|}
  $\boxed{\bullet}$ & $\bullet$ & \\ \hline
  & $\bullet$ & $\boxed{\bullet}$ \\
  $\bullet$ & $\boxed{\bullet}$ &  \\      
  & $\boxed{\bullet}$ & 
\end{tabular}
\ \times\ 
\begin{tabular}{|c c c|}
  $\boxed{\bullet}$ & $\bullet$ & \\ \hline
  $\bullet$ & $\boxed{\bullet}$ &  \\      
  & $\bullet$ & $\boxed{\bullet}$ \\
  & $\boxed{\bullet}$ & 
\end{tabular}
\ \times \
\begin{tabular}{|c c c|}\label{minmax_phi21_end}
  $\boxed{\bullet}$ & $\bullet$ & \\ \hline
  $\bullet$ & $\boxed{\bullet}$ &  \\      
  $\bullet$ & $\boxed{\bullet}$ &   \\    
  & & $\boxed{\bullet}$
\end{tabular}
\end{eqnarray}
The boxed monomials are those, which remain frozen (always non-zero) when doing detalization, i.e. we do not cross them out. We obtain decompositions of each binomial diagram (of the three in the upper row) into four pieces (forming a column), such that all of them along with $\mu_{2,1}$ define the corresponding extremal monomial $\phi_{2,1}^{\mathrm{max}}$. For the sake of completeness, we also give a formula for $\mu_{2,1}$:
\begin{equation}
\begin{aligned}
(a_0):\ 
\begin{tabular}{|c c c|}
  $\bullet$ &  & \\ \hline
  $\bullet$ & $\bullet$ & \\
  $\bullet$ & $\bullet$ &  \\      
  $\bullet$ & $\bullet$ & 
\end{tabular}
\quad
(a_1):\ 
\begin{tabular}{|c c c|}
  & $\bullet$  & \\ \hline
  & $\bullet$ & $\bullet$ \\
  & $\bullet$ & $\bullet$  \\      
  & $\bullet$ & $\bullet$ 
\end{tabular}\quad 
\begin{tabular}{|c c c|}
  & $\bullet$  & \\ \hline
  $\bullet$ & $\bullet$ & \\
  & $\bullet$ & $\bullet$  \\      
  & $\bullet$ & $\bullet$ 
\end{tabular}_{(\pi)}
\quad 
(b_{*,0}):\
\begin{tabular}{|c c c|}
  $\bullet$ & $\bullet$  & \\ \hline
  $\bullet$ & & \\
  & $\bullet$ & $\bullet$  \\      
  & $\bullet$ & $\bullet$ 
\end{tabular}_{(\pi)}
\end{aligned}
\end{equation}
Now we do the same thing for $\phi_{1,2}$:
\begin{equation}
\phi_{1,2} \sim \qquad   
\begin{tabular}{|c c c|}
  $\bullet$ & $\boxed{\bullet}$ & \\ \hline
  $\boxed{\bullet}$ & $\bullet$ & \\
  & $\boxed{\bullet}$ & $\bullet$  \\      
  & $\boxed{\bullet}$ & $\bullet$       
\end{tabular}
\ \times\ 
\begin{tabular}{|c c c|}
  $\bullet$ & $\boxed{\bullet}$ & \\ \hline
  & $\boxed{\bullet}$ & $\bullet$  \\      
  $\boxed{\bullet}$ & $\bullet$ & \\
  & $\boxed{\bullet}$ & $\bullet$       
\end{tabular}
\ \times\ 
\begin{tabular}{|c c c|}
  $\bullet$ & $\boxed{\bullet}$ & \\ \hline
  & $\boxed{\bullet}$ & $\bullet$  \\      
  & $\boxed{\bullet}$ & $\bullet$   \\    
  $\boxed{\bullet}$ & $\bullet$ &
\end{tabular}
\end{equation}
and its detalization which gives the minimal weight:
\begin{eqnarray}\label{minmax_phi12}
&\ \begin{tabular}{|c c c|}
  & $\boxed{\bullet}$ & \\ \hline
  $\boxed{\bullet}$ & $\bullet$ & \\
  & $\boxed{\bullet}$ & $\bullet$  \\      
  & $\boxed{\bullet}$ & $\bullet$       
\end{tabular}
\ \times\ 
\begin{tabular}{|c c c|}
  & $\boxed{\bullet}$ & \\ \hline
  & $\boxed{\bullet}$ & $\bullet$  \\      
  $\boxed{\bullet}$ & $\bullet$ & \\
  & $\boxed{\bullet}$ & $\bullet$       
\end{tabular}
\ \times\ 
\begin{tabular}{|c c c|}
  & $\boxed{\bullet}$ & \\ \hline
  & $\boxed{\bullet}$ & $\bullet$  \\      
  & $\boxed{\bullet}$ & $\bullet$   \\    
  $\boxed{\bullet}$ & $\bullet$ &
\end{tabular}
\\ \ \\
&\ \begin{tabular}{|c c c|}
  $\bullet$ & $\boxed{\bullet}$ & \\ \hline
  $\boxed{\bullet}$ & & \\
  & $\boxed{\bullet}$ & $\bullet$  \\      
  & $\boxed{\bullet}$ & $\bullet$       
\end{tabular}
\ \times\ 
\begin{tabular}{|c c c|}
  $\bullet$ & $\boxed{\bullet}$ & \\ \hline
  & $\boxed{\bullet}$ &  \\      
  $\boxed{\bullet}$ & $\bullet$ & \\
  & $\boxed{\bullet}$ & $\bullet$       
\end{tabular}
\ \times\ 
\begin{tabular}{|c c c|}
  $\bullet$ & $\boxed{\bullet}$ & \\ \hline
  & $\boxed{\bullet}$ &  \\      
  & $\boxed{\bullet}$ & $\bullet$   \\    
  $\boxed{\bullet}$ & $\bullet$ &
\end{tabular}
\\ \ \\
&\ \begin{tabular}{|c c c|}
  $\bullet$ & $\boxed{\bullet}$ & \\ \hline
  $\boxed{\bullet}$ & $\bullet$ & \\
  & $\boxed{\bullet}$ &  \\      
  & $\boxed{\bullet}$ & $\bullet$       
\end{tabular}
\ \times\ 
\begin{tabular}{|c c c|}
  $\bullet$ & $\boxed{\bullet}$ & \\ \hline
  & $\boxed{\bullet}$ & $\bullet$  \\      
  $\boxed{\bullet}$ & & \\
  & $\boxed{\bullet}$ & $\bullet$       
\end{tabular}
\ \times\ 
\begin{tabular}{|c c c|}
  $\bullet$ & $\boxed{\bullet}$ & \\ \hline
  & $\boxed{\bullet}$ & $\bullet$  \\      
  & $\boxed{\bullet}$ &   \\    
  $\boxed{\bullet}$ & $\bullet$ &
\end{tabular}
\\ \ \\
&\ \begin{tabular}{|c c c|}
  $\bullet$ & $\boxed{\bullet}$ & \\ \hline
  $\boxed{\bullet}$ & $\bullet$ & \\
  & $\boxed{\bullet}$ & $\bullet$  \\      
  & $\boxed{\bullet}$ &       
\end{tabular}
\ \times\ 
\begin{tabular}{|c c c|}
  $\bullet$ & $\boxed{\bullet}$ & \\ \hline
  & $\boxed{\bullet}$ & $\bullet$  \\      
  $\boxed{\bullet}$ & $\bullet$ & \\
  & $\boxed{\bullet}$ &       
\end{tabular}
\ \times\ 
\begin{tabular}{|c c c|}\label{minmax_phi12_end}
  $\bullet$ & $\boxed{\bullet}$ & \\ \hline
  & $\boxed{\bullet}$ & $\bullet$  \\      
  & $\boxed{\bullet}$ & $\bullet$   \\    
  $\boxed{\bullet}$ & &
\end{tabular}
\end{eqnarray}
The monomial $\mu_{1,2}$:
\begin{equation}
\begin{aligned}
(a_0):\ 
\begin{tabular}{|c c c|}
  $\bullet$ &  & \\ \hline
  $\bullet$ & $\bullet$ & \\
  $\bullet$ & $\bullet$ &  \\      
  $\bullet$ & $\bullet$ & 
\end{tabular}\quad 
\begin{tabular}{|c c c|}
  $\bullet$ &  & \\ \hline
  & $\bullet$ & $\bullet$ \\
  $\bullet$ & $\bullet$ &  \\      
  $\bullet$ & $\bullet$ & 
\end{tabular}_{(\pi)}
\quad
(a_1):\ 
\begin{tabular}{|c c c|}
  & $\bullet$  & \\ \hline
  & $\bullet$ & $\bullet$ \\
  & $\bullet$ & $\bullet$  \\      
  & $\bullet$ & $\bullet$ 
\end{tabular}
\quad 
(b_{*,2}):\
\begin{tabular}{|c c c|}
  $\bullet$ & $\bullet$  & \\ \hline
  & & $\bullet$ \\
  $\bullet$ & $\bullet$ &  \\      
  $\bullet$ & $\bullet$ &
\end{tabular}_{(\pi)}
\end{aligned}
\end{equation}
We see that the binomial counterparts of both $\phi_{1,2}^{\mathrm{max}}$ and $\phi_{2,1}^{\mathrm{min}}$ have the identical collections of $b_{*,1}$-perfograms. Moreover, it immediately extends to any $\phi_{p,q}$, since detalizing any of $b_{*,1}$ corresponds to taking a row of the form $[\bullet \bullet \ ]$ or $[\ \bullet \bullet]$. So, in order to get the maximum (minimum), we remove the left (right) neighbouring ``$\bullet$'', which results in the same perfogram. 
Next, comparing the $a_0$- and $a_1$-perfograms, we see that those ones, which are in the binomial part of $\phi_{2,1}^{\mathrm{min}}$, coincide with the $\mu$-part in $\phi_{2,1}^{\mathrm{max}}$, and vice versa.
This is also true for $b_{*,2}$ counterpart (follows from Proposition \ref{initial_forms_diag_arbirary}).
Therefore, there is an ``exchange relation'' between the two collections of perfograms, resulting into identical extremal monomials. Moreover, this rule extends to any $p,q$, hence the claim of the proposition. $\square$

Returning to the proof of Theorem \ref{main_prop_diagonal},
what is left to show is that when sending $a_0$ to $y$ and $b_{i,2}$ to $x$, for $i=1\dots m$,
the $x$- and $y$-degrees of each of monomial for $\phi_{p,q}$ grow linearly. That is, after the principal specialization they project onto the same line segment, if and only if $\alpha_i$ are all equal.

First, recall that all $\phi_{p,q}$ are simple, meaning that all its monomials are extreme monomials of $N(\mathcal{R})$, and its dimension equals to the number of distinct binomial factors. We already described now to compute its monomials with minimal and maximal powers of $a_0$. Now how we do it for all other monomials? The answer is simple (and was in fact already given in \cite{Stu}): we have to take all possible combinations of min and max applied to a particular binomial, in such way to obtain its full detalization (e.g. Figure \ref{fig:delatization_map} for $m = 2$), and the resulting mixed decomposition would give us the extreme monomial of $\phi_{p,q}$, and then changing the min max configuration will give another extremal monomial, and so on.

Consider, for example, $\phi_{2,1}$ versus $\phi_{1,2}$. Each of them contains three distinct binomial factors -- denote them as $\mathcal{H}_{p,q}^{(i)},\ (p,q)=(2,1)$ or $(1,2)$. Therefore, their monomials are given by the triples (\textcolor{blue}{min},\textcolor{blue}{min},\textcolor{blue}{min}) ... (\textcolor{red}{max},\textcolor{red}{max},\textcolor{red}{max}) (\ref{H_mixmax_comb}).
\begin{equation}\label{H_mixmax_comb}
\begin{tabular}{|c|c|c|}
\hline
 $\mathcal{H}_{p,q}^{(1)}$ & $\mathcal{H}_{p,q}^{(2)}$ & $\mathcal{H}_{p,q}^{(3)}$ \\
 \hline
 \textcolor{blue}{min} & \textcolor{blue}{min} & \textcolor{blue}{min}  \\
 \textcolor{blue}{min} & \textcolor{blue}{min} & \textcolor{red}{max}  \\
 \textcolor{blue}{min} & \textcolor{red}{max} & \textcolor{blue}{min}  \\
 \textcolor{blue}{min} & \textcolor{red}{max} & \textcolor{red}{max}  \\
 \textcolor{red}{max} & \textcolor{blue}{min} & \textcolor{blue}{min}  \\
 \textcolor{red}{max} & \textcolor{blue}{min} & \textcolor{red}{max}  \\
 \textcolor{red}{max} & \textcolor{red}{max} & \textcolor{blue}{min}  \\
 \textcolor{red}{max} & \textcolor{red}{max} & \textcolor{red}{max} \\
 \hline
\end{tabular}
\end{equation}
In total there are 8 monomials for $\phi_{2,1}$ (or $\phi_{1,2}$). We have the following rule for ``\textcolor{blue}{min}'' (``\textcolor{red}{max}''): make the $a_1$ ($a_0$) bullet frozen: $\boxed{\bullet}$, along with all the leftmost (rightmost) $b$-type bullets, as shown on (\ref{minmax_phi21}-\ref{minmax_phi21_end}) and (\ref{minmax_phi12}-\ref{minmax_phi12_end}). Then, duplicate the perfogram by removing each non-frozen bullet, to obtain a collection of perfograms corresponding to a single coefficient in a non-negative power. Therefore, the whole collection will now define some extreme monomial.

\begin{figure}[h!]
    \centering
    \begin{tikzpicture}[x=0.75pt,y=0.75pt,yscale=-1,xscale=1]

\draw    (187,1724.75) -- (187,1784.75) ;
\draw [line width=1.5]    (347,1564.75) -- (187,1724.75) ;
\draw    (427,1544.75) -- (347,1564.75) ;
\draw  [fill={rgb, 255:red, 208; green, 2; blue, 27 }  ,fill opacity=1 ] (183,1724.75) .. controls (183,1722.54) and (184.79,1720.75) .. (187,1720.75) .. controls (189.21,1720.75) and (191,1722.54) .. (191,1724.75) .. controls (191,1726.96) and (189.21,1728.75) .. (187,1728.75) .. controls (184.79,1728.75) and (183,1726.96) .. (183,1724.75) -- cycle ;
\draw  [fill={rgb, 255:red, 208; green, 2; blue, 27 }  ,fill opacity=1 ] (343,1564.75) .. controls (343,1562.54) and (344.79,1560.75) .. (347,1560.75) .. controls (349.21,1560.75) and (351,1562.54) .. (351,1564.75) .. controls (351,1566.96) and (349.21,1568.75) .. (347,1568.75) .. controls (344.79,1568.75) and (343,1566.96) .. (343,1564.75) -- cycle ;
\draw  [fill={rgb, 255:red, 65; green, 117; blue, 5 }  ,fill opacity=1 ] (233,1673.75) .. controls (233,1671.54) and (234.79,1669.75) .. (237,1669.75) .. controls (239.21,1669.75) and (241,1671.54) .. (241,1673.75) .. controls (241,1675.96) and (239.21,1677.75) .. (237,1677.75) .. controls (234.79,1677.75) and (233,1675.96) .. (233,1673.75) -- cycle ;
\draw  [fill={rgb, 255:red, 65; green, 117; blue, 5 }  ,fill opacity=1 ] (283,1624.75) .. controls (283,1622.54) and (284.79,1620.75) .. (287,1620.75) .. controls (289.21,1620.75) and (291,1622.54) .. (291,1624.75) .. controls (291,1626.96) and (289.21,1628.75) .. (287,1628.75) .. controls (284.79,1628.75) and (283,1626.96) .. (283,1624.75) -- cycle ;
\draw    (188.5,1692.25) -- (208.59,1672.16) ;
\draw [shift={(210,1670.75)}, rotate = 495] [color={rgb, 255:red, 0; green, 0; blue, 0 }  ][line width=0.75]    (10.93,-3.29) .. controls (6.95,-1.4) and (3.31,-0.3) .. (0,0) .. controls (3.31,0.3) and (6.95,1.4) .. (10.93,3.29)   ;

\draw (191,1724.75) node [anchor=north west][inner sep=0.75pt]  [font=\scriptsize]  {$\mathrm{\{\textcolor{blue}{min},\textcolor{blue}{min},\textcolor{blue}{min}\}}$};
\draw (347,1568.75) node [anchor=north west][inner sep=0.75pt]  [font=\scriptsize]  {$\mathrm{\{\textcolor{red}{max},\textcolor{red}{max},\textcolor{red}{max}\}}$};
\draw (241,1673.75) node [anchor=north west][inner sep=0.75pt]  [font=\scriptsize]  {$\mathrm{\{\textcolor{blue}{min},\textcolor{blue}{min},\textcolor{red}{max}\}}$};
\draw (291,1624.75) node [anchor=north west][inner sep=0.75pt]  [font=\scriptsize]  {$\mathrm{\{\textcolor{blue}{min},\textcolor{red}{max},\textcolor{red}{max}\}}$};
\draw (139,1701.75) node [anchor=north west][inner sep=0.75pt]    {$\phi ^{\mathrm{\textcolor{blue}{min}}}_{2,1}$};
\draw (287,1521.75) node [anchor=north west][inner sep=0.75pt]    {$\phi ^{\mathrm{\textcolor{red}{max}}}_{2,1} =\phi ^{\mathrm{\textcolor{blue}{min}}}_{1,2}$};
\draw (358,1680.75) node [anchor=north west][inner sep=0.75pt]  [font=\LARGE]  {$N(A)$};

\end{tikzpicture}
    \caption{The $(x,y)$-projection of $\phi_{2,1}$ onto the edge of $N(A)$ gives the four nodes, and the min/max rule}
    \label{fig:increments}
\end{figure}
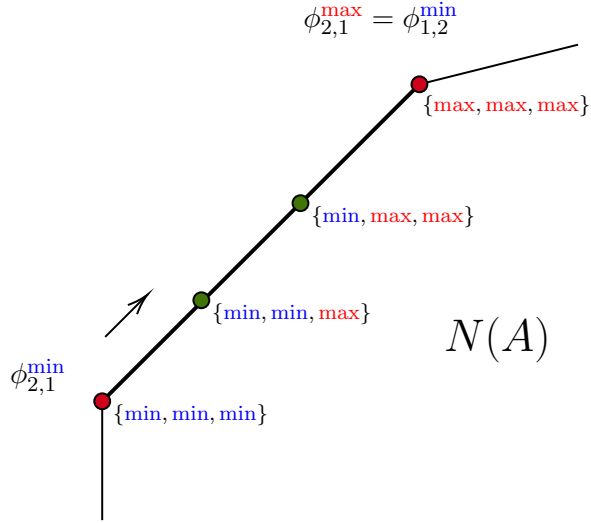

Recall that the binomial counterpart of $\phi_{2,1}^{\mathrm{min}}$ does not depend on neither $a_0$ nor $b_{i,2}$, which  means that the $(x,y)$-coordinates of the minimal monomial are completely fixed by $\mu_{2,1}$. Moving to the next order gives an increment to both $x$- and $y$- degrees of $\mu_{2,1}$ (which we denote as $\tilde{\mu}_x$ and $\tilde{\mu}_y$). For the fist increment, we replace a single ``\textcolor{blue}{min}'' by ``\textcolor{red}{max}'', say, in $\mathcal{H}_s := \mathcal{H}^{(s)}_{p,q}$ for $s\in \{1,2,3\}$.
This amounts to changing the frozen configuration, so that the $a_0$-degree gets the increment $+\prod_{j\in K'_s}(\alpha_j-1)$, where $K'_s$ is attached to $\mathcal{H}_s$. If we do that again, we modify yet another factor $\mathcal{H}_{s'}$, getting the increment: $a_0\mapsto a_0 + \prod_{j\in K'_{s'}}(\alpha_j-1)$, and so on, until we reach $\phi_{2,1}^{\mathrm{max}} \simeq (\mathrm{max},\mathrm{max},\mathrm{max})$.
Analogously, for $b_{*,2}$ the increment at $\mathcal{H}_s$ be like: $+\prod_{j\in K'_s\setminus \{*\}}(\alpha_j-1)$. The total $x$-degree is given by summing up the latter expression. We obtain the sequence of increments (Figure \ref{fig:increments}).

Therefore, each time by changing ``\textcolor{blue}{min}'' of $\mathcal{H}_s$ to ``\textcolor{red}{max}'' (\ref{H_mixmax_comb}), we get the increments for the $(x,y)$-coordinates of a monomial on the edge of $N(A)$:
\begin{equation}\label{incr_form}
    x \mapsto x + \sum_{r=1\dots m}\prod_{j\in K'_s\setminus \{r\}}(\alpha_j-1), \quad y \mapsto y + \prod_{j\in K'_s}(\alpha_j-1),
\end{equation}
We see that the increment, being a function of $(\alpha_1,\dots,\alpha_m)$, is non-linear. Now it becomes clear by looking at (\ref{incr_form}) that it is linear only when $\alpha_i$ are all equal. 
Of course the $(x,y)$-degree of the monomial does not depend on permutation (i.e. $(\textcolor{blue}{\mathrm{min}},\textcolor{blue}{\mathrm{min}},\textcolor{red}{\mathrm{max}})$ and $(\textcolor{red}{\mathrm{max}},\textcolor{blue}{\mathrm{min}},\textcolor{blue}{\mathrm{min}})$ are indistinguishable after the $(x,y)$-projection). Therefore, eight monomials of $\phi_{2,1}$ (or $\phi_{1,2}$) are mapped onto four points on the edge of $N(A)$, see Figure \ref{fig:increments}.
The endpoints are extremal, and in one-to-one correspondence with their preimages, while the intermediate nodes (green) correspond to classes of permutations. The picture is drawn when $\alpha$'s are all equal, which guarantees linearity of the increments, therefore a single line segment being a projection slope.
Vice versa, linearity of the increment forces all $\alpha_i$ to be equal, since any pair $\alpha_i \neq \alpha_j$ will result in a change of slope on Figure \ref{fig:increments} between any of the two nodes. This calculation is absolutely analogous for any $m$ and $p,q$, so that the result does not depend on the size of perfograms.

Finally, we have to clarify the following: if $\alpha_i$ are all equal, there are no intermediate monomials on the edges, except from the projection of $\phi_{p,q}$.
This follows from the fact that the minimally allowed $y$-increment is equal to $(\alpha-1)$. For $m = 2$ this holds trivially, since there are no integer lattice points on each edge of $N(A)$, which are not the monomials of $\phi_{p,q}$ for some $p$ and $q$. Then, for $\alpha>2$ there are integer points between red - green, and green-green nodes (Figure \ref{fig:increments}). But in order to have the corresponding monomials in $A(x,y)$, one has to apply $y$-increment which is smaller than ($\alpha-1$), among all the initial forms $\{\phi_{p,q}\}$, which is of course forbidden. Those integer lattice points would stay unoccupied, hence the claim.

At this point, we have finally completed the proof of the main Theorem \ref{main_prop_diagonal}. $\square$

\newpage	

\appendix
	
\section{Experimental data}\label{append_a}

\subsection{Diagonal quivers}\label{append_a_1}

Here we provide some examples of quivers with diagonal matrix
$C$, in order to support Conjecture (\ref{3vertex_conj}). All computations were performed by using Maple.
Since their quiver polynomials are quite huge, we use the notation $[c_{ijkl},i,j,k,l]$ for each monomial
in $A(x_1,x_2,x_3,y)$:
\begin{equation}
     A(x_1,x_2,x_3,y) = \sum_{i,j,k,l} c_{ijkl}x_1^ix_2^jx_3^ky^l
\end{equation}

\subsubsection*{$\mathrm{diag}(2,2,2)$}

\small
\begin{verbatim}
[1,4,4,4,8],[1,3,3,3,7],[-4,3,3,3,6],[-2,3,3,2,6],[-2,3,2,3,6],[-2,2,
3,3,6],[-1,3,2,2,6],[-1,2,3,2,6],[-1,2,2,3,6],[5,2,2,2,5],[1,2,2,1,5],
[1,2,1,2,5],[1,1,2,2,5],[6,2,2,2,4],[4,2,2,1,4],[4,2,1,2,4],[4,1,2,2,4
],[1,2,2,0,4],[1,2,0,2,4],[1,0,2,2,4],[-1,1,1,1,4],[-5,1,1,1,3],[-1,1,
1,0,3],[-1,1,0,1,3],[-1,0,1,1,3],[-4,1,1,1,2],[-2,1,1,0,2],[-2,1,0,1,2
],[-2,0,1,1,2],[-1,1,0,0,2],[-1,0,1,0,2],[-1,0,0,1,2],[-1,0,0,0,1],[1,0
,0,0,0]
\end{verbatim}
\normalsize

\begin{figure}[h!]
    \centering
    \includegraphics[scale=0.4]{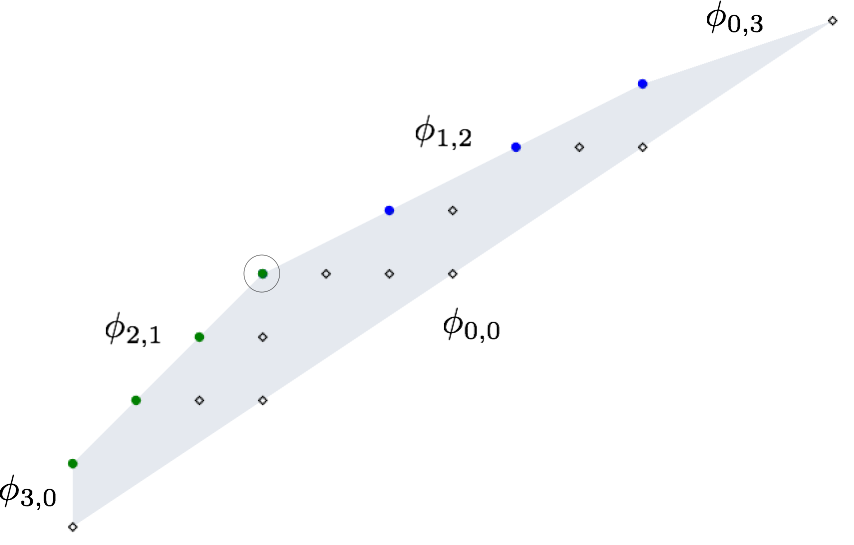}
    \caption{Newton polygon and the initial forms $\phi_{p,q}$ forming its edges for $\mathrm{diag}(2,2,2)$ quiver}
    \label{fig:diag222}
\end{figure}

On Figure \ref{fig:diag222} and henceforth, the green and blue nodes are (the projections of) $\phi_{2,1}$ and $\phi_{1,2}$, correspondingly. The circled point is where $\phi_{2,1}$ meets $\phi_{1,2}$, i.e. $\phi_{2,1}^{\mathrm{max}} = \phi_{1,2}^{\mathrm{min}}$. This is the $m = 3$ case of Figure \ref{fig:newton_projection_phi}.

\subsubsection*{$\mathrm{diag}(2,2,3)$}

\small
\begin{verbatim}
[[1,6,6,4,12],[-4,5,5,3,10],[-2,5,4,3,10],[-2,4,5,3,10],[-1,4,4,3,10],
[9,4,4,3,9],[3,4,3,3,9],[3,3,4,3,9],[1,3,3,3,9],[6,4,4,2,8],[4,4,3,2,8
],[4,3,4,2,8],[1,4,2,2,8],[1,2,4,2,8],[-11,3,3,2,7],[-5,3,2,2,7],[-5,2
,3,2,7],[-2,3,3,2,6],[-1,3,1,2,7],[-1,1,3,2,7],[-4,3,3,1,6],[-9,3,2,2,
6],[-9,2,3,2,6],[-2,3,2,1,6],[-6,3,1,2,6],[-2,2,3,1,6],[-6,1,3,2,6],[-
1,3,0,2,6],[-1,2,2,1,6],[-1,0,3,2,6],[3,2,2,1,5],[2,2,1,1,5],[2,1,2,1,
5],[4,2,2,1,4],[1,1,1,1,5],[1,2,2,0,4],[8,2,1,1,4],[8,1,2,1,4],[2,2,0,
1,4],[7,1,1,1,4],[2,0,2,1,4],[1,1,0,1,4],[1,0,1,1,4],[9,1,1,1,3],[-1,1
,1,0,3],[3,1,0,1,3],[3,0,1,1,3],[1,0,0,1,3],[-2,1,1,0,2],[-1,1,0,0,2],
[-1,0,1,0,2],[-1,0,0,0,1],[1,0,0,0,0]]
\end{verbatim}
\normalsize

\begin{figure}[h!]
    \centering
    \includegraphics[scale=0.4]{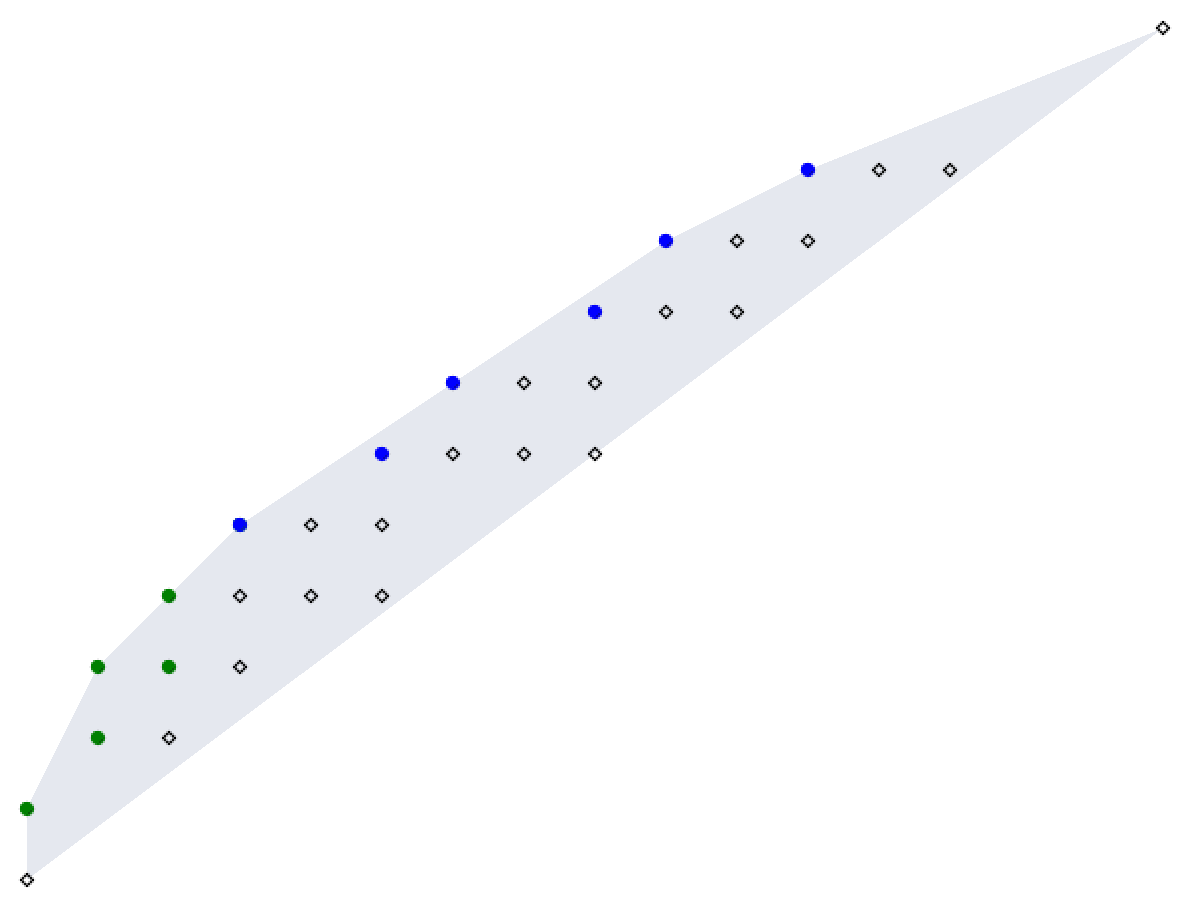}
    \caption{Newton polygon for $\mathrm{diag}(2,2,3)$ quiver}
    \label{fig:diag223}
\end{figure}
On Figure \ref{fig:diag223} (unlike for $\mathrm{diag}(2,2,2)$, Figure \ref{fig:diag222}) we can see that some of the monomials of $\phi_{2,1}$ and $\phi_{1,2}$ project onto the interior of $N(A)$. Nevertheless, those initial forms still fully cover the edges, and binomiality of the face polynomials is preserved.

\subsubsection*{$\mathrm{diag}(2,2,4)$}

\small
\begin{verbatim}
[[1,8,8,4,16],[9,6,6,3,13],[3,6,5,3,13],[3,5,6,3,13],[-4,6,6,3,12],[1,
5,5,3,13],[-8,6,5,3,12],[-8,5,6,3,12],[-2,6,4,3,12],[-16,5,5,3,12],[-2
,4,6,3,12],[-4,5,4,3,12],[-4,4,5,3,12],[-1,4,4,3,12],[-2,5,5,2,10],[-9
,5,4,2,10],[-9,4,5,2,10],[-6,5,3,2,10],[-6,3,5,2,10],[-1,5,2,2,10],[-1
,2,5,2,10],[13,4,4,2,9],[14,4,3,2,9],[14,3,4,2,9],[6,4,4,2,8],[7,4,2,2
,9],[7,2,4,2,9],[16,4,3,2,8],[16,3,4,2,8],[1,4,1,2,9],[1,1,4,2,9],[20,
4,2,2,8],[20,2,4,2,8],[8,4,1,2,8],[8,1,4,2,8],[1,4,0,2,8],[1,0,4,2,8],
[9,3,3,1,7],[3,3,2,1,7],[3,2,3,1,7],[4,3,3,1,6],[1,2,2,1,7],[2,3,2,1,6
],[2,2,3,1,6],[-8,2,2,1,6],[-3,2,1,1,6],[-3,1,2,1,6],[-21,2,2,1,5],[-1
,1,1,1,6],[-17,2,1,1,5],[-17,1,2,1,5],[-4,2,2,1,4],[-3,2,0,1,5],[-9,1,
1,1,5],[-3,0,2,1,5],[1,2,2,0,4],[-8,2,1,1,4],[-8,1,2,1,4],[-1,1,0,1,5]
,[-1,0,1,1,5],[-2,2,0,1,4],[-16,1,1,1,4],[-2,0,2,1,4],[-4,1,0,1,4],[-4
,0,1,1,4],[-1,0,0,1,4],[-1,1,1,0,3],[-2,1,1,0,2],[-1,1,0,0,2],[-1,0,1,0
,2],[-1,0,0,0,1],[1,0,0,0,0]]
\end{verbatim}
\normalsize

\begin{figure}[h!]
    \centering
    \includegraphics[scale=0.4]{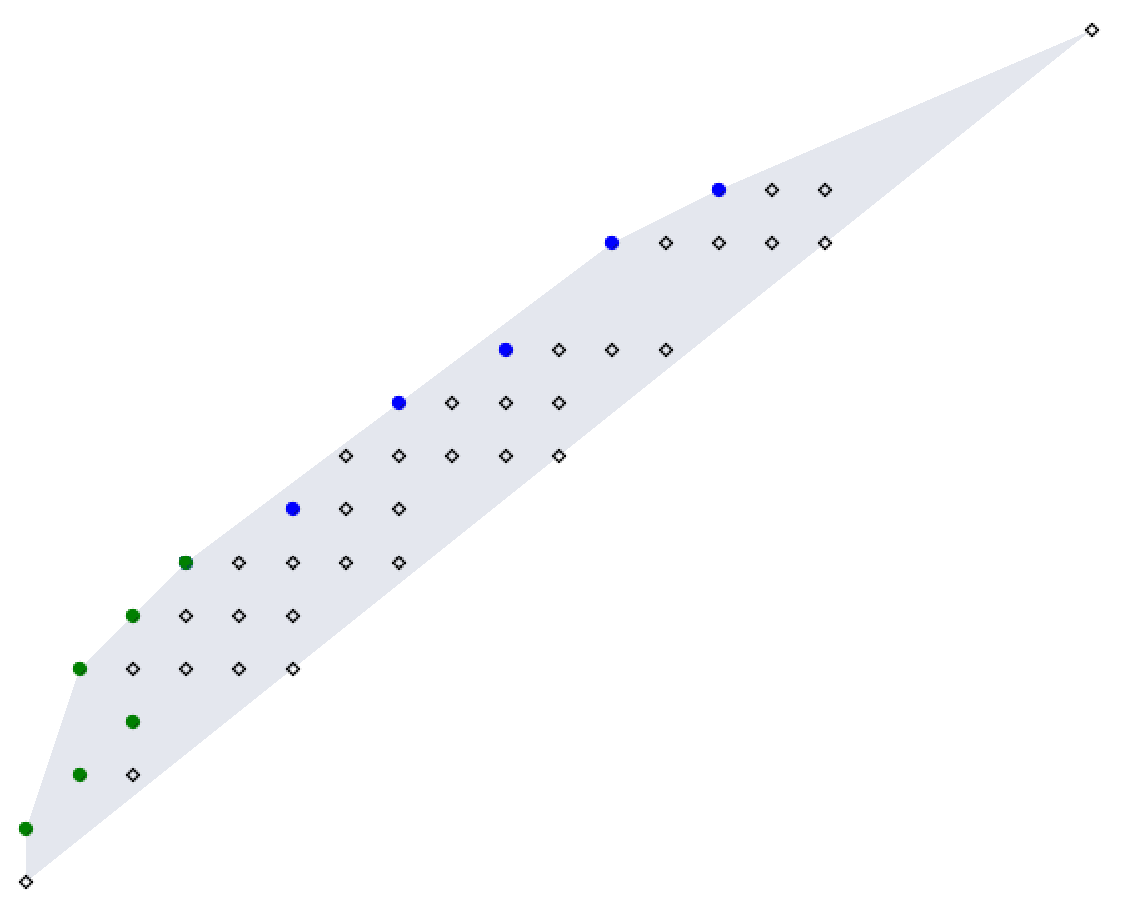}
    \caption{Newton polygon for $\mathrm{diag}(2,2,4)$ quiver}
    \label{fig:diag224}
\end{figure}

\subsubsection*{$\mathrm{diag}(2,3,3)$}

\small
\begin{verbatim}
[[1,9,6,6,18],[-4,8,5,5,16],[-2,7,5,5,16],[9,7,5,5,15],[3,6,5,5,15],[6
,7,4,4,14],[4,6,4,4,14],[1,5,4,4,14],[-11,6,4,4,13],[-5,5,4,4,13],[-3,
6,4,4,12],[-1,4,4,4,13],[-2,6,4,3,12],[-2,6,3,4,12],[27,5,4,4,12],[-4,
6,3,3,12],[-9,5,4,3,12],[-9,5,3,4,12],[18,4,4,4,12],[-2,5,3,3,12],[-6,
4,4,3,12],[-6,4,3,4,12],[3,3,4,4,12],[-1,3,4,3,12],[-1,3,3,4,12],[3,5,
3,3,11],[2,4,3,3,11],[-4,5,3,3,10],[4,5,3,2,10],[4,5,2,3,10],[-5,4,3,3
,10],[1,5,2,2,10],[8,4,3,2,10],[8,4,2,3,10],[14,3,3,3,10],[-18,4,3,3,9
],[2,3,3,2,10],[2,3,2,3,10],[8,2,3,3,10],[9,4,3,2,9],[9,4,2,3,9],[21,3
,3,3,9],[1,1,3,3,10],[-1,4,2,2,9],[3,3,3,2,9],[3,3,2,3,9],[27,2,3,3,9]
,[9,1,3,3,9],[1,4,2,2,8],[1,0,3,3,9],[-2,4,2,1,8],[-2,4,1,2,8],[-15,3,
2,2,8],[-1,3,2,1,8],[-1,3,1,2,8],[-12,2,2,2,8],[-26,3,2,2,7],[-2,1,2,2
,8],[-1,3,2,1,7],[-1,3,1,2,7],[-43,2,2,2,7],[3,3,2,2,6],[-17,1,2,2,7],
[-6,3,2,1,6],[-6,3,1,2,6],[-27,2,2,2,6],[-2,0,2,2,7],[1,3,2,0,6],[4,3,
1,1,6],[1,3,0,2,6],[-18,1,2,2,6],[4,2,1,1,6],[-3,0,2,2,6],[1,1,1,1,6],
[8,2,1,1,5],[6,1,1,1,5],[8,2,1,1,4],[1,0,1,1,5],[-2,2,1,0,4],[-2,2,0,1
,4],[7,1,1,1,4],[-1,1,1,0,4],[-1,1,0,1,4],[1,0,1,1,4],[9,1,1,1,3],[-3,
1,1,0,3],[-3,1,0,1,3],[3,0,1,1,3],[-1,0,1,0,3],[-1,0,0,1,3],[1,1,0,0,2
],[1,0,0,0,1],[-1,0,0,0,0]]
\end{verbatim}
\normalsize

\begin{figure}[h!]
    \centering
    \includegraphics[scale=0.4]{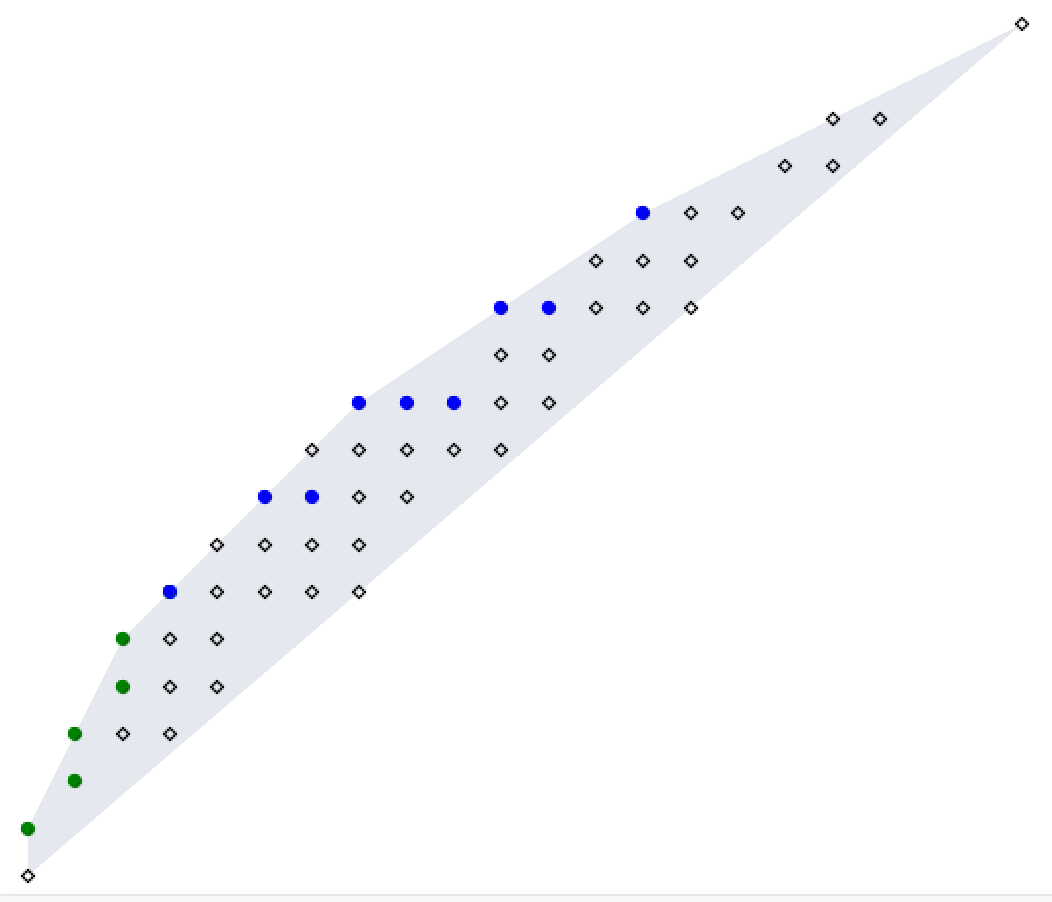}
    \includegraphics[scale=0.4]{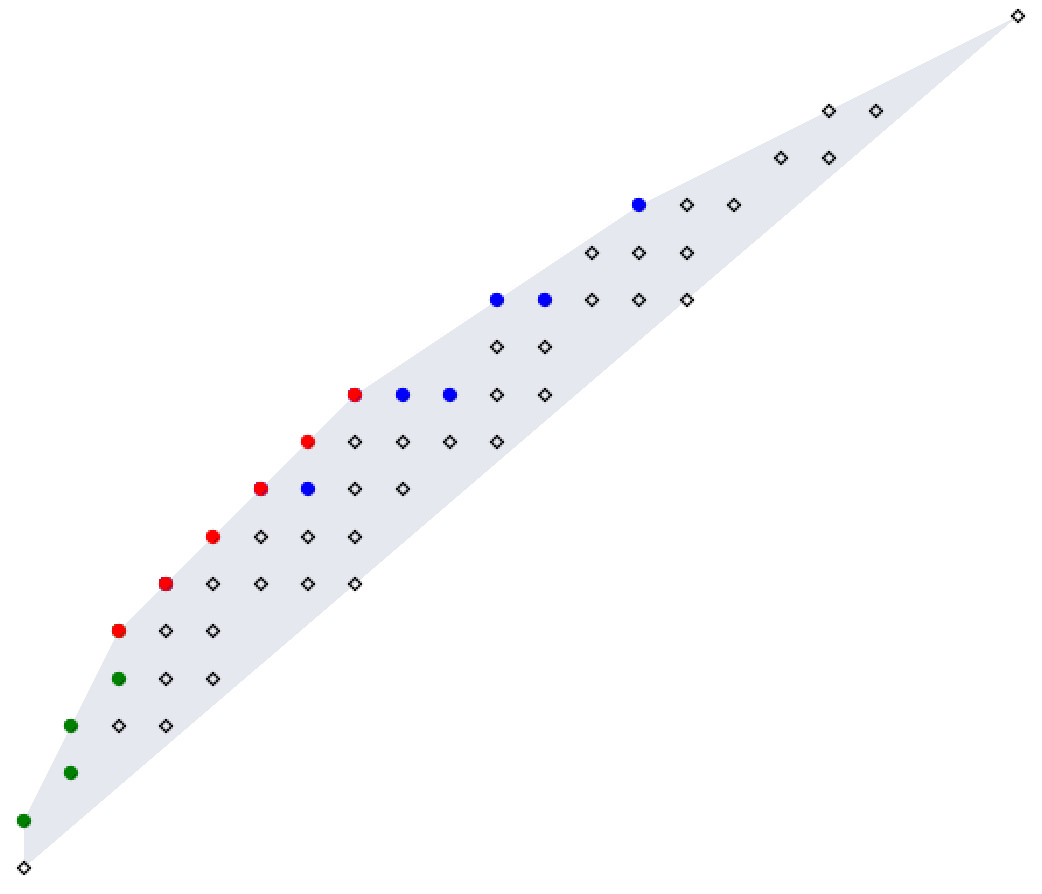}
    \caption{Newton polygon for $\mathrm{diag}(2,3,3)$ quiver: without and with $init_{r'}$ highlighted (red)}
    \label{fig:diag233}
\end{figure}

Calculating the initial form $init_{r'}$ (\ref{3vertex_init_rr}), we get:
\begin{equation}
\begin{aligned}
    init_{r'} = &\ {a_{{0}}}^{5}{a_{{1}}}^{8}{b_{{1}}}^{4}{c_{{0}}}^{2}c_{{2}}{d_{{0}}}^{
2}d_{{2}}\cdot \mathrm{GCD}\left(\left( {a_{{0}}}^{4}{b_{{1}}}^{4}{c_{{2}}}^{2}{d_{{2}}}^{2}-
{a_{{1}}}^{4}{b_{{0}}}^{4}{c_{{1}}}^{2}{d_{{1}}}^{2} \right)  \left( a
_{{0}}b_{{2}}c_{{1}}d_{{1}}-a_{{1}}b_{{1}}c_{{0}}d_{{0}} \right)\right) = \\
&\ {a_{{0}}}^{5}{a_{{1}}}^{8}{b_{{1}}}^{4}{c_{{0}}}^{2}c_{{2}}{d_{{0}}}^{
2}d_{{2}} \left( {a_{{0}}}^{2}{b_{{1}}}^{2}c_{{2}}d_{{2}}-{a_{{1}}}^{2
}{b_{{0}}}^{2}c_{{1}}d_{{1}} \right) ^{2} \left( a_{{0}}b_{{2}}c_{{1}}
d_{{1}}-a_{{1}}b_{{1}}c_{{0}}d_{{0}} \right).
\end{aligned}
\end{equation}
After specialization (\ref{quiver_A_poly_def}), it takes form:
\begin{equation}
    init_{r'} = {y}^{5}x_{{2}}x_{{3}} \left( {y}^{2}x_{{2}}x_{{3}}-1 \right) ^{2}
 \left( yx_{{1}}-1 \right) 
\end{equation}
Its projection onto $N(A)$ is shown on Figure \ref{fig:diag233}.

On Figure \ref{fig:diag233}, left, one can see that $\phi_{1,2}$ does not fully cover the edge: some of the monomials are missing. However, by introducing $init_{r'}$, we get rid of this problem (Figure \ref{fig:diag233}, right).

\subsubsection*{$\mathrm{diag(2,4,4)}$}

\small
\begin{verbatim}
[[1,16,8,8,32],[9,14,7,7,29],[3,13,7,7,29],[-8,14,7,7,28],[-16,13,7,7,
28],[-4,12,7,7,28],[-3,13,6,6,26],[27,12,6,6,26],[18,11,6,6,26],[-23,
12,6,6,25],[3,10,6,6,26],[-70,11,6,6,25],[28,12,6,6,24],[-35,10,6,6,25
],[96,11,6,6,24],[-5,9,6,6,25],[120,10,6,6,24],[48,9,6,6,24],[6,8,6,6,
24],[-18,11,5,5,23],[21,10,5,5,23],[4,11,5,5,22],[27,9,5,5,23],[23,10,
5,5,22],[9,8,5,5,23],[-35,9,5,5,22],[-19,10,5,5,21],[1,7,5,5,23],[-35,
8,5,5,22],[-15,9,5,5,21],[-56,10,5,5,20],[-10,7,5,5,22],[59,8,5,5,21],
[-2,10,5,4,20],[-2,10,4,5,20],[-240,9,5,5,20],[-1,6,5,5,22],[44,7,5,5,
21],[3,10,4,4,20],[-36,9,5,4,20],[-36,9,4,5,20],[-444,8,5,5,20],[11,6,
5,5,21],[-27,9,4,4,20],[-105,8,5,4,20],[-105,8,4,5,20],[-448,7,5,5,20]
,[1,5,5,5,21],[-18,8,4,4,20],[-112,7,5,4,20],[-112,7,4,5,20],[-216,6,5
,5,20],[-14,9,4,4,19],[-3,7,4,4,20],[-54,6,5,4,20],[-54,6,4,5,20],[-48
,5,5,5,20],[22,8,4,4,19],[13,9,4,4,18],[-12,5,5,4,20],[-12,5,4,5,20],[
-4,4,5,5,20],[17,7,4,4,19],[6,8,4,4,18],[-1,4,5,4,20],[-1,4,4,5,20],[3
,6,4,4,19],[-36,7,4,4,18],[125,8,4,4,17],[-20,6,4,4,18],[27,8,4,3,17],
[27,8,3,4,17],[444,7,4,4,17],[70,8,4,4,16],[-3,5,4,4,18],[9,8,3,3,17],
[90,7,4,3,17],[90,7,3,4,17],[674,6,4,4,17],[8,8,4,3,16],[8,8,3,4,16],[
320,7,4,4,16],[3,7,3,3,17],[81,6,4,3,17],[81,6,3,4,17],[554,5,4,4,17],
[4,8,3,3,16],[64,7,4,3,16],[64,7,3,4,16],[656,6,4,4,16],[27,5,4,3,17],
[27,5,3,4,17],[287,4,4,4,17],[-7,7,3,3,16],[80,6,4,3,16],[80,6,3,4,16]
,[800,5,4,4,16],[3,4,4,3,17],[3,4,3,4,17],[90,3,4,4,17],[-3,6,3,3,16],
[32,5,4,3,16],[32,5,3,4,16],[676,4,4,4,16],[6,7,3,3,15],[15,2,4,4,17],
[4,4,4,3,16],[4,4,3,4,16],[352,3,4,4,16],[10,6,3,3,15],[-32,7,3,3,14],
[1,1,4,4,17],[104,2,4,4,16],[3,5,3,3,15],[-6,7,3,2,14],[-6,7,2,3,14],[
-202,6,3,3,14],[16,1,4,4,16],[-1,7,2,2,14],[-27,6,3,2,14],[-27,6,2,3,
14],[-382,5,3,3,14],[-165,6,3,3,13],[1,0,4,4,16],[-18,5,3,2,14],[-18,5
,2,3,14],[-348,4,3,3,14],[-25,6,3,2,13],[-25,6,2,3,13],[-587,5,3,3,13]
,[-56,6,3,3,12],[-3,4,3,2,14],[-3,4,2,3,14],[-162,3,3,3,14],[1,6,2,2,
13],[-25,5,3,2,13],[-25,5,2,3,13],[-874,4,3,3,13],[-12,6,3,2,12],[-12,
6,2,3,12],[-240,5,3,3,12],[-36,2,3,3,14],[-5,4,3,2,13],[-5,4,2,3,13],[
-644,3,3,3,13],[-1,6,2,2,12],[-24,5,3,2,12],[-24,5,2,3,12],[-444,4,3,3
,12],[-3,1,3,3,14],[-239,2,3,3,13],[-1,5,2,2,12],[-6,4,3,2,12],[-6,4,2
,3,12],[-448,3,3,3,12],[-43,1,3,3,13],[-216,2,3,3,12],[38,5,2,2,11],[-
3,0,3,3,13],[-48,1,3,3,12],[3,5,2,1,11],[3,5,1,2,11],[94,4,2,2,11],[23
,5,2,2,10],[-4,0,3,3,12],[1,4,2,1,11],[1,4,1,2,11],[81,3,2,2,11],[2,5,
2,1,10],[2,5,1,2,10],[159,4,2,2,10],[27,2,2,2,11],[1,4,2,1,10],[1,4,1,
2,10],[250,3,2,2,10],[91,4,2,2,9],[3,1,2,2,11],[145,2,2,2,10],[1,4,2,1
,9],[1,4,1,2,9],[234,3,2,2,9],[28,4,2,2,8],[35,1,2,2,10],[177,2,2,2,9]
,[8,4,2,1,8],[8,4,1,2,8],[96,3,2,2,8],[3,0,2,2,10],[51,1,2,2,9],[1,4,2
,0,8],[1,4,0,2,8],[120,2,2,2,8],[5,0,2,2,9],[-9,3,1,1,8],[48,1,2,2,8],
[-6,2,1,1,8],[6,0,2,2,8],[-12,3,1,1,7],[-1,1,1,1,8],[-19,2,1,1,7],[-4,
3,1,1,6],[-8,1,1,1,7],[-13,2,1,1,6],[-1,0,1,1,7],[-7,1,1,1,6],[-17,2,1
,1,5],[-1,0,1,1,6],[-3,2,1,0,5],[-3,2,0,1,5],[-9,1,1,1,5],[-8,2,1,1,4]
,[-1,1,1,0,5],[-1,1,0,1,5],[-1,0,1,1,5],[-2,2,1,0,4],[-2,2,0,1,4],[-16
,1,1,1,4],[-4,1,1,0,4],[-4,1,0,1,4],[-4,0,1,1,4],[-1,0,1,0,4],[-1,0,0,
1,4],[-1,1,0,0,2],[-1,0,0,0,1],[1,0,0,0,0]]
\end{verbatim}
\normalsize

\begin{equation}
    \begin{aligned}
        init_{r'} = &\ {a_{{0}}}^{7}{a_{{1}}}^{15}{b_{{1}}}^{6}{c_{{0}}}^{3}c_{{2}}{d_{{0}}}^
{3}d_{{2}}\cdot \mathrm{GCD}\left( \left( {a_{{0}}}^{9}{b_{{1}}}^{9}{c_{{2}}}^{3}{d_{{2}}}^{3}
-{a_{{1}}}^{9}{b_{{0}}}^{9}{c_{{1}}}^{3}{d_{{1}}}^{3} \right)  \left( 
a_{{0}}b_{{2}}c_{{1}}d_{{1}}-a_{{1}}b_{{1}}c_{{0}}d_{{0}} \right)\right) =
\\
&\ {a_{{0}}}^{7}{a_{{1}}}^{15}{b_{{1}}}^{6}{c_{{0}}}^{3}c_{{2}}{d_{{0}}}^
{3}d_{{2}} \left( {a_{{0}}}^{3}{b_{{1}}}^{3}c_{{2}}d_{{2}}-{a_{{1}}}^{
3}{b_{{0}}}^{3}c_{{1}}d_{{1}} \right) ^{3} \left( a_{{0}}b_{{2}}c_{{1}
}d_{{1}}-a_{{1}}b_{{1}}c_{{0}}d_{{0}} \right). 
    \end{aligned}
\end{equation}
After specialization (\ref{quiver_A_poly_def}), it takes form:
\begin{equation}
    init_{r'} = -{y}^{7}x_{{2}}x_{{3}} \left( -{y}^{3}x_{{2}}x_{{3}}+1 \right) ^{3}
 \left( yx_{{1}}-1 \right) 
\end{equation}
Now we take its counterpart : $-{y}^{7}x_{{2}}x_{{3}} \left( -{y}^{3}x_{{2}}x_{{3}}+1 \right) ^{3}$. Its projection is shown on Figure 
\ref{fig:diag244}.

\begin{figure}[h!]
    \centering
    \includegraphics[scale=0.4]{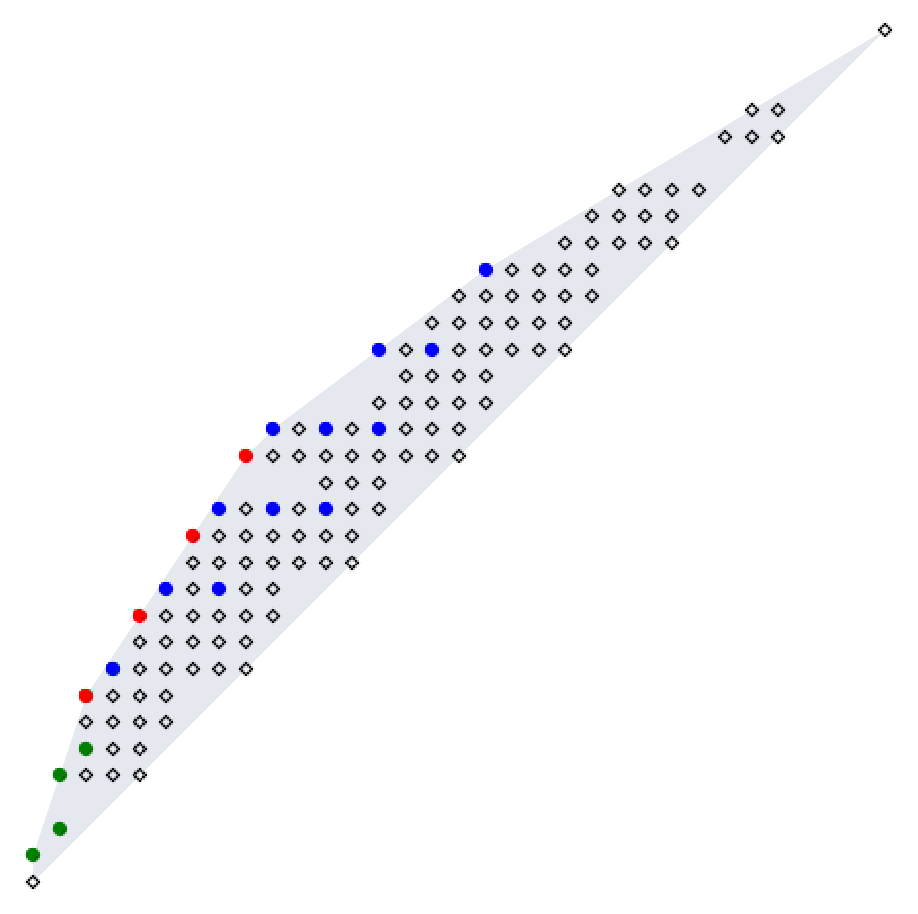}
    \caption{Newton polygon and the initial form $init_{r'}$ (red nodes) for $\mathrm{diag}(2,4,4)$ quiver}
    \label{fig:diag244}
\end{figure}

\subsubsection*{$\mathrm{diag}(3,3,4)$}

\small
\begin{verbatim}
[[1,12,12,9,36],[9,11,11,8,33],[-4,10,10,8,32],[36,10,10,7,30],[-3,10,
9,7,30],[-3,9,10,7,30],[-2,9,9,7,30],[13,9,9,7,29],[-32,9,9,7,28],[-8,
9,8,7,28],[-8,8,9,7,28],[6,8,8,7,28],[84,9,9,6,27],[-18,9,8,6,27],[-18
,8,9,6,27],[9,8,8,6,27],[72,8,8,6,26],[-18,8,7,6,26],[-18,7,8,6,26],[-
167,8,8,6,25],[4,7,7,6,26],[49,8,7,6,25],[49,7,8,6,25],[-3,8,8,6,24],[
-21,7,7,6,25],[126,8,8,5,24],[-24,8,7,6,24],[-24,7,8,6,24],[-45,8,7,5,
24],[-2,8,6,6,24],[-45,7,8,5,24],[-64,7,7,6,24],[-2,6,8,6,24],[3,8,6,5
,24],[36,7,7,5,24],[16,7,6,6,24],[3,6,8,5,24],[16,6,7,6,24],[-6,7,6,5,
24],[-6,6,7,5,24],[-4,6,6,6,24],[79,7,7,5,23],[1,6,6,5,24],[-7,7,6,5,
23],[-7,6,7,5,23],[-357,7,7,5,22],[-1,6,6,5,23],[162,7,6,5,22],[162,6,
7,5,22],[-18,7,7,5,21],[-18,7,5,5,22],[-84,6,6,5,22],[-18,5,7,5,22],[
126,7,7,4,21],[-66,7,6,5,21],[-66,6,7,5,21],[13,6,5,5,22],[13,5,6,5,22
],[-60,7,6,4,21],[27,7,5,5,21],[-60,6,7,4,21],[-128,6,6,5,21],[27,5,7,
5,21],[-2,5,5,5,22],[9,7,5,4,21],[35,6,6,4,21],[18,6,5,5,21],[9,5,7,4,
21],[18,5,6,5,21],[124,6,6,5,20],[-3,6,5,4,21],[-3,5,6,4,21],[-1,5,5,5
,21],[-4,6,6,4,20],[-128,6,5,5,20],[-128,5,6,5,20],[15,6,5,4,20],[20,6
,4,5,20],[15,5,6,4,20],[64,5,5,5,20],[20,4,6,5,20],[-400,6,6,4,19],[-2
,6,4,4,20],[-8,5,4,5,20],[-2,4,6,4,20],[-8,4,5,5,20],[172,6,5,4,19],[
172,5,6,4,19],[-45,6,6,4,18],[1,4,4,5,20],[-19,6,4,4,19],[-76,5,5,4,19
],[-19,4,6,4,19],[84,6,6,3,18],[-45,6,5,4,18],[-45,5,6,4,18],[6,5,4,4,
19],[6,4,5,4,19],[-45,6,5,3,18],[27,6,4,4,18],[-45,5,6,3,18],[-17,5,5,
4,18],[27,4,6,4,18],[9,6,4,3,18],[-6,6,3,4,18],[9,5,5,3,18],[-23,5,4,4
,18],[9,4,6,3,18],[-23,4,5,4,18],[-6,3,6,4,18],[310,5,5,4,17],[-1,6,3,
3,18],[4,5,3,4,18],[-1,3,6,3,18],[4,3,5,4,18],[-45,5,5,3,17],[-187,5,4
,4,17],[-187,4,5,4,17],[16,5,5,4,16],[6,5,4,3,17],[17,5,3,4,17],[6,4,5
,3,17],[47,4,4,4,17],[17,3,5,4,17],[-250,5,5,3,16],[92,5,4,4,16],[92,4
,5,4,16],[-3,4,3,4,17],[-3,3,4,4,17],[100,5,4,3,16],[-64,5,3,4,16],[
100,4,5,3,16],[-63,4,4,4,16],[-64,3,5,4,16],[-60,5,5,3,15],[-19,5,3,3,
16],[8,5,2,4,16],[-14,4,4,3,16],[16,4,3,4,16],[-19,3,5,3,16],[16,3,4,4
,16],[8,2,5,4,16],[36,5,5,2,15],[21,5,4,3,15],[21,4,5,3,15],[2,5,2,3,
16],[-2,4,2,4,16],[2,2,5,3,16],[-2,2,4,4,16],[-18,5,4,2,15],[-8,5,3,3,
15],[-18,4,5,2,15],[71,4,4,3,15],[-8,3,5,3,15],[3,5,3,2,15],[3,5,2,3,
15],[-12,4,3,3,15],[3,3,5,2,15],[-12,3,4,3,15],[3,2,5,3,15],[304,4,4,3
,14],[-16,4,4,2,14],[-72,4,3,3,14],[-72,3,4,3,14],[22,4,4,3,13],[2,4,3
,2,14],[11,4,2,3,14],[2,3,4,2,14],[1,3,3,3,14],[11,2,4,3,14],[-87,4,4,
2,13],[95,4,3,3,13],[95,3,4,3,13],[3,4,4,3,12],[-1,4,1,3,14],[-1,1,4,3
,14],[39,4,3,2,13],[-14,4,2,3,13],[39,3,4,2,13],[-64,3,3,3,13],[-14,2,
4,3,13],[-45,4,4,2,12],[-40,4,3,3,12],[-40,3,4,3,12],[-6,4,2,2,13],[-1
,4,1,3,13],[6,3,2,3,13],[-6,2,4,2,13],[6,2,3,3,13],[-1,1,4,3,13],[9,4,
4,1,12],[33,4,3,2,12],[42,4,2,3,12],[33,3,4,2,12],[64,3,3,3,12],[42,2,
4,3,12],[-3,4,3,1,12],[-9,4,2,2,12],[-12,4,1,3,12],[-3,3,4,1,12],[40,3
,3,2,12],[-16,3,2,3,12],[-9,2,4,2,12],[-16,2,3,3,12],[-12,1,4,3,12],[1
,4,0,3,12],[1,3,3,1,12],[-4,3,2,2,12],[-4,2,3,2,12],[4,2,2,3,12],[1,0,
4,3,12],[121,3,3,2,11],[1,3,3,1,11],[-30,3,2,2,11],[-30,2,3,2,11],[-3,
3,3,2,10],[3,3,1,2,11],[3,1,3,2,11],[-17,3,3,1,10],[3,3,2,2,10],[3,2,3
,2,10],[9,3,3,2,9],[6,3,2,1,10],[3,3,1,2,10],[6,2,3,1,10],[-28,2,2,2,
10],[3,1,3,2,10],[-18,3,3,1,9],[-54,3,2,2,9],[-54,2,3,2,9],[-2,2,2,1,
10],[2,2,1,2,10],[2,1,2,2,10],[1,3,3,0,9],[9,3,2,1,9],[27,3,1,2,9],[9,
2,3,1,9],[-20,2,2,2,9],[27,1,3,2,9],[-3,3,0,2,9],[-5,2,2,1,9],[2,2,1,2
,9],[2,1,2,2,9],[-3,0,3,2,9],[-56,2,2,2,8],[4,2,2,1,8],[16,2,1,2,8],[
16,1,2,2,8],[-3,2,1,1,8],[-2,2,0,2,8],[-3,1,2,1,8],[-2,0,2,2,8],[-8,2,
2,1,7],[1,1,1,1,8],[-2,2,2,0,7],[-3,2,1,1,7],[-3,1,2,1,7],[9,2,2,1,6],
[2,1,1,1,7],[-3,2,2,0,6],[-18,2,1,1,6],[-18,1,2,1,6],[3,2,0,1,6],[11,1
,1,1,6],[3,0,2,1,6],[-1,1,0,1,6],[-1,0,1,1,6],[9,1,1,1,5],[1,1,1,0,5],
[-1,1,0,1,5],[-1,0,1,1,5],[16,1,1,1,4],[1,1,1,0,4],[-4,1,0,1,4],[-4,0,
1,1,4],[1,0,0,1,4],[3,1,1,0,3],[-1,1,0,0,3],[-1,0,1,0,3],[1,0,0,0,1],[
-1,0,0,0,0]]
\end{verbatim}
\normalsize

\begin{figure}[h!]
    \centering
    \includegraphics[scale=0.4]{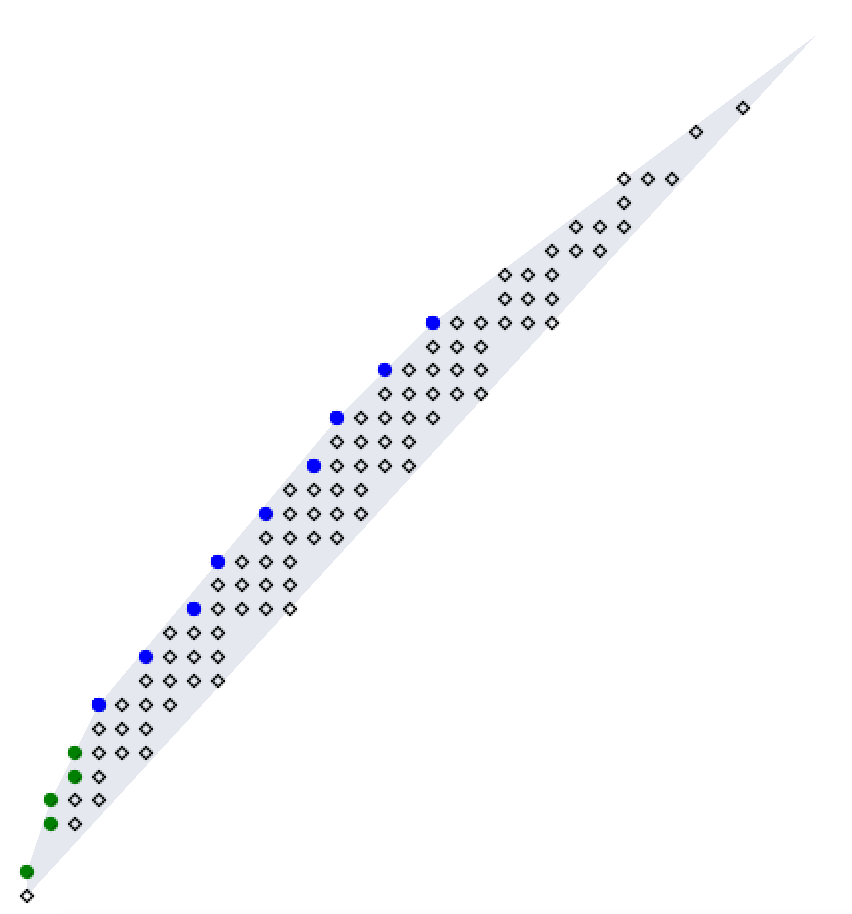}
    \caption{Newton polygon for $\mathrm{diag}(3,3,4)$ quiver}
    \label{fig:diag334}
\end{figure}

\subsubsection*{$\mathrm{diag}(3,4,4)$}

\begin{figure}[h!]
    \centering
    \includegraphics[scale=0.4]{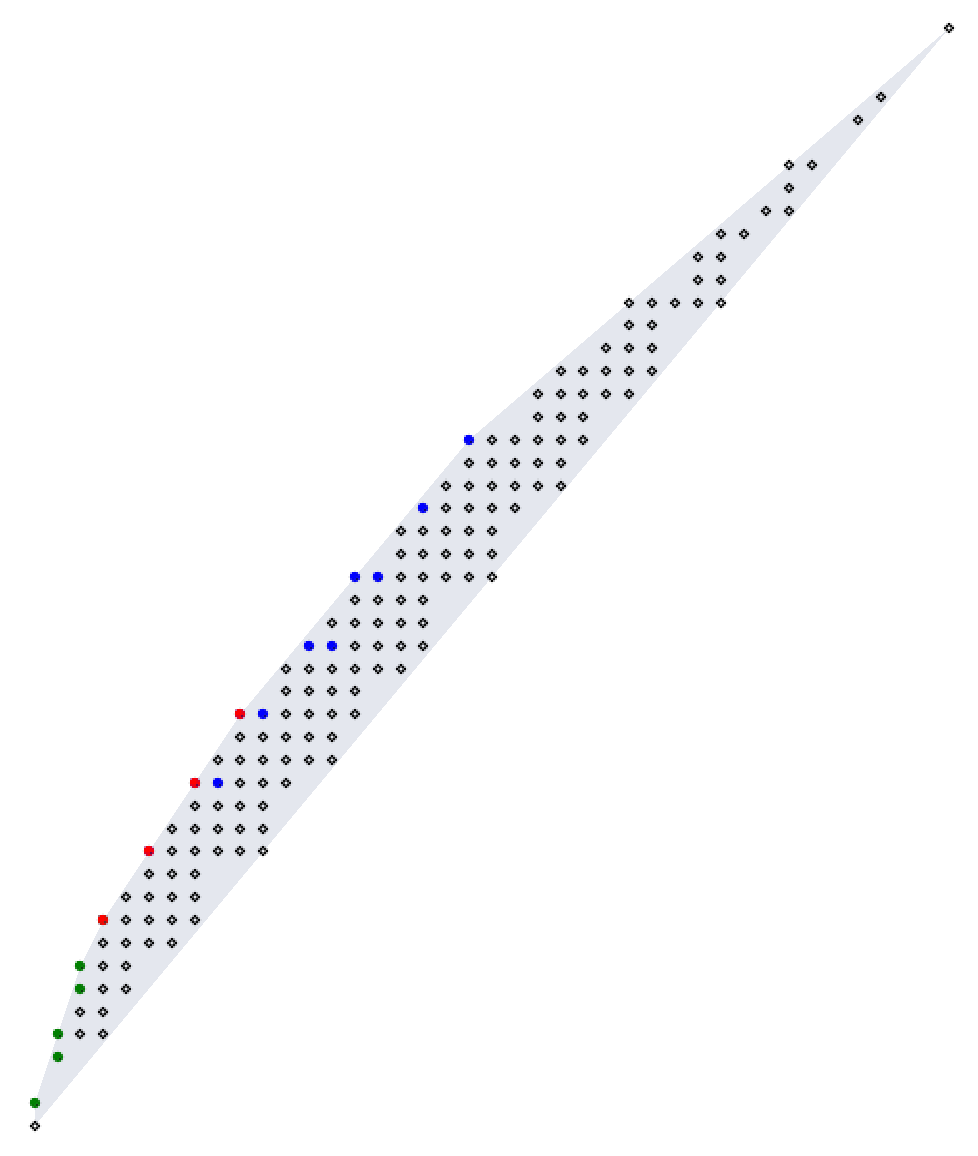}
    \caption{Newton polygon and the initial form $init_{r'}$ (red nodes) for $\mathrm{diag}(3,4,4)$ quiver}
    \label{fig:diag344}
\end{figure}

\small
\begin{verbatim}
[[1,16,12,12,48],[9,15,11,11,45],[-8,14,11,11,44],[36,14,10,10,42],[-3
,13,10,10,42],[-23,13,10,10,41],[-16,13,10,10,40],[28,12,10,10,40],[84
,13,9,9,39],[-18,12,9,9,39],[3,12,9,9,38],[4,11,9,9,38],[-23,12,9,9,37
],[-19,11,9,9,37],[-4,12,9,9,36],[-3,12,9,8,36],[-3,12,8,9,36],[96,11,
9,9,36],[126,12,8,8,36],[-24,11,9,8,36],[-24,11,8,9,36],[-56,10,9,9,36
],[-45,11,8,8,36],[-2,10,9,8,36],[-2,10,8,9,36],[3,10,8,8,36],[80,11,8
,8,35],[-14,10,8,8,35],[72,11,8,8,34],[-85,10,8,8,34],[9,11,8,8,33],[
13,9,8,8,34],[-18,11,8,7,33],[-18,11,7,8,33],[-138,10,8,8,33],[126,11,
7,7,33],[-66,10,8,7,33],[-66,10,7,8,33],[125,9,8,8,33],[120,10,8,8,32]
,[-60,10,7,7,33],[27,9,8,7,33],[27,9,7,8,33],[-136,10,8,7,32],[-136,10
,7,8,32],[-240,9,8,8,32],[9,9,7,7,33],[110,10,7,7,32],[70,8,8,8,32],[-
39,9,7,7,32],[8,8,8,7,32],[8,8,7,8,32],[199,10,7,7,31],[4,8,7,7,32],[-
71,9,7,7,31],[45,10,7,7,30],[6,8,7,7,31],[-45,10,7,6,30],[-45,10,6,7,
30],[-173,9,7,7,30],[84,10,6,6,30],[-45,9,7,6,30],[-45,9,6,7,30],[86,8
,7,7,30],[-329,9,7,7,29],[-45,9,6,6,30],[27,8,7,6,30],[27,8,6,7,30],[-
32,7,7,7,30],[-360,9,7,6,29],[-360,9,6,7,29],[523,8,7,7,29],[48,9,7,7,
28],[9,8,6,6,30],[-6,7,7,6,30],[-6,7,6,7,30],[57,9,6,6,29],[151,8,7,6,
29],[151,8,6,7,29],[-165,7,7,7,29],[-152,9,7,6,28],[-152,9,6,7,28],[-
444,8,7,7,28],[-1,7,6,6,30],[-20,8,6,6,29],[-25,7,7,6,29],[-25,7,6,7,
29],[176,9,6,6,28],[-170,8,7,6,28],[-170,8,6,7,28],[320,7,7,7,28],[1,7
,6,6,29],[-52,8,6,6,28],[96,7,7,6,28],[96,7,6,7,28],[-56,6,7,7,28],[40
,9,6,6,27],[9,7,6,6,28],[-12,6,7,6,28],[-12,6,6,7,28],[-60,9,6,5,27],[
-60,9,5,6,27],[207,8,6,6,27],[-1,6,6,6,28],[36,9,5,5,27],[21,8,6,5,27]
,[21,8,5,6,27],[-152,7,6,6,27],[-39,8,6,6,26],[-18,8,5,5,27],[-8,7,6,5
,27],[-8,7,5,6,27],[38,6,6,6,27],[-320,8,6,5,26],[-320,8,5,6,26],[360,
7,6,6,26],[-243,8,6,6,25],[3,7,5,5,27],[3,6,6,5,27],[3,6,5,6,27],[7,8,
5,5,26],[157,7,6,5,26],[157,7,5,6,26],[-94,6,6,6,26],[-288,8,6,5,25],[
-288,8,5,6,25],[1018,7,6,6,25],[6,8,6,6,24],[1,7,5,5,26],[-21,6,6,5,26
],[-21,6,5,6,26],[23,5,6,6,26],[63,8,5,5,25],[-55,7,6,5,25],[-55,7,5,6
,25],[-508,6,6,6,25],[-48,8,6,5,24],[-48,8,5,6,24],[-448,7,6,6,24],[2,
5,6,5,26],[2,5,5,6,26],[-22,7,5,5,25],[26,6,6,5,25],[26,6,5,6,25],[91,
5,6,6,25],[3,8,6,4,24],[9,8,5,5,24],[3,8,4,6,24],[-512,7,6,5,24],[-512
,7,5,6,24],[656,6,6,6,24],[3,6,5,5,25],[1,5,6,5,25],[1,5,5,6,25],[-45,
8,5,4,24],[-45,8,4,5,24],[-40,7,6,4,24],[159,7,5,5,24],[-40,7,4,6,24],
[384,6,6,5,24],[384,6,5,6,24],[-240,5,6,6,24],[9,8,4,4,24],[33,7,5,4,
24],[33,7,4,5,24],[42,6,6,4,24],[-35,6,5,5,24],[42,6,4,6,24],[-96,5,6,
5,24],[-96,5,5,6,24],[28,4,6,6,24],[611,7,5,5,23],[-3,7,4,4,24],[-9,6,
5,4,24],[-9,6,4,5,24],[-12,5,6,4,24],[-9,5,5,5,24],[-12,5,4,6,24],[8,4
,6,5,24],[8,4,5,6,24],[-106,7,5,4,23],[-106,7,4,5,23],[-329,6,5,5,23],
[234,7,5,5,22],[1,4,6,4,24],[1,4,4,6,24],[-2,7,4,4,23],[33,6,5,4,23],[
33,6,4,5,23],[44,5,5,5,23],[-150,7,5,4,22],[-150,7,4,5,22],[1032,6,5,5
,22],[-45,7,5,5,21],[-6,5,5,4,23],[-6,5,4,5,23],[-12,4,5,5,23],[8,7,4,
4,22],[37,6,5,4,22],[37,6,4,5,22],[-450,5,5,5,22],[-36,7,5,4,21],[-36,
7,4,5,21],[1348,6,5,5,21],[-3,6,4,4,22],[-11,5,5,4,22],[-11,5,4,5,22],
[39,4,5,5,22],[9,7,5,3,21],[18,7,4,4,21],[9,7,3,5,21],[-366,6,5,4,21],
[-366,6,4,5,21],[-530,5,5,5,21],[-216,6,5,5,20],[-4,3,5,5,22],[-18,7,4
,3,21],[-18,7,3,4,21],[-54,6,5,3,21],[-35,6,4,4,21],[-54,6,3,5,21],[95
,5,5,4,21],[95,5,4,5,21],[111,4,5,5,21],[-450,6,5,4,20],[-450,6,4,5,20
],[800,5,5,5,20],[1,7,3,3,21],[9,6,4,3,21],[9,6,3,4,21],[27,5,5,3,21],
[27,5,4,4,21],[27,5,3,5,21],[11,4,5,4,21],[11,4,4,5,21],[-17,3,5,5,21]
,[-48,6,5,3,20],[331,6,4,4,20],[-48,6,3,5,20],[328,5,5,4,20],[328,5,4,
5,20],[-444,4,5,5,20],[-3,4,5,3,21],[-3,4,3,5,21],[-3,3,5,4,21],[-3,3,
4,5,21],[-12,6,4,3,20],[-12,6,3,4,20],[32,5,5,3,20],[-126,5,4,4,20],[
32,5,3,5,20],[-115,4,5,4,20],[-115,4,4,5,20],[96,3,5,5,20],[365,6,4,4,
19],[6,5,4,3,20],[6,5,3,4,20],[-4,4,5,3,20],[36,4,4,4,20],[-4,4,3,5,20
],[24,3,5,4,20],[24,3,4,5,20],[-8,2,5,5,20],[-17,6,4,3,19],[-17,6,3,4,
19],[176,5,4,4,19],[99,6,4,4,18],[-2,2,5,4,20],[-2,2,4,5,20],[1,6,3,3,
19],[13,5,4,3,19],[13,5,3,4,19],[-33,4,4,4,19],[21,6,4,3,18],[21,6,3,4
,18],[1065,5,4,4,18],[12,3,4,4,19],[9,6,4,2,18],[26,6,3,3,18],[9,6,2,4
,18],[17,5,4,3,18],[17,5,3,4,18],[-300,4,4,4,18],[919,5,4,4,17],[-3,6,
3,2,18],[-3,6,2,3,18],[-18,5,4,2,18],[-27,5,3,3,18],[-18,5,2,4,18],[-
54,4,4,3,18],[-54,4,3,4,18],[-8,3,4,4,18],[-73,5,4,3,17],[-73,5,3,4,17
],[150,4,4,4,17],[-48,5,4,4,16],[3,4,4,2,18],[3,4,2,4,18],[9,3,4,3,18]
,[9,3,3,4,18],[15,2,4,4,18],[-25,5,4,2,17],[55,5,3,3,17],[-25,5,2,4,17
],[53,4,4,3,17],[53,4,3,4,17],[-211,3,4,4,17],[-176,5,4,3,16],[-176,5,
3,4,16],[676,4,4,4,16],[-1,1,4,4,18],[-2,5,3,2,17],[-2,5,2,3,17],[5,4,
4,2,17],[-36,4,3,3,17],[5,4,2,4,17],[-42,3,4,3,17],[-42,3,3,4,17],[38,
2,4,4,17],[-24,5,4,2,16],[28,5,3,3,16],[-24,5,2,4,16],[140,4,4,3,16],[
140,4,3,4,16],[-448,3,4,4,16],[6,2,4,3,17],[6,2,3,4,17],[-1,1,4,4,17],
[-3,5,3,2,16],[-3,5,2,3,16],[6,4,4,2,16],[-15,4,3,3,16],[6,4,2,4,16],[
-64,3,4,3,16],[-64,3,3,4,16],[120,2,4,4,16],[10,5,3,3,15],[-12,3,3,3,
16],[8,2,4,3,16],[8,2,3,4,16],[-16,1,4,4,16],[3,5,3,2,15],[3,5,2,3,15]
,[114,4,3,3,15],[1,0,4,4,16],[3,5,3,1,15],[9,5,2,2,15],[3,5,1,3,15],[
43,4,3,2,15],[43,4,2,3,15],[55,3,3,3,15],[180,4,3,3,14],[-1,4,3,1,15],
[-1,4,1,3,15],[-9,3,3,2,15],[-9,3,2,3,15],[-36,2,3,3,15],[47,4,3,2,14]
,[47,4,2,3,14],[266,3,3,3,14],[267,4,3,3,13],[3,1,3,3,15],[-1,4,3,1,14
],[12,4,2,2,14],[-1,4,1,3,14],[21,3,3,2,14],[21,3,2,3,14],[-101,2,3,3,
14],[63,4,3,2,13],[63,4,2,3,13],[393,3,3,3,13],[-4,4,3,3,12],[-6,2,3,2
,14],[-6,2,2,3,14],[7,1,3,3,14],[-1,4,3,1,13],[9,4,2,2,13],[-1,4,1,3,
13],[32,3,3,2,13],[32,3,2,3,13],[-236,2,3,3,13],[-30,4,3,2,12],[-30,4,
2,3,12],[352,3,3,3,12],[4,3,2,2,13],[-10,2,3,2,13],[-10,2,2,3,13],[43,
1,3,3,13],[-12,4,3,1,12],[-36,4,2,2,12],[-12,4,1,3,12],[48,3,3,2,12],[
48,3,2,3,12],[-216,2,3,3,12],[-3,0,3,3,13],[-1,4,3,0,12],[-9,4,2,1,12]
,[-9,4,1,2,12],[-1,4,0,3,12],[-33,3,2,2,12],[-12,2,3,2,12],[-12,2,2,3,
12],[48,1,3,3,12],[3,3,2,1,12],[3,3,1,2,12],[27,2,2,2,12],[-4,0,3,3,12
],[-99,3,2,2,11],[-3,1,2,2,12],[-3,3,2,1,11],[-3,3,1,2,11],[72,2,2,2,
11],[-90,3,2,2,10],[2,2,2,1,11],[2,2,1,2,11],[-8,1,2,2,11],[3,3,2,1,10
],[3,3,1,2,10],[160,2,2,2,10],[27,3,2,2,9],[2,2,2,1,10],[2,2,1,2,10],[
-41,1,2,2,10],[27,3,2,1,9],[27,3,1,2,9],[134,2,2,2,9],[3,0,2,2,10],[3,
3,2,0,9],[9,3,1,1,9],[3,3,0,2,9],[2,2,2,1,9],[2,2,1,2,9],[-51,1,2,2,9]
,[104,2,2,2,8],[-6,2,1,1,9],[5,0,2,2,9],[16,2,2,1,8],[16,2,1,2,8],[-48
,1,2,2,8],[1,1,1,1,9],[2,2,2,0,8],[-6,2,1,1,8],[2,2,0,2,8],[6,0,2,2,8]
,[2,1,1,1,8],[-15,2,1,1,7],[9,1,1,1,7],[-18,2,1,1,6],[-1,0,1,1,7],[-3,
2,1,0,6],[-3,2,0,1,6],[11,1,1,1,6],[1,1,1,0,6],[1,1,0,1,6],[-1,0,1,1,6
],[9,1,1,1,5],[1,1,1,0,5],[1,1,0,1,5],[-1,0,1,1,5],[16,1,1,1,4],[4,1,1
,0,4],[4,1,0,1,4],[-4,0,1,1,4],[-1,0,1,0,4],[-1,0,0,1,4],[1,1,0,0,3],[
-1,0,0,0,1],[1,0,0,0,0]]
\end{verbatim}
\normalsize

\begin{equation}
    \begin{aligned}
        init_{r'} = &\ {a_{{0}}}^{7}{a_{{1}}}^{30}{b_{{1}}}^{6}{c_{{0}}}^{6}c_{{2}}{d_{{0}}}^
{6}d_{{2}} \left( {a_{{0}}}^{3}{b_{{1}}}^{3}c_{{2}}d_{{2}}-{a_{{1}}}^{
3}{b_{{0}}}^{3}c_{{1}}d_{{1}} \right) ^{3} \left( {a_{{0}}}^{2}b_{{2}}
{c_{{1}}}^{2}{d_{{1}}}^{2}+{a_{{1}}}^{2}b_{{1}}{c_{{0}}}^{2}{d_{{0}}}^
{2} \right).
    \end{aligned}
\end{equation}
After specialization, it takes form:
\begin{equation}
    init_{r'} = {y}^{7}x_{{2}}x_{{3}} \left( -{y}^{3}x_{{2}}x_{{3}}+1 \right) ^{3}
 \left( {y}^{2}x_{{1}}-1 \right)
\end{equation}
We take its counterpart: ${y}^{9}x_{{2}}x_{{3}} \left( -{y}^{3}x_{{2}}x_{{3}}+1 \right) ^{3}x_{{
1}}$ and project it onto $N(A)$ (Figure \ref{fig:diag344}).

\subsubsection*{$\mathrm{diag}(2,3,4)$}

\small
\begin{verbatim}
[[1,12,8,6,24],[9,10,7,5,21],[3,9,7,5,21],[-4,10,6,5,20],[-8,9,6,5,20]
,[-2,8,6,5,20],[-3,9,6,4,18],[-2,9,5,4,18],[27,8,6,4,18],[-9,8,5,4,18]
,[18,7,6,4,18],[-6,7,5,4,18],[3,6,6,4,18],[13,8,5,4,17],[-1,6,5,4,18],
[14,7,5,4,17],[-8,8,5,4,16],[7,6,5,4,17],[6,8,4,4,16],[-64,7,5,4,16],[
1,5,5,4,17],[16,7,4,4,16],[-80,6,5,4,16],[20,6,4,4,16],[-32,5,5,4,16],
[-18,7,5,3,15],[8,5,4,4,16],[-4,4,5,4,16],[9,7,4,3,15],[21,6,5,3,15],[
1,4,4,4,16],[3,6,4,3,15],[27,5,5,3,15],[-18,7,4,3,14],[9,4,5,3,15],[4,
7,3,3,14],[-18,6,4,3,14],[1,3,5,3,15],[2,6,3,3,14],[-3,5,4,3,14],[49,6
,4,3,13],[-21,6,3,3,13],[5,5,4,3,13],[-2,6,4,3,12],[-17,5,3,3,13],[-65
,4,4,3,13],[3,6,4,2,12],[16,6,3,3,12],[-36,5,4,3,12],[-3,4,3,3,13],[-
44,3,4,3,13],[-6,6,3,2,12],[-4,6,2,3,12],[-27,5,4,2,12],[32,5,3,3,12],
[-105,4,4,3,12],[-11,2,4,3,13],[1,6,2,2,12],[-8,5,2,3,12],[-18,4,4,2,
12],[8,4,3,3,12],[-112,3,4,3,12],[-1,1,4,3,13],[-2,4,2,3,12],[-3,3,4,2
,12],[-54,2,4,3,12],[-7,5,3,2,11],[-12,1,4,3,12],[-1,5,2,2,11],[-10,4,
3,2,11],[-18,5,3,2,10],[-1,0,4,3,12],[-2,3,3,2,11],[13,5,2,2,10],[36,4
,3,2,10],[-2,5,1,2,10],[7,4,2,2,10],[60,3,3,2,10],[27,4,3,2,9],[-1,4,1
,2,10],[24,2,3,2,10],[9,4,3,1,9],[18,4,2,2,9],[90,3,3,2,9],[3,1,3,2,10
],[-3,4,2,1,9],[-1,4,1,2,9],[3,3,3,1,9],[28,3,2,2,9],[81,2,3,2,9],[20,
4,2,2,8],[-1,3,2,1,9],[14,2,2,2,9],[27,1,3,2,9],[-2,4,2,1,8],[-8,4,1,2
,8],[32,3,2,2,8],[2,1,2,2,9],[3,0,3,2,9],[1,4,0,2,8],[-1,3,2,1,8],[40,
2,2,2,8],[16,1,2,2,8],[-19,3,2,1,7],[2,0,2,2,8],[6,3,1,1,7],[-15,2,2,1
,7],[-6,3,2,1,6],[5,2,1,1,7],[-3,1,2,1,7],[-1,3,2,0,6],[4,3,1,1,6],[-
27,2,2,1,6],[1,1,1,1,7],[13,2,1,1,6],[-18,1,2,1,6],[7,1,1,1,6],[-3,0,2
,1,6],[17,2,1,1,5],[1,0,1,1,6],[-3,2,0,1,5],[9,1,1,1,5],[8,2,1,1,4],[-
1,1,0,1,5],[1,0,1,1,5],[2,2,1,0,4],[-2,2,0,1,4],[16,1,1,1,4],[1,1,1,0,
4],[-4,1,0,1,4],[4,0,1,1,4],[-1,0,0,1,4],[3,1,1,0,3],[1,0,1,0,3],[-1,1
,0,0,2],[-1,0,0,0,1],[1,0,0,0,0]]
\end{verbatim}
\normalsize

Initial form $init_r$ (\ref{3vertex_init_r}):

\begin{equation}
    init_r = {a_{{0}}}^{6}{a_{{1}}}^{12}{b_{{1}}}^{6}{c_{{0}}}^{4}c_{{2}}{d_{{0}}}^
{3}d_2 \left( {a_{{0}}}^{6}{b_{{1}}}^{6}{c_{{2}}}^{3}{d_{{2}}}^{2}+{a_{{1
}}}^{6}{b_{{0}}}^{6}{c_{{1}}}^{3}{d_{{1}}}^{2} \right)
\end{equation}

After specialization:

\begin{equation}
    init_r = {y}^{6}x_{{2}}x_{{3}} \left( -{y}^{6}{x_{{2}}}^{3}{x_{{3}}}^{2}+1
 \right) 
\end{equation}

\begin{figure}[h!]
    \centering
    \includegraphics[scale=0.4]{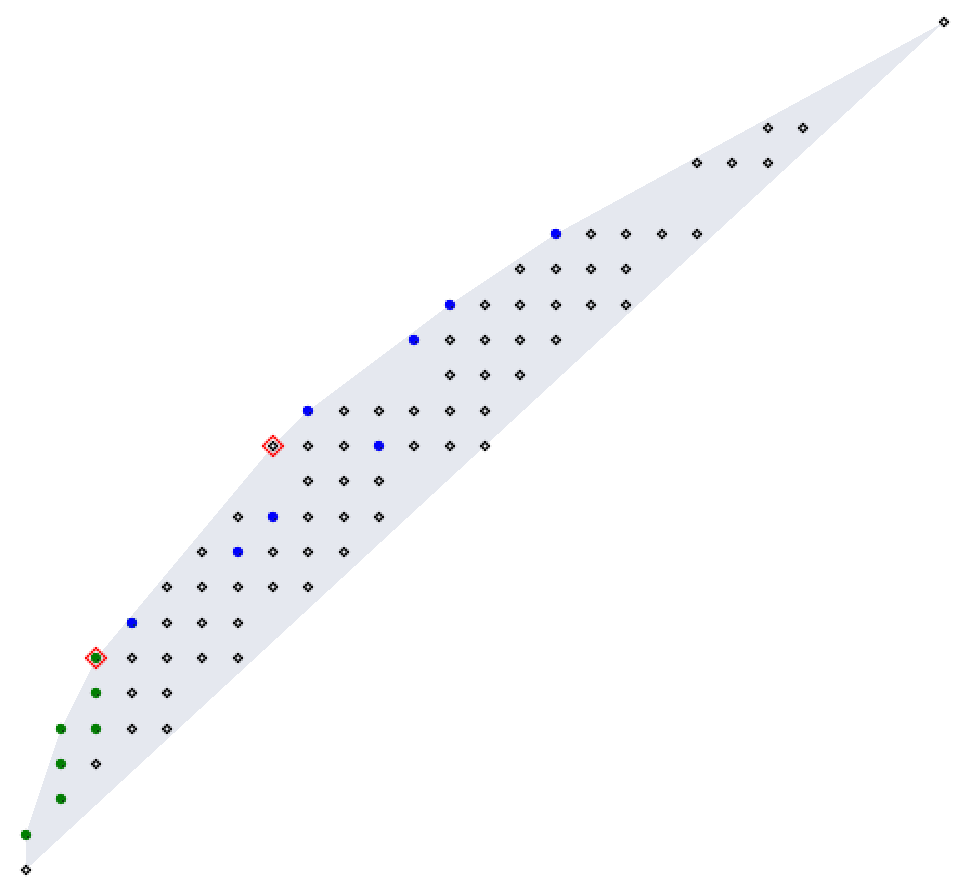}
    \caption{Newton polygon and the initial form $init_r$ (red nodes) for $\mathrm{diag}(2,3,4)$ quiver}
    \label{fig:diag244}
\end{figure}

\newpage

\subsection{Non-diagonal quivers}\label{append_a_2}

Here we provide some examples of quivers with non-diagonal matrix
$C$ and polynomials $A(x_1,\dots,x_m,y)$, satisfying the K-theoretic condition.

    \begin{gather*} a)\
         \left[ \begin {array}{ccc} 1&1&1\\ \noalign{\medskip}1&0&0
\\ \noalign{\medskip}1&0&0\end {array} \right], 
\\ {y}^{3}{x_{{1}}}^{3}x_{{2}}x_{{3}}+3\,{y}^{2}{x_{{1}}}^{2}x_{{2}}x_{{3
}}-{y}^{2}{x_{{1}}}^{2}x_{{2}}-{y}^{2}{x_{{1}}}^{2}x_{{3}}+3\,yx_{{1}}
x_{{2}}x_{{3}} \\
 -2\,yx_{{1}}x_{{2}}-2\,yx_{{1}}x_{{3}}+yx_{{1}}+x_{{2}}x_{{3}}-y-x_{{2
}}-x_{{3}}+1.
    \end{gather*}

\begin{gather*} b)\
    \left[ \begin {array}{ccc} 1&1&1\\ \noalign{\medskip}1&1&0
\\ \noalign{\medskip}1&0&0\end {array} \right], \\
{y}^{2}{x_{{1}}}^{2}x_{{3}}-{y}^{2}x_{{1}}x_{{2}}+2\,yx_{{1}}x_{{3}}-y
x_{{1}}-yx_{{2}}+y+x_{{3}}-1.
\end{gather*}

\begin{gather*} c)\
     \left[ \begin {array}{ccc} 2&1&1\\ \noalign{\medskip}1&1&0
\\ \noalign{\medskip}1&0&0\end {array} \right], \\
{y}^{3}{x_{{1}}}^{2}-{y}^{2}x_{{1}}x_{{2}}+2\,{y}^{2}x_{{1}}-yx_{{1}}-
yx_{{2}}+y+x_{{3}}-1.
\end{gather*}

\begin{gather*} d)\
    \left[ \begin {array}{ccc} 2&2&1\\ \noalign{\medskip}2&2&1
\\ \noalign{\medskip}1&1&0\end {array} \right], \\
{y}^{6}{x_{{1}}}^{2}{x_{{2}}}^{2}-{y}^{4}{x_{{1}}}^{2}x_{{2}}-{y}^{4}x
_{{1}}{x_{{2}}}^{2}+{y}^{3}{x_{{1}}}^{2}x_{{3}}+{y}^{3}{x_{{2}}}^{2}x_
{{3}}-{y}^{3}x_{{1}}x_{{2}}+{y}^{2}x_{{1}}x_{{2}} \\
+2\,{y}^{2}x_{{1}}x_{{3}}+2\,{y}^{2}x_{{2}}x_{{3}}-yx_{{1}}x_{{3}}-yx_{
{2}}x_{{3}}+yx_{{3}}+{x_{{3}}}^{2}-x_{{3}}.
\end{gather*}

\begin{gather*} e)\
    \left[ \begin {array}{ccc} 3&2&1\\ \noalign{\medskip}2&2&1
\\ \noalign{\medskip}1&1&0\end {array} \right], \\
{y}^{4}{x_{{1}}}^{2}+{y}^{3}{x_{{2}}}^{2}x_{{3}}-{y}^{3}x_{{1}}x_{{2}}
-2\,{y}^{2}x_{{1}}x_{{3}}+2\,{y}^{2}x_{{2}}x_{{3}}-{y}^{2}x_{{1}} \\
-yx_{{2}}x_{{3}}+yx_{{1}}+yx_{{3}}+{x_{{3}}}^{2}-x_{{3}}.
\end{gather*}

\newpage

\begin{figure}[h!]
\input{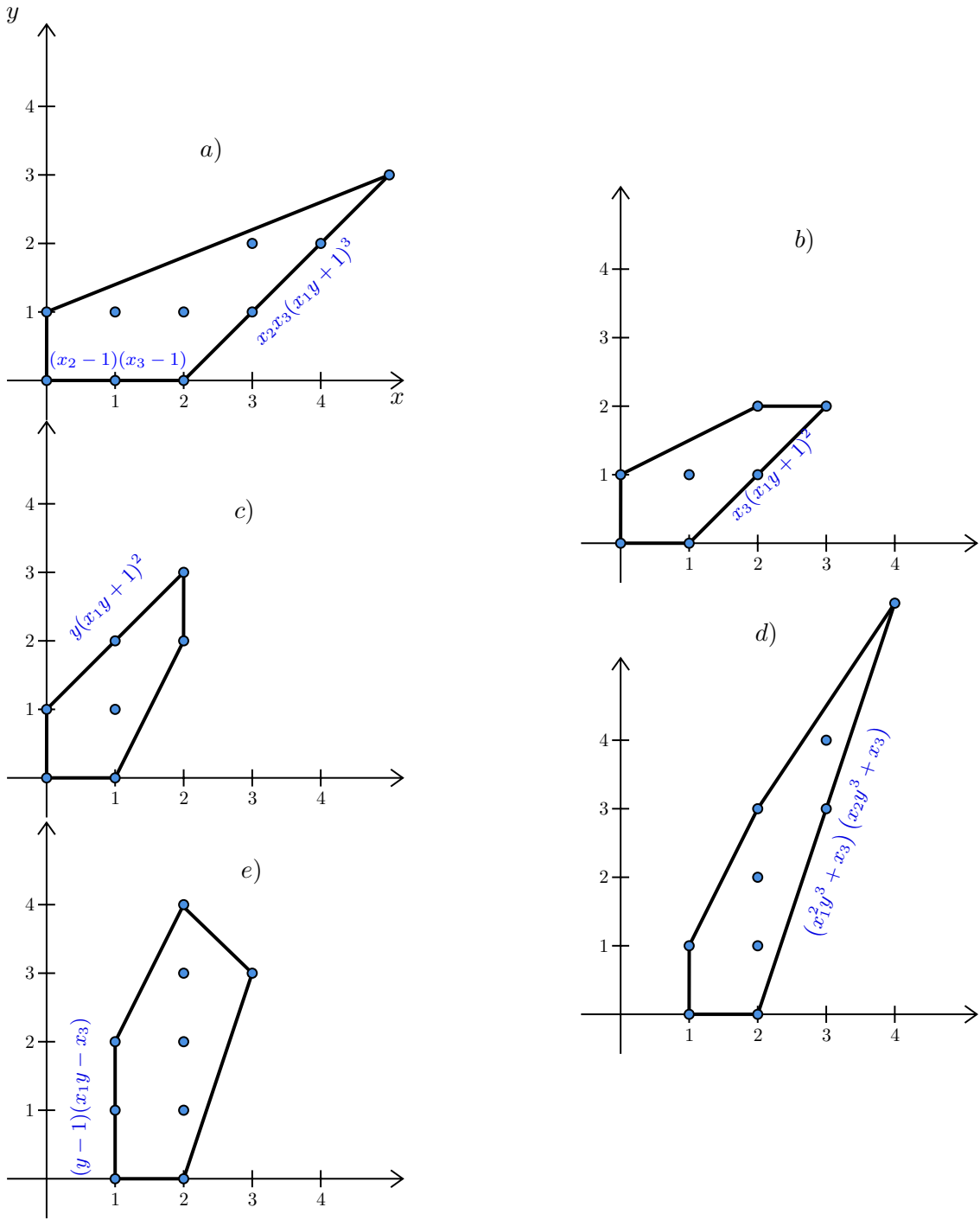}
\caption{Newton polygons and face polynomials for the examples $a),b),c),d),e)$}
\label{fig:examples_a}
\end{figure}

\newpage

\begin{gather*}
    \left[ \begin {array}{cccc} 1&1&1&1\\ \noalign{\medskip}1&0&0&0
\\ \noalign{\medskip}1&0&0&0\\ \noalign{\medskip}1&0&0&0\end {array}
 \right], \\
{y}^{4}{x_{{1}}}^{4}x_{{2}}x_{{3}}x_{{4}}+4\,{y}^{3}{x_{{1}}}^{3}x_{{2
}}x_{{3}}x_{{4}}-{y}^{3}{x_{{1}}}^{3}x_{{2}}x_{{3}}-{y}^{3}{x_{{1}}}^{
3}x_{{2}}x_{{4}}-{y}^{3}{x_{{1}}}^{3}x_{{3}}x_{{4}}+6\,{y}^{2}{x_{{1}}
}^{2}x_{{2}}x_{{3}}x_{{4}}
 \\
 -3\,{y}^{2}{x_{{1}}}^{2}x_{{2}}x_{{3}}-3\,{y}^{2}{x_{{1}}}^{2}x_{{2}}x
_{{4}}-3\,{y}^{2}{x_{{1}}}^{2}x_{{3}}x_{{4}}+{y}^{2}{x_{{1}}}^{2}x_{{2
}}+{y}^{2}{x_{{1}}}^{2}x_{{3}}+{y}^{2}{x_{{1}}}^{2}x_{{4}} \\
+4\,yx_{{1}}x_{{2}}x_{{3}}x_{{4}}-3\,yx_{{1}}x_{{2}}x_{{3}}-3\,yx_{{1}}
x_{{2}}x_{{4}}-3\,yx_{{1}}x_{{3}}x_{{4}}+2\,yx_{{1}}x_{{2}}+2\,yx_{{1}
}x_{{3}} \\
+2\,yx_{{1}}x_{{4}}+x_{{2}}x_{{3}}x_{{4}}-yx_{{1}}-x_{{2}}x_{{3}}-x_{{2
}}x_{{4}}-x_{{3}}x_{{4}}+y+x_{{2}}+x_{{3}}+x_{{4}}-1.
\end{gather*}

\begin{gather*}
    \left[ \begin {array}{cccc} 1&1&1&1\\ \noalign{\medskip}1&1&1&1
\\ \noalign{\medskip}1&1&0&0\\ \noalign{\medskip}1&1&0&0\end {array}
 \right], \\
{y}^{6}{x_{{1}}}^{3}{x_{{2}}}^{3}x_{{3}}x_{{4}}+3\,{y}^{5}{x_{{1}}}^{3
}{x_{{2}}}^{2}x_{{3}}x_{{4}}+3\,{y}^{5}{x_{{1}}}^{2}{x_{{2}}}^{3}x_{{3
}}x_{{4}}+3\,{y}^{4}{x_{{1}}}^{3}x_{{2}}x_{{3}}x_{{4}}+9\,{y}^{4}{x_{{
1}}}^{2}{x_{{2}}}^{2}x_{{3}}x_{{4}} \\
+ 3\,{y}^{4}x_{{1}}{x_{{2}}}^{3}x_{{3}}x_{{4}}-{y}^{4}{x_{{1}}}^{2}{x_{{
2}}}^{2}x_{{3}}-{y}^{4}{x_{{1}}}^{2}{x_{{2}}}^{2}x_{{4}}+{y}^{3}{x_{{1
}}}^{3}x_{{3}}x_{{4}}+9\,{y}^{3}{x_{{1}}}^{2}x_{{2}}x_{{3}}x_{{4}}+9\,
{y}^{3}x_{{1}}{x_{{2}}}^{2}x_{{3}}x_{{4}} \\
+ {y}^{3}{x_{{2}}}^{3}x_{{3}}x_{{4}}-2\,{y}^{3}{x_{{1}}}^{2}x_{{2}}x_{{3
}}-2\,{y}^{3}{x_{{1}}}^{2}x_{{2}}x_{{4}}-2\,{y}^{3}x_{{1}}{x_{{2}}}^{2
}x_{{3}}-2\,{y}^{3}x_{{1}}{x_{{2}}}^{2}x_{{4}}+3\,{y}^{2}{x_{{1}}}^{2}
x_{{3}}x_{{4}} \\
+9\,{y}^{2}x_{{1}}x_{{2}}x_{{3}}x_{{4}}+3\,{y}^{2}{x_{{2}}}^{2}x_{{3}}x
_{{4}}-{y}^{2}{x_{{1}}}^{2}x_{{3}}-{y}^{2}{x_{{1}}}^{2}x_{{4}}-4\,{y}^
{2}x_{{1}}x_{{2}}x_{{3}}-4\,{y}^{2}x_{{1}}x_{{2}}x_{{4}}-{y}^{2}{x_{{2
}}}^{2}x_{{3}} \\
-{y}^{2}{x_{{2}}}^{2}x_{{4}}+{y}^{2}x_{{1}}x_{{2}}+3\,yx_{{1}}x_{{3}}x
_{{4}}+3\,yx_{{2}}x_{{3}}x_{{4}}-2\,yx_{{1}}x_{{3}}-2\,yx_{{1}}x_{{4}}
-2\,yx_{{2}}x_{{3}} \\
-2\,yx_{{2}}x_{{4}}+yx_{{1}}+yx_{{2}}+x_{{3}}x_{{4}}-y-x_{{3}}-x_{{4}}
+1.
\end{gather*}

\begin{gather*}
    \left[ \begin {array}{cccc} 2&2&2&2\\ \noalign{\medskip}2&2&1&1
\\ \noalign{\medskip}2&1&0&0\\ \noalign{\medskip}2&1&0&0\end {array}
 \right], \\
 {y}^{10}{x_{{1}}}^{5}x_{{3}}x_{{4}}-{y}^{10}{x_{{1}}}^{3}{x_{{2}}}^{3}
+{y}^{9}{x_{{1}}}^{4}x_{{2}}x_{{3}}+{y}^{9}{x_{{1}}}^{4}x_{{2}}x_{{4}}
-5\,{y}^{8}{x_{{1}}}^{4}x_{{3}}x_{{4}}+{y}^{8}{x_{{1}}}^{3}{x_{{2}}}^{
2}+3\,{y}^{8}{x_{{1}}}^{2}{x_{{2}}}^{3} \\
-4\,{y}^{7}{x_{{1}}}^{3}x_{{2}}x_{{3}}-4\,{y}^{7}{x_{{1}}}^{3}x_{{2}}x
_{{4}}+3\,{y}^{7}{x_{{1}}}^{2}{x_{{2}}}^{2}+10\,{y}^{6}{x_{{1}}}^{3}x_
{{3}}x_{{4}}-{y}^{6}{x_{{1}}}^{3}x_{{3}}-{y}^{6}{x_{{1}}}^{3}x_{{4}}-3
\,{y}^{6}{x_{{1}}}^{2}{x_{{2}}}^{2} \\
-3\,{y}^{6}x_{{1}}{x_{{2}}}^{3}+6\,{y}^{5}{x_{{1}}}^{2}x_{{2}}x_{{3}}+
6\,{y}^{5}{x_{{1}}}^{2}x_{{2}}x_{{4}}-2\,{y}^{5}{x_{{1}}}^{2}x_{{2}}-6
\,{y}^{5}x_{{1}}{x_{{2}}}^{2}-10\,{y}^{4}{x_{{1}}}^{2}x_{{3}}x_{{4}}+3
\,{y}^{4}{x_{{1}}}^{2}x_{{3}} \\
+3\,{y}^{4}{x_{{1}}}^{2}x_{{4}}+3\,{y}^{4}x_{{1}}{x_{{2}}}^{2}+{y}^{4}{
x_{{2}}}^{3}-3\,{y}^{4}x_{{1}}x_{{2}}-4\,{y}^{3}x_{{1}}x_{{2}}x_{{3}}-
4\,{y}^{3}x_{{1}}x_{{2}}x_{{4}}+4\,{y}^{3}x_{{1}}x_{{2}}+3\,{y}^{3}{x_
{{2}}}^{2} \\
+5\,{y}^{2}x_{{1}}x_{{3}}x_{{4}}-3\,{y}^{2}x_{{1}}x_{{3}}-3\,{y}^{2}x_{
{1}}x_{{4}}-{y}^{2}{x_{{2}}}^{2}+{y}^{2}x_{{1}}+3\,{y}^{2}x_{{2}}+yx_{
{2}}x_{{3}} \\
+yx_{{2}}x_{{4}}-2\,yx_{{2}}-x_{{3}}x_{{4}}+y+x_{{3}}+x_{{4}}-1.
\end{gather*}

\begin{figure}[h!]
\input{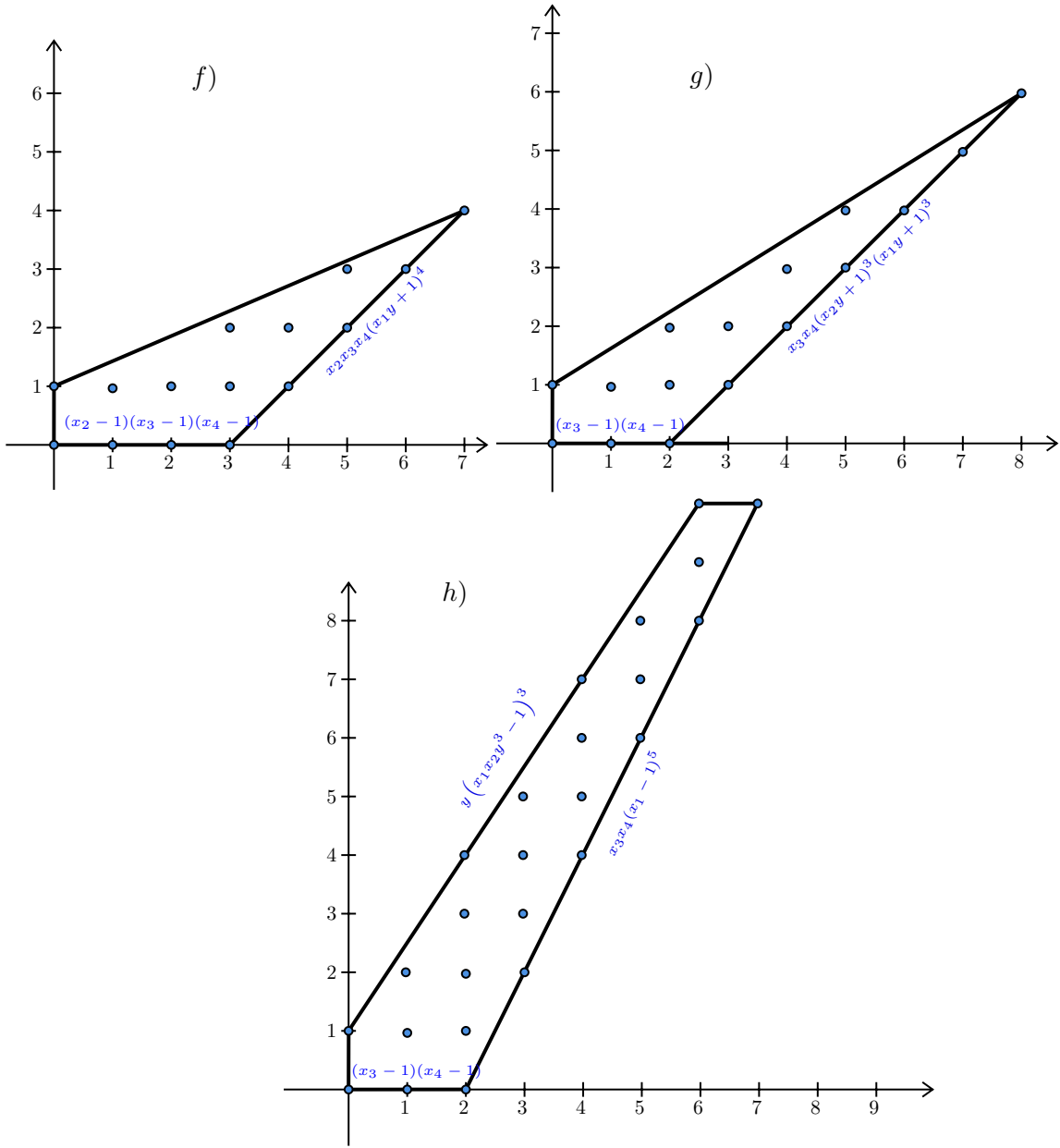}
\caption{Newton polygons and face polynomials for the examples $f),g),h)$}
\label{fig:examples_b}
\end{figure}
On Figures \ref{fig:examples_a} and \ref{fig:examples_b} only those face polynomials are shown (even more, their preimages before setting $x_i=a_ix$), which consist of three or more monomials. For every binomial we can re-label its nodes using the parameter $\tau$ along the edge, so that it would give a factor $(\tau+1)$. Therefore, any product of binomials maps into $(\tau+1)^k$, where the integer $k\geq 1$ varies from one edge to another. According to this, all roots of all face polynomials are equal to 1 in these examples, which means that $A(x,y)$ is tempered.

\newpage

\section{Canny-Emiris matrix for the diagonal quiver}\label{append_b} 

We construct the square matrix of size $\left(\prod_{i=1}^m \alpha_i + \sum_{j'=1}^m \prod_{j\neq j'} \alpha_j\right)$, which determinant equals to the sparse mixed resultant $\mathcal{R}_{\mathbf{A}}$ for the diagonal quiver with $C = \mathrm{diag}(\alpha_1,\dots,\alpha_m)$. Note that, however, this method suits for any set of supports $\mathbf{A}$, and has been introduced by J. Canny and I. Emiris in \cite{CE} and further studied in \cite{Stu,AJS}.

In fact, for every vertex of $N(\mathcal{R}_{\mathbf{A}})$ there exist a version of such matrix, where the letters from the corresponding extremal monomial sit on its main diagonal (matrices are different, but having the same determinant, up to redundant monomial prefactors, which we usually ignore). Therefore, we may say that the matrices are labelled by those TCMDs, which are attached to the vertices of the Newton polytope.
Adapting the simplest scenario, we choose the following
\begin{equation}\label{canny_diag_TCMD}
\mathrm{TCMD}: \qquad 
\begin{tabular}{|c c c|}
  $\bullet$ &  & \\ \hline 
  $\bullet$ & & $\bullet$ \\
  $\bullet$ & & $\bullet$ \\      
  \vdots & \vdots & \vdots \\
  $\bullet$ & & $\bullet$      
\end{tabular}
\rightarrow a_0^{\prod\alpha_i},\quad
\begin{tabular}{|c c c|}
  $\bullet$ & $\bullet$ & \\ \hline 
  & & $\bullet$ \\
  $\bullet$ & & $\bullet$ \\      
  \vdots & \vdots & \vdots \\
  $\bullet$ & & $\bullet$      
\end{tabular}_{(\pi)}
\rightarrow b_{j',2}^{\prod_{j\neq j'}\alpha_j}
\end{equation}
(recall that the $i$-th row in a perfogram stands for $F_i$ in (\ref{Nahm_eqs})), and denote $\mathcal{M}$ the corresponding matrix, such that $|\mathcal{M}| = \mathcal{R}_{\mathbf{A}}$.
For diagonal quivers after specialization (\ref{quiver_A_poly_def}), it would have the letters $x_1,\dots,x_m,y$ only on the main diagonal, whereas the off-diagonal entries will be equal to $0$ or $1$. Therefore, the choice (\ref{canny_diag_TCMD}) seems to be quite interesting (still, this is the property of diagonal quivers). We want to emphasise that the advantage of this method is that it can immediately extended to any dimension $m$, and the structure of the matrix would be somewhat similar and nicely structured. 

Denote by $Q_{\delta}$ the translation of $Q = \sum_{i=0}^m\mathrm{conv}(\mathrm{supp}(F_i))$ by some integer vector $\overline{\delta} = (\delta,\dots,\delta)$. It is chosen such that the number of integer lattice points of the intersection $Q_{\delta} \cap \mathbb{Z}^m$ is minimally possible. The entries of $\mathcal{M}$ are attached to pairs $(p,p')\in Q_{\delta} \cap \mathbb{Z}^m$. We have a decomposition induced by (\ref{canny_diag_TCMD}):
\begin{equation}
    Q_{\delta} \cap \mathbb{Z}^m = \mathcal{E}_{a_0} \cup \mathcal{E}_{b_{1,2}} \cup \dots \cup \mathcal{E}_{b_{m,2}}
\end{equation}
where each $\mathcal{E}_{k}$ is associated to a cell in (\ref{canny_diag_TCMD}).
Each entry is labelled by a pair of lattice points $p=(i,j),p'=(i',j')$, and is calculated by the following rule:
\begin{equation}\label{canny-emiris_def}
    \mathcal{M}_{p,p'} := \mathrm{coeff}(z_1^{i}z_2^{j}F_s(z_1,z_2),z_1^{i'}z_2^{j'})
\end{equation}
where $s$ is an indicator, depending on whether $p$ belongs to $\mathcal{E}_{a_0}$ ($s = 0$), or to $\mathcal{E}_{b_{i,2}}$ ($s = i$). Therefore, $\mathcal{M}$ has a block structure, where each block represents ``interaction'' of the cells in a chosen TCMD. Below we proceed with the explicit construction of $\mathcal{M}$ for the two-vertex quiver.

\subsection*{The case $m = 2$} 
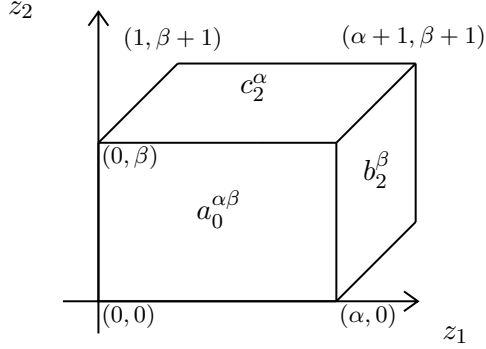
\begin{figure}[h!]
    \centering
    \tikzset{every picture/.style={line width=0.75pt}} 

\begin{tikzpicture}[x=0.75pt,y=0.75pt,yscale=-1,xscale=1]

\draw    (449,5066) -- (569,5066) ;
\draw    (569,5066) -- (609,5026) ;
\draw    (449,5066) -- (449,4986) ;
\draw    (609,5026) -- (609,4946) ;
\draw    (449,4986) -- (489,4946) ;
\draw    (489,4946) -- (609,4946) ;
\draw    (449,4986) -- (529,4986) ;
\draw    (569,4986) -- (529,4986) ;
\draw    (569,5066) -- (569,4986) ;
\draw    (569,4986) -- (609,4946) ;
\draw  (431.1,5066) -- (610.1,5066)(449,4919.9) -- (449,5082.23) (603.1,5061) -- (610.1,5066) -- (603.1,5071) (444,4926.9) -- (449,4919.9) -- (454,4926.9)  ;

\draw (497,5008.67) node [anchor=north west][inner sep=0.75pt]  [color={rgb, 255:red, 0; green, 0; blue, 0 }  ,opacity=1 ]  {$a^{\alpha \beta }_{0}$};
\draw (581,4988.67) node [anchor=north west][inner sep=0.75pt]  [color={rgb, 255:red, 0; green, 0; blue, 0 }  ,opacity=1 ]  {$b^{\beta }_{2}$};
\draw (519,4949.67) node [anchor=north west][inner sep=0.75pt]  [color={rgb, 255:red, 0; green, 0; blue, 0 }  ,opacity=1 ]  {$c^{\alpha }_{2}$};
\draw (449,5066) node [anchor=north west][inner sep=0.75pt]  [font=\footnotesize]  {$( 0,0)$};
\draw (569,5066) node [anchor=north west][inner sep=0.75pt]  [font=\footnotesize]  {$( \alpha ,0)$};
\draw (449,4986) node [anchor=north west][inner sep=0.75pt]  [font=\footnotesize]  {$( 0,\beta )$};
\draw (460,4926) node [anchor=north west][inner sep=0.75pt]  [font=\footnotesize]  {$( 1,\beta +1)$};
\draw (570,4925) node [anchor=north west][inner sep=0.75pt]  [font=\footnotesize]  {$( \alpha +1,\beta +1)$};
\draw (621,5077.33) node [anchor=north west][inner sep=0.75pt]    {$z_{1}$};
\draw (402,4913.33) node [anchor=north west][inner sep=0.75pt]    {$z_{2}$};

\end{tikzpicture}
    \caption{The TCMD (\ref{TCMD_ce_m2}) divides $Q$ into the three cells: $a_0^{\alpha\beta}$, $c_2^{\alpha}$ and $b_2^{\beta}$, producing the extreme monomial $a_0^{\alpha\beta}b_2^{\beta}c_2^{\alpha}$}
    \label{fig:bitmap}
\end{figure}

The Nahm equations for $C = \mathrm{diag}(\alpha,\beta),\ \alpha,\beta \geq 2$:
\begin{equation}
\begin{tabular}{l l}
  $F_0$ = & $a_0+a_1z_1z_2$ \\
  \hline
  $F_1$ = & $b_0+b_1z_1+b_2z_1^{\alpha}$ \\      
  $F_2$ = & $c_0+c_1z_2+c_2z_2^{\beta}$
\end{tabular}
\end{equation}
The TCMD (\ref{canny_diag_TCMD}) is given by the three 2-dimensional cells:
\begin{equation}\label{TCMD_ce_m2}
\begin{tabular}{|c c c|}
  $\bullet$ &  & \\ \hline 
  $\bullet$ & & $\bullet$ \\
  $\bullet$ & & $\bullet$      
\end{tabular}
\times
\begin{tabular}{|c c c|}
  $\bullet$ & $\bullet$ & \\ \hline 
   & & $\bullet$ \\
  $\bullet$ & & $\bullet$      
\end{tabular}
\times
\begin{tabular}{|c c c|}
  $\bullet$ & $\bullet$ & \\ \hline 
  $\bullet$ & & $\bullet$ \\
  & & $\bullet$      
\end{tabular}
\end{equation}
for $a_0,b_2:=b_{1,2}$ and $c_2:=b_{2,2}$, correspondingly (Figure \ref{fig:bitmap}).
Therefore, after shifting by $\overline{\delta}$ equal to, say, $(\frac{1}{3},\frac{1}{3})$, the number of integer lattice points in each block will be $\alpha\beta$, $\beta$ and $\alpha$, for $a_0$, $b_2$ and $c_2$, correspondingly, which is indeed minimal. The matrix $\mathcal{M}$ is then a block matrix of size $\alpha\beta+\alpha+\beta$:
\begin{equation}\label{canny_diag_matrix}
\mathcal{M} = 
\begin{pmatrix}
    \mathcal{E}_{a_0} \times \mathcal{E}_{a_0} & \mathcal{E}_{a_0} \times \mathcal{E}_{b_0} & \mathcal{E}_{a_0}\times \mathcal{E}_{c_0} \\
         \mathcal{E}_{b_2} \times \mathcal{E}_{a_0} & \mathcal{E}_{b_2} \times \mathcal{E}_{b_2} & \mathcal{E}_{b_2}\times \mathcal{E}_{c_2} \\
              \mathcal{E}_{c_2} \times \mathcal{E}_{a_0} & \mathcal{E}_{c_2} \times \mathcal{E}_{b_2} & \mathcal{E}_{c_2}\times \mathcal{E}_{c_2} \\
\end{pmatrix}
\end{equation}
For a better presentation we need to enumerate the 2d lattice points with a single index. We shall get $\mathcal{M}_{k,l}$, where $k,l$ are integers labelling the pair of points  $(p,p')$, and each $p=(i,j),p'=(i',j')$ is given by its $\mathbb{Z}^2$-coordinates. For $m = 2$, we can act as follows: from left to right, down to top; $\mathcal{E}_{a_0}$ goes first, then $\mathcal{E}_{b_2}$ and $\mathcal{E}_{c_2}$. We know that the main diagonal shall consist of several letters $a_0$ for $\mathcal{E}_{a_0} \times \mathcal{E}_{a_0}$, $b_2$ for $\mathcal{E}_{b_2} \times \mathcal{E}_{b_2}$ and $c_2$ for $\mathcal{E}_{c_2} \times \mathcal{E}_{c_2}$.
Calculating the ``interactions'' by the formula (\ref{canny-emiris_def}) leads to the following result for all non-diagonal entries:
\begin{equation}
\begin{aligned}
\mathcal{E}_{a_0} \times \mathcal{E}_{a_0}: &\quad
\mathcal{M}_{k=1\dots\alpha,l=1\dots\beta} & = & 
\begin{cases}
               a_1;\ l = k + \alpha +1,\ k \Mod \alpha\neq 0\\
               0
\end{cases}
\\
\mathcal{E}_{a_0} \times \mathcal{E}_{b_2}: &\quad
\mathcal{M}_{\substack{k=1\dots\alpha\beta,\\ l=\alpha\beta+1\dots\alpha\beta+\beta}} & = &
\begin{cases}
               a_1;\ l = \alpha\beta+\frac{k}{\alpha},\ k \Mod \alpha = 0 \\
               0
\end{cases}
\\
\mathcal{E}_{a_0} \times \mathcal{E}_{c_2}: &\quad 
\mathcal{M}_{\substack{k=1\dots\alpha\beta, \\ l=\alpha\beta+\beta+1\dots\alpha\beta+\alpha+\beta}} & = &
\begin{cases}
               a_1;\ k\geq \alpha\beta-\alpha+1, l = k+\alpha+\beta+1 \\
               0
\end{cases}
\\
\mathcal{E}_{b_2} \times \mathcal{E}_{a_0}: &\quad
\mathcal{M}_{\substack{k=\alpha\beta+1\dots \alpha\beta+\beta,\\ l=1\dots\alpha\beta}} & = & 
\begin{cases}
               b_0;\ l = (k-\alpha\beta)\alpha + 1 \\
               b_1;\ l = (k-\alpha\beta)\alpha + 2 \\
               0
\end{cases}
\\
\mathcal{E}_{b_2} \times \mathcal{E}_{c_2}: &\quad
\mathcal{M}_{\alpha\beta+\beta,\alpha\beta+\beta+1} = b_0, \\ &\
\mathcal{M}_{\alpha\beta+\beta,\alpha\beta+\beta+2} = b_1
\\
\mathcal{E}_{c_2} \times \mathcal{E}_{a_0}: &\quad
\mathcal{M}_{\substack{k=\alpha\beta+\beta+1\dots \alpha\beta+\alpha+\beta,\\ l=1\dots\alpha\beta}} & = & 
\begin{cases}
               c_0;\ l = k-\alpha\beta-\beta \\
               c_1;\ l = k-\alpha\beta-\beta+\alpha \\
               0
\end{cases}
\\
\mathcal{E}_{c_2} \times \mathcal{E}_{b_2}: &\quad
\mathcal{M}_{\substack{k=\alpha\beta+\beta+1 \dots \alpha\beta+\alpha+\beta,\\ l=\alpha\beta+1 \dots \alpha\beta+\beta}} & = & 
\begin{cases}
               c_1;\ k = \alpha\beta+\alpha+\beta+1, l = \alpha\beta+1 \\
               0
\end{cases}
\end{aligned}
\end{equation}
All other entries are zero, in particular, the off-diagonal blocks $\mathcal{E}_{b_2} \times \mathcal{E}_{b_2}$ and $\mathcal{E}_{c_2} \times \mathcal{E}_{c_2}$ all vanish. For example, $(\alpha,\beta) = (2,2)$ gives
\begin{equation}
\begin{aligned}
\mathcal{R}_{\mathbf{A}} = & \left| \begin {array}{cccccccc} a_{0}&&&a_{{1}}&&&&
\\ \noalign{\medskip}&a_{0}&&&a_{{1}}&&&\\ \noalign{\medskip}
&&a_{0}&&&&&a_{{1}}\\ \noalign{\medskip}&&&a_{0}&&a_{{1}}
&&\\ \noalign{\medskip}&&b_{0}&b_{{1}}&b_{{2}}&&&
\\ \noalign{\medskip}&&&&&b_{{2}}&b_{0}&b_{{1}}
\\ \noalign{\medskip}c_{0}&&c_{{1}}&&&&c_{{2}}&
\\ \noalign{\medskip}&c_{0}&&c_{{1}}&&&&c_{{2}}\end {array}
 \right|
\end{aligned}
\end{equation}
(unoccupied entries are zeroes), which reproduces the expression (\ref{diag22_resultant_unspec}) from Section \ref{section_two_dimensional}.


\newpage

\bibliographystyle{JHEP}
\bibliography{refs}

\end{document}